\documentclass[12pt]{article} %

\usepackage{PRIMEarxiv}

\usepackage{amsfonts,amsmath,amssymb,authblk,fullpage,graphicx,tabularx,threeparttable,tikz,pgfplots}

\usepackage[utf8]{inputenc}  
\usepackage[T1]{fontenc}     
\usepackage{hyperref}       
\usepackage{url}             
\usepackage{booktabs}        
\usepackage{amsfonts}        
\usepackage{nicefrac}        
\usepackage{microtype}      
\usepackage{lipsum}
\usepackage{graphicx}
\usepackage{bm}
\usepackage{amsmath}
\usepackage{amsmath}
\usepackage{enumerate}
\usepackage{subcaption}
\usepackage{epstopdf}
\usepackage{ragged2e}
\usepackage{grffile}
\usepackage[capitalise, noabbrev]{cleveref}
\usepackage{caption}
\usepackage{amsthm}
\usepackage{adjustbox}
\usepackage{pgfplots}
\usetikzlibrary{calc}
\usepackage{tikz-dimline}
\usetikzlibrary{patterns}
\pgfplotsset{compat = newest}
\usepackage[symbol]{footmisc}

\newtheorem{definition}{Definition}[subsection] 
\newtheorem{remark}[definition]{Remark}

\graphicspath{{media/}}

\newcommand{\magn}[1]{\parallel \bm{#1} \parallel}

\newcommand{\cosIA}{$\cos[\sphericalangle (\hat{\bm{n}}^1_1,\hat{\bm{n}}^1_p)]$}
\newcommand{\Int}{\mathrm{Int}}

\title{Structure-Preserving Invariant Interpolation Schemes for Invertible Second-Order Tensors}

\author{ Abhiroop Satheesh, Christoph P. Schmidt, Wolfgang A. Wall, Christoph Meier \\
Institute for Computational Mechanics  \\
Technical University of Munich  \\
Garching, Germany, 85748\\
\texttt{\{abhiroop.satheesh, christoph.schmidt, wolfgang.a.wall, christoph.anton.meier\}@tum.de}} 

\begin{document}
\maketitle

\begin{abstract}
  Tensor interpolation is an essential step for tensor data analysis in various fields of application and scientific disciplines. In the present work, novel interpolation schemes for general, i.e., symmetric or non-symmetric, invertible square tensors are proposed. Critically, the proposed schemes rely on a combined polar and spectral decomposition of the tensor data $\bm{T}\!\!=\!\!\bm{R}\bm{Q}^T \!\! \bm{\Lambda} \bm{Q}$, followed by an individual interpolation of the two rotation tensors~$\bm{R}$ and~$\bm{Q}$ and the positive definite diagonal eigenvalue tensor~$\bm{\Lambda}$ resulting from this decomposition. Two different schemes are considered for a consistent rotation interpolation within the special orthogonal group $\mathbb{SO}(3)$, either based on relative rotation vectors or quaternions. For eigenvalue interpolation, three different schemes, either based on the logarithmic weighted average, moving least squares or logarithmic moving least squares, are considered. It is demonstrated that the proposed interpolation procedure preserves the structure of a tensor, i.e.,~$\bm{R}$ and~$\bm{Q}$ remain orthogonal tensors and~$\bm{\Lambda}$ remains a positive definite diagonal tensor during interpolation, as well as scaling and rotational invariance (objectivity). Based on selected numerical examples considering the interpolation of either symmetric or non-symmetric tensors, the proposed schemes are compared to existing approaches such as Euclidean, Log-Euclidean, Cholesky and Log-Cholesky interpolation. In contrast to these existing methods, the proposed interpolation schemes result in smooth and monotonic evolutions of tensor invariants such as determinant, trace, fractional anisotropy (FA), and Hilbert's anisotropy (HA). Moreover, a consistent spatial convergence behavior is confirmed for first- and second-order realizations of the proposed schemes. The present work is mainly motivated by the frequently occurring necessity for remeshing or mesh adaptivity when applying the finite element method to complex problems of nonlinear continuum mechanics with inelastic constitutive behavior, which requires the consistent interpolation of tensor-valued history data for the transfer between different meshes. However, the proposed schemes are very general in nature and suitable for the interpolation of general invertible second-order square tensors independent of the specific application.
\end{abstract}

\keywords{tensor interpolation \and invertible tensors \and symmetric and non-symmetric tensors \and polar decomposition \and spectral decomposition \and large rotations}
\section{Introduction}
Tensor interpolation is an essential step for tensor data analysis in various fields of application and scientific disciplines, e.g., in medicine, computer vision, general physics, or continuum
mechanics. For example, diffusion tensors are second-order symmetric positive definite tensors, which describe
anisotropic diffusion behavior, e.g., visualized by diffusion tensor imaging (DTI) in the field of medicine. In
continuum mechanics, the deformation gradient is a second-order non-symmetric invertible tensor, whose polar decomposition in a rotation tensor and a symmetric positive definite tensor describes the local rotation and stretch of material fibers. The  present  work is  mainly  motivated  by  adaptive  finite  element discretizations for problems of nonlinear continuum mechanics with inelastic constitutive behavior (see, e.g.,~\cite{prakash2015multiscale,frydrych2019solution,proell12b,proell12c,dittmann2020phase}), which requires the consistent interpolation of tensor-valued history data (e.g., the deformation gradient associated with the inelastic part of the deformation) for the transfer between coarse and fine mesh. However, the proposed schemes are very general in nature and suitable for the interpolation of general invertible second-order square tensors independent of the specific application.
Strictly speaking, the term interpolation refers to interpolation functions that pass through the data values at given data points (interpolation property). For simplicity, throughout this work, the notion tensor interpolation includes also the more general case of tensor approximation, where an approximation function approximates the data values at given data points without exactly representing them.  

There is a considerable amount of literature on tensor interpolation methods, predominantly for second-order symmetric positive definite  tensor as outlined in the following. A direct approach is an Euclidean interpolation, where the tensor components, i.e., the coordinates when expressed in a specific coordinate frame, are individually interpolated. In principle, arbitrary scalar interpolation functions, e.g., linear or bilinear Lagrange interpolation functions, can be applied. However, as consequence of this simple interpolation approach, the invariants of the tensor cannot be controlled. For instance, the so-called swelling effect~\cite{arsigny2006log}, i.e., a strongly non-monotonic evolution of determinant, trace and fractional anisotropy (FA), has been observed for this class of interpolation schemes. Wang et al.~\cite{wang2004constrained} proposed
a scheme for symmetric positive definite tensors utilizing a Cholesky decomposition to interpolate the resulting lower triangular tensors. Even though it preserves positive definiteness
and symmetry, it does not provide control over the remaining invariants. Batchelor et al.
\cite{batchelor2005rigorous}, Pennec et al. \cite{pennec2006riemannian}, Lenglet et al.
\cite{lenglet2006statistics}, and Fletcher and Joshi \cite{fletcher2007riemannian} derived interpolation schemes from a tensor metric defined on the Riemannian manifold of positive definite tensors, resulting in an affine invariant interpolation. Riemannian approaches can evade the swelling effect while retaining positive definiteness. However, these schemes typically go along with significant computational costs and in some cases even lack a closed-form representation of the interpolation. Also, an extension to an arbitrary number of data points and to higher-order interpolation is not straight-forward for most of these schemes. A close approximation of the Riemannian approaches, which are not exactly affine invariant however, is denoted as Log-Euclidean interpolation and was proposed by Arsigny \cite{arsigny2006log}. In this context, the notion Log-Euclidean refers to the Euclidean, i.e, componentwise, interpolation of tensor logarithms. While this scheme perseveres positive definiteness and a monotonic evolution of the determinant, it typically results in a non-monotonic evolution of the trace. Lin \cite{lin2019riemannian}
proposed a new Riemannian metric via Cholesky decomposition termed as Log-Cholesky metric. The corresponding tensor interpolation is realized via Cholesky decomposition followed by a logarithmic transformation of the diagonal terms of the resulting lower triangular matrix. Also for this approach, the trace of the tensor typically evolves in a non-monotonic manner.

Spectral decomposition followed by a separate interpolation of the resulting rotation and eigenvalue tensors was proposed
by Yang et al. \cite{yang2012feature} for symmetric positive definite tensors. Accordingly, the final tensor is reconstructed from interpolated Euler angles or quaternions and logarithmically transformed eigenvalues. This method
guarantees a monotonic evolution of determinant, trace, and FA interpolation, while preserving symmetry and positive
definiteness. The interpolation between multiple data points is approximated by employing spherical linear interpolation (slerp), designed for rotation interpolation between two data points. Instead, Collard et al.
\cite{collard2014anisotropy} proposed to use a linear weighted quaternion scheme for rotation interpolation of problems with multiple data points, which is applicable for small to moderate relative rotations between the data points. Wang et al. \cite{wang2022spectrum} employed the so-called spectrum-sine interpolation. Accordingly, the rotation interpolation is carried out based on the relative angles between the eigenvectors.

In summary, there are two main classes of interpolation approaches for symmetric positive definite tensors that preserve positive definiteness and result in a monotonic evolution of invariants such as determinant, trace and fractional anisotropy (FA): schemes based on either (i) Riemannian metrics or on a (ii) spectral decomposition followed by a separate interpolation of rotations and eigenvalues. However, to the best of our knowledge, none of the existing approaches fulfills all of the following requirements, which are relevant for many practical applications:
\begin{enumerate}
    \item Interpolation/approximation of an arbitrary number of data points
    \item Suitability for higher-order interpolation
    \item Scaling and rotational invariance (objectivity)
\end{enumerate}
Moreover, relevant applications often involve non-symmetric tensors, while the aforementioned methods can only be applied to symmetric tensors. Only a few interpolation approaches for non-symmetric tensors can be found in literature, e.g., the works by Prakash et al.~\cite{prakash2015multiscale} and Frydrych et al.~\cite{frydrych2019solution} in the context of adaptive finite element discretizations for crystal plasticity (CPFEM). These approaches rely on a polar decomposition of a non-symmetric tensor (the deformation gradient) in a rotation tensor and a symmetric positive definite tensor (the stretch tensor) followed by a separate interpolation of rotation and stretch tensor. However, a standard Euclidean (component-wise) interpolation is applied to the stretch tensor, i.e., positive definiteness and a monotonic evolution of tensor invariants such as determinant, trace and fractional anisotropy (FA) cannot be guaranteed.

The present contribution aims to close this gap by proposing novel interpolation schemes for general, i.e., symmetric or non-symmetric, invertible square tensors, which preserve positive definiteness and monotonicity of invariants. Moreover, the proposed schemes are suitable for data sets of arbitrary size, i.e., \textit{not} limited to the interpolation between only two tensors and suitable for higher-order interpolation. Critically, the proposed schemes rely on a combined polar and spectral decomposition of the tensor data $\bm{T}\!\!=\!\!\bm{R}\bm{Q}^T \!\! \bm{\Lambda} \bm{Q}$, followed by an individual interpolation of the two rotation tensors $\bm{R}$ and $\bm{Q}$ and the positive definite diagonal eigenvalue tensor $\bm{\Lambda}$ resulting from this decomposition. Two different schemes are considered for a consistent rotation interpolation within the special orthogonal group $\mathbb{SO}(3)$: The first scheme is a moving least squares approximation of relative rotation vectors and can be identified as an extension of the approach proposed by Crisfield and Jelenic~\cite{Crisfield1999} from 1D interpolation to multidimensional approximation. The second scheme represents the so-called spherical weighted average (SWA) proposed by Buss and Fillmore~\cite{buss2001spherical} based on quaternions. For eigenvalue interpolation, three different schemes are considered: The first one is the logarithmic weighted average investigated by Yang et al.~\cite{yang2012feature} and Collard et al.~\cite{collard2014anisotropy}, which can be identified with a weighted geometric mean. As second scheme, a moving least squares approximation of the eigenvalues is utilized. Apart from these two well-established methods, a new approach based on a moving least squares approximation of logarithmic eigenvalues is proposed. It will be demonstrated that this approach, denoted as logarithmic moving least squares, combines the advantages of the aforementioned two schemes, namely preservation of positive definiteness and suitability for higher-order approximation. 

It is demonstrated that the proposed interpolation procedure based on a combined polar and spectral decomposition preserves the structure of a tensor $\bm{T}$, in the sense that $\bm{R}$ and $\bm{Q}$ remain orthogonal tensors and $\bm{\Lambda}$ remains a positive definite diagonal tensor during interpolation, as well as scaling and rotational invariance (objectivity). Based on selected numerical examples considering the interpolation of either symmetric or non-symmetric tensors, the proposed schemes are compared to existing approaches such as Euclidean, Log-Euclidean, Cholesky and Log-Cholesky interpolation. In contrast to these existing methods, the proposed interpolation schemes result in smooth and monotonic evolutions of tensor invariants such as determinant, trace, fractional anisotropy (FA), and Hilbert's anisotropy (HA). Moreover, a consistent spatial convergence behavior is confirmed for first- and second-order realizations of the proposed schemes.

The paper is organized as follows: In Section~\ref{sec:tensor_decomposition}, the basics of polar and spectral decomposition are presented. In Sections~\ref{sec:rotation_tensor_parameterization} and~\ref{sec:paramateirzed-interpolation}, the fundamentals of rotation tensor
parameterization and the proposed schemes for rotation interpolation, either based on relative rotation vectors or quaternions, are presented. The different approaches for eigenvalue interpolation are proposed and discussed in Section~\ref{sec:eigenvalue_interpolation}. Eventually, the overall tensor interpolation procedure and the resulting properties are discussed in Section~\ref{sec:tensor_interpolation}. Selected numerical examples and test cases are presented and analyzed in Section~\ref{sec:Results}. A summary and concluding remarks are given in Section~\ref{sec:Conclusion}.
\section{Methodology}\label{sec:headings} This section describes the necessary mathematical preliminaries and delineate the proposed tensor interpolation methods.

\subsection{Tensor decomposition}
\label{sec:tensor_decomposition} For any invertible tensor $\bm{T} \in
  \mathbb{R}^{3\times3}$, the unique polar decomposition is defined according to
\begin{align}
  \bm{T}=\bm{R}\bm{U}, \label{polar-decomposition}
\end{align}
where $\bm{R}$ is a rotation tensor and $\bm{U}\in
  \mathbb{R}^{3\times3}$ is a symmetric, positive definite tensor.
  When $\bm{T}$ is a symmetric tensor, the rotation part $\bm{R}$ reduces to an identity tensor. The spectral decomposition of $\bm{U}$ reads $ \bm{U} =  \sum_i \lambda^i \hat{\bm{n}}^i \otimes
  \hat{\bm{n}}^i$, where $\hat{\bm{n}}^i $ are the unit eigenvectors and $\lambda^i$ are the eigenvalues of $\bm{U}$. Equivalently, $\bm{U}$ can be represented by the tensor product
  \begin{align}
    \bm{U} = \bm{Q}^T \bm{\Lambda} \bm{Q}, \label{spectral-decomposition}
\end{align}
 with the diagonal eigenvalue tensor $\bm{\Lambda}=\text{diag}(\lambda^1,\lambda^2,\lambda^3)$ and the rotation tensor $\bm{Q}$ spanned by the eigenvectors of~$\bm{U}$ according to $\bm{Q}^T=[\hat{\bm{n}}^1,\hat{\bm{n}}^2,\hat{\bm{n}}^3]$. The eigendecomposition of $\bm{U}$ is not unique, as $\bm{Q}$ depends on the choice of eigenvectors. When the order of the eigenvectors is fixed, it is unique up to sign change. Combining~\eqref{polar-decomposition} and~\eqref{spectral-decomposition}, any invertible non-symmetric tensor $\bm{T} \in \mathbb{R}^{3\times3}$ can be decomposed into
two rotation tensors $\bm{R}$ and $\bm{Q}$ and three scalar eigenvalues $\lambda^1,\lambda^2,\lambda^3$. In the tensor interpolation strategy proposed in this work, the rotations and eigenvalues will be interpolated individually.
\subsection{Rotation tensor parameterization}
\label{sec:rotation_tensor_parameterization}

Various different parametrizations for rotation tensors have been proposed in literature~\cite{Simo1986b, Cardona1988, Ibrahimbegovic1995, Romero2004, Meier2019}. A
three-component parameterization is achieved, e.g., by rotation (pseudo) vectors or Euler angles. Alternatively, unit-quaternions or angle-axis parameterization give rise to four rotation parameters. This work focuses on rotation parametrization either by rotation vectors $\bm{\theta}$ or by unit-quaternions $\Hat{\bm{q}}$.

A rotation tensor $\bm{R}$ can be considered as element of the special orthogonal group $\mathbb{SO}(3)$, which is a smooth differential manifold, i.e. a compact Lie group with dimension 3, according to:
\begin{align}
  \mathbb{SO}(3) := \{\bm{R} \in  \mathbb{R}^{3 \times 3}: \bm{R}^T \bm{R} = \bm{I}_3, \det(\bm{R})=1 \}, \label{SO3_definition}
\end{align}
where $\bm{I}_3$ is the second-order identity tensor in $\mathbb{R}^3$. The Lie algebra associated with the Lie group is denoted as $\mathfrak{so}(3)$, and represents the tangent space to the Lie group at identity:
\begin{align}
  \mathfrak{so}(3) := T_{\bm{I}} \mathbb{SO}(3) := \{\bm{S}(\bm{a}) : \bm{S}(\bm{a})= -\bm{S}(\bm{a})^T \ \forall \bm{a}\in \mathbb{R}^3  \},
\end{align}
where $\bm{S}(\bm{a}) \in \mathbb{R}^{3 \times 3}$ is a  skew-symmetric tensor with
$\bm{S}(\bm{a}) \bm{b} = \bm{a} \times \bm{b} \,\,\, \forall \bm{a}, \bm{b} \in \mathbb{R}^3$.
The skew-symmetric tensor $\bm{S}(\bm{a})$  maps vectors from $\mathbb{R}^3$ to the Lie algebra $\mathfrak{so}(3)$ such that,
\begin{align}
  \bm{S}(\bm{a}) = \left[\begin{array}{lll}
      0    & -a_{3} & a_{2} \\
      a_3  & 0      & -a_1  \\
      -a_2 & a_1    & 0
    \end{array}\right] \quad \text{with} \quad  \bm{a} =  \left[\begin{array}{l}
      a_1 \\
      a_2 \\
      a_3
    \end{array}\right]. \label{gen-skew-symm}
\end{align}
Formally, the map from Lie algebra to Lie group is given by the \textit{exponential map} $\bm{R}=\exp\left( \bm{S}(\bm{a}) \right):\mathfrak{so}(3) \to \mathbb{SO}(3)$, based on the power series expansion of the exponential function.
\subsubsection{ Rotation vectors}\label{sec:rotation_vec}

 A closed-form parametrization of the rotation tensor, and the exponential map, by means of a rotation vector $\bm{\theta} \in \mathbb{R}^3$ is given by the
so-called Rodrigues formula:
\begin{align}
  \bm{R}(\bm{\theta})= \exp\left(\bm{S}(\bm{\theta})\right)  =  \bm{I}_3 + \sin (\theta) \bm{S}(\bm{e}_{\bm{\theta}}) + (1-\cos (\theta)) \ \bm{S}(\bm{e}_{\bm{\theta}}) \bm{S}(\bm{e}_{\bm{\theta}}), \label{Rodrigues-Formula}
\end{align}
where $\theta$ represents the scalar rotation angle and
$\bm{e}_{\bm{\theta}}$ the axis of rotation associated with the rotation vector $\bm{\theta} = \theta \bm{e}_{\bm{\theta}}$. The unique computation of a rotation vector from a given rotation tensor is possible  within $\theta \in \,\, ]-\pi, \pi]$, e.g., by applying Spurrier's algorithm (see \cite{spurrier1978comment} and \cite{Crisfield1999}).
Formally, this inverse mapping from rotation
tensor to rotation vector is denoted as $\bm{\theta}(\bm{R})\hspace{-2pt}: \mathbb{SO}(3) \mapsto \mathbb{R}^3$, governed by
$\exp\left(\bm{S}(\bm{\theta})\right) = \bm{R}$.

Finally, two rotation tensors $\bm{R}_1(\bm{\theta}_1)$ and $\bm{R}_2(\bm{\theta}_2)$, with corresponding rotation vectors $\bm{\theta}_1$ and $\bm{\theta}_2$, can be related by the relative rotation tensor $\bm{R}_{21}(\bm{\theta}_{21})$ according to:
\begin{align}
  \bm{R}_2(\bm{\theta}_2)=\bm{R}_{1}(\bm{\theta}_{1})\bm{R}_{21}(\bm{\theta}_{21})
  \Longleftrightarrow 
  \bm{R}_{21}(\bm{\theta}_{21})=\bm{R}_{1}(\bm{\theta}_{1})^T \bm{R}_2(\bm{\theta}_2),
  \label{compound-rotation}
\end{align}
with the identity $\bm{R}^T=\bm{R}^{-1}$ for all elements of $\mathbb{SO}(3)$. For given rotation vectors $\bm{\theta}_1$ and $\bm{\theta}_{21}$, the resulting compound rotation tensor $\bm{R}_{2}(\bm{\theta}_{2})$ can be calculated according to~\eqref{compound-rotation}, and the associated rotation vector $\bm{\theta}_{2}$ can be extracted using, e.g., Spurrier's algorithm. Note, that these rotation vectors are non-additive, i.e., $\bm{\theta}_{2} \neq \bm{\theta}_{1}+\bm{\theta}_{21}$.

\begin{remark}
  Equation~\eqref{compound-rotation} describes the rotation update from $\bm{R}_1$ to $\bm{R}_2$ via right-multiplication with $\bm{R}_{21}$. Alternatively, an update procedure based on left-multiplication can be defined according to
  \begin{align}
  \bm{R}_2(\bm{\theta}_2)=\bm{R}_{12}(\bm{\theta}_{12}) \bm{R}_{1}(\bm{\theta}_{1})
  \Longleftrightarrow 
  \bm{R}_{12}(\bm{\theta}_{12})= \bm{R}_2(\bm{\theta}_2)\bm{R}_{1}(\bm{\theta}_{1})^T.
  \label{compound-rotation-left}
  \end{align}
  It can be shown that the relations $\bm{R}_{12} = \bm{R}_{1} \bm{R}_{21} \bm{R}_{1}^T$ and $\bm{\theta}_{12}=\bm{R}_{1} \bm{\theta}_{21}$ hold between the relative rotations based on either right- or left-multiplication.
\end{remark}
\subsubsection{Quaternions} \label{sec:quaternions}

A quaternion $\hat{\bm{q}} \in \mathbb{H}$ is considered as element of the 4-dimensional vector space ${\mathbb{H}} = \{ \hat{\bm{q}} = a+ b \hat{\bm{i}} + c \hat{\bm{j}}+ d \hat{\bm{k}}:
  a,b,c,d \in \mathbb{R} \}$ with basis $\{1,\hat{\bm{i}},
  \hat{\bm{j}},\hat{\bm{k}}\}$. The scalar (or real) and vector (or imaginary) part of the quaternion shall be denoted as $q = a$ and ${\bm{q}} =b \hat{\bm{i}} + c \hat{\bm{j}}+ d \hat{\bm{k}}$ such that $\hat{\bm{q}} = q+\bm{q}$ with $q \in \mathbb{R}$. The multiplication of two quaternions $\hat{\bm{q}}$ and $\hat{\bm{p}}$, often denoted as Hamilton or quaternion product, is defined as
  \begin{align}
  \hat{\bm{q}} \hat{\bm{p}} = q p - \bm{q} \cdot \bm{p} + q \bm{p} + p \bm{q} + \bm{q} \times \bm{p}, \label{quaternion-product}
\end{align}
 which is non-commutative because $\hat{\bm{p}} \hat{\bm{q}} = q p - \bm{q} \cdot \bm{p} + q \bm{p} + p \bm{q} - \bm{q} \times \bm{p}$. Here, the dot product $\cdot$ and the cross-product $\times$ for the vector part of the quaternion are inherited from the Euclidian vector space $\mathbb{R}^3$. A \textit{unit} quaternion $\hat{\bm{q}} \in \mathbb{H}^1$ defined via $||\hat{\bm{q}}||=\sqrt{a^2+b^2+c^2+d^2}=\sqrt{q^2+\bm{q} \cdot \bm{q}}=1$ can be interpreted as element of the 3-sphere $\mathbb{S}^3$
\begin{align}
  \mathbb{S}^3 = \{ \hat{\bm{q}} \in \mathbb{H}: \ \magn{\hat{\bm{q}}}=1\}.
\end{align}
Based on the scalar rotation angle $\theta$ and the axis of rotation $\bm{e}_{\bm{\theta}}$ (with $||\bm{e}_{\bm{\theta}}||=1$) as defined in Section~\ref{sec:rotation_vec}, a unit quaternion $\hat{\bm{q}}$ and its inverse $\hat{\bm{q}}^{-1}$ are defined according to:
\begin{align}
  \hat{\bm{q}}(\theta,\bm{e}_{\bm{\theta}}) = \cos (\theta/2) + \ \sin (\theta/2) \bm{e}_{\bm{\theta}},\quad 
  \hat{\bm{q}}^{-1}(\theta,\bm{e}_{\bm{\theta}})=\hat{\bm{q}}(-\theta,\bm{e}_{\bm{\theta}}) = \cos (\theta/2) - \ \sin (\theta/2) \bm{e}_{\bm{\theta}}. \label{unit-quaternion}
\end{align}
From~\eqref{unit-quaternion}, it can be verified by trigonometric manipulations that an alternative parametrization of the rotation tensor~\eqref{Rodrigues-Formula} can be stated as: 
\begin{align}
  \bm{R}(\hat{\bm{q}}) = \bm{I}_3 + 2 q \bm{S}(\bm{q}) + 2  \bm{S}(\bm{q})\bm{S}(\bm{q}) \quad \textrm{with} \quad q=\cos (\theta/2), \quad \bm{q}=\ \sin (\theta/2) \bm{e}_{\bm{\theta}}, \label{quat-rot-to-quat-map}
\end{align}
Note, that $\bm{R}(\hat{\bm{q}})=\bm{R}(-\hat{\bm{q}})$, i.e., the quaternions $\hat{\bm{q}}$ and $-\hat{\bm{q}}$ represent the same rotation. In contrast, the rotation by a negative angle $-\theta$ is given by the inverse quaternion $\hat{\bm{q}}^{-1}$ according to~\eqref{unit-quaternion}. Again, Spurrier's algorithm \cite{spurrier1978comment} can be employed to extract a unique quaternion from a given rotation tensor. The inverse mapping from rotation tensor to quaternion is formally denoted as $\hat{\bm{q}}(\bm{R}):\mathbb{SO}(3) \mapsto \mathbb{H}^1$. In case of a compound rotation according to~\eqref{compound-rotation}, the corresponding quaternions can be related in a straightforward manner using the quaternion product~\eqref{quaternion-product} (see~\cite{jia2008quaternions}):
\begin{align}
  \hat{\bm{q}}_2= \hat{\bm{q}}_{1} \hat{\bm{q}}_{21}.
  \label{compound-quaternion}
\end{align}
\begin{remark}
  If left-multiplication according to~\eqref{compound-rotation-left} is used for the rotation update, the quaternion update reads
  \begin{align}
  \hat{\bm{q}}_2= \hat{\bm{q}}_{12} \hat{\bm{q}}_{1}.
  \label{compound-quaternion-left}
\end{align}
\end{remark}
\subsection{Rotation tensor interpolation}\label{sec:paramateirzed-interpolation}

Consider a set of $N$ rotation tensors $\bm{R}_1, \dots, \bm{R}_N \in \mathbb{SO}(3)$ distributed in
space with corresponding position vectors  $\bm{x}_1, \dots, \bm{x}_N \in \Omega$. The associated rotation vectors and unit quaternions shall be denoted as $\bm{\theta}_{1} , \dots, \bm{\theta}_{N} \in \mathbb{R}^3$ and $\hat{\bm{q}}_{1} , \dots, \hat{\bm{q}}_{N} \in \mathbb{H}^1$, respectively. This section aims to construct a smooth rotation tensor field based on this $N$ data points that allows to approximate the rotation tensor $\bm{R}_p$ at a given location $\bm{x}_p$ within the domain $\Omega$. 

As shown in~\cite{Crisfield1999}, a direct interpolation of the rotation vectors $\bm{\theta}_{j}$ for $j=1,...,N$ would result in a non-objective interpolation scheme. Instead, in order to achieve an objective, i.e., frame-invariant, interpolation scheme, relative rotation vectors need to be considered for interpolation. Thereto, in a first step, a reference rotation tensor $\bm{R}^{0}$ is defined, which can be conveniently chosen as the rotation tensor $\bm{R}_j$ associated with one of the given data points $\bm{x}_j$. Within this work, the data point that is closest to the interpolation point $\bm{x}_p$ is chosen for this purpose. Now, for every rotation tensor $\bm{R}_j$ with $j=1,...,N$, the relative rotation tensor $\bm{R}^\mathrm{r}_j$ with respect to the reference rotation tensor $\bm{R}^{0}$ can be calculated according to~\eqref{compound-rotation}:
\begin{align}
  \bm{R}^\mathrm{r}_j=\bm{R}^{0T} \bm{R}_j.
  \label{reference-rotation}
\end{align}
From $\bm{R}^\mathrm{r}_j$ according to~\eqref{reference-rotation}, the corresponding relative rotation vector $\bm{\theta}^\mathrm{r}_j$ or relative quaternion $\hat{\bm{q}}^\mathrm{r}_j$ can be extracted. According to~\eqref{compound-quaternion}, the relative quaternion can also be directly calculated as $\hat{\bm{q}}_j= \hat{\bm{q}}^{0} \hat{\bm{q}}^\mathrm{r}_{j}$. In the following sections, two different interpolation strategies based on either relative rotation vectors ${\bm{\theta}}^\mathrm{r}_j$ or relative quaternions $\hat{\bm{q}}^\mathrm{r}_j$ are presented, resulting in a relative rotation vector ${\bm{\theta}}^\mathrm{r}_p$ or relative quaternion $\hat{\bm{q}}^\mathrm{r}_p$ at the interpolation point $P$ with associated relative rotation tensor $\bm{R}^\mathrm{r}_p$. In both cases, the rotation tensor at the interpolation point $P$ can be recovered from
\begin{align}
  \bm{R}_p = \bm{R}^{0} \bm{R}^\mathrm{r}_p.
  \label{final-interpolated-tensor}
\end{align}
\subsubsection{Rotation vector interpolation}\label{sec-rotation-vector-interpolation}

In order to approximate the relative rotation vector at the point $\bm{x}_p$, a moving least square scheme is employed to construct a continuous relative rotation vector field $\bm{\theta}^\mathrm{r}(\bm{x})$ based on $N$ data points at positions $\bm{x}_j$ with corresponding relative rotation vectors $\bm{\theta}^\mathrm{r}_{j}$ for $j=1,...,N$. Thereto, $\bm{\theta}^\mathrm{r}(\bm{x})$ is approximated component-wise as polynomial according to
\begin{align}
  \bm{\theta}^{\mathrm{r}i}(\bm{x}) := \bm{p}(\bm{x}) \bm{a}^i,
  \label{LS-basis-interpolation}
\end{align}
where the index $i=1,2,3$ represents the three components of the relative rotation vector, $\bm{p}(\bm{x}) \in \mathbb{R}^m$ is the polynomial basis function vector of order $m$, and $\bm{a}^i \in \mathbb{R}^m$ are the corresponding vectors of coefficients. For example, a complete quadratic basis in 3D with corresponding coefficients is given by
\begin{align}
  \bm{p} = [1\ x\ y\ z\ x^2\ y^2\ z^2\ xy\ yz\ xz], \quad \bm{a}^{iT} = [a^i_0\ a^i_1\ a^i_2\ a^i_3\ a^i_4\ a^i_5\ a^i_6\ a^i_7\ a^i_8\ a^i_9],\label{LS-basis-qudartic}
\end{align}
with $\bm{x}=(x,y,z)^T \in \mathbb{R}^3$. The unknown coefficient vectors $\bm{a}^i$ are obtained by minimizing the weighted residual
\begin{align}
  r = \sum_{j=1}^N \tilde{w}(\bm{x}_j) \bigg[  \big(\bm{p} (\bm{x}_j) \bm{a}^1 - \bm{\theta}^{\mathrm{r}1}_j \big)^2 + \big(\bm{p} (\bm{x}_j) \bm{a}^2 - \bm{\theta}^{\mathrm{r}2}_j \big)^2 + \big(\bm{p} (\bm{x}_j) \bm{a}^3 - \bm{\theta}^{\mathrm{r}3}_j \big)^2 \bigg], \label{LS-definition}
\end{align}
where  $\tilde{w}(\bm{x}_j)$ is a normalized weighting function. Within our work, we employ the normalized weights according to:
\begin{align}
  \tilde{w}(\bm{x}_j) := \frac{{w}(\bm{x}_j)}{\sum_{j=1}^N {w}(\bm{x}_j)} \quad \textrm{such that} \quad \sum_j \tilde{w}(\bm{x}_j) =1. \label{exponential-weighting-function-SWA}
\end{align}
Here the weighting function ${w}(\bm{x}_j)$ can be any monotonic continuous function that decreases as it moves away from the interpolation point $\bm{x}_p$. For example, an exponential weighting function, with control parameter $c$, reads
\begin{align}
  {w}(\bm{x}_j) = \exp\left(-c || \bm{x}_j - \bm{x}_p||^2\right). \label{exponential-weighting-function}
\end{align}

Minimization of the residual $r$ according to~\eqref{LS-definition} leads to the following result for the unknown coefficient vectors: 
\begin{align}
  \bm{a}^i=\bm{P}^{-1} \bm{b}^i \quad \text{with} \quad \bm{P} = \sum_{j=1}^N \tilde{w}(\bm{x}_j)  \bm{p}(\bm{x}_j)^T \bm{p}(\bm{x}_j) \quad \textrm{and} \quad \bm{b}^i =  \sum_{j=1}^N \tilde{w}(\bm{x}_j) \bm{p}(\bm{x}_j)^T  {\bm{\theta}}_j^{\mathrm{r}i}.
\end{align}
The condition number of $\bm{P} \in \mathbb{R}^{m \times m}$ depends on the number and location of the data points $\bm{x}_j$ within the domain~$\Omega$. The necessary condition for a non-singular matrix $\bm{P}$ is  $N \geq m$. Once the unknown coefficient vectors $\bm{a}^i$ have been calculated, the components of the relative rotation vector at position $\bm{x}_p$ can be determined by evaluating~\eqref{LS-basis-interpolation} for $\bm{x}=\bm{x}_p$:
\begin{align}
  \bm{\theta}^{\mathrm{r}i}_p := \bm{p}(\bm{x}_p) \bm{a}^i.
  \label{LS-basis-interpolation_xp}
\end{align}
\begin{remark}
  When the order of the polynomial basis equals the number of data points ($m=N$), the moving least squares method reduces to the least squares method. In this case, the objective function in~\eqref{LS-definition} simplifies to 
  \begin{align}
     r = \sum_{j=1}^N \bigg[  \big(\bm{p} (\bm{x}_j) \bm{a}^1 - \bm{\theta}^{\mathrm{r}1}_j \big)^2 + \big(\bm{p} (\bm{x}_j) \bm{a}^2 - \bm{\theta}^{\mathrm{r}2}_j \big)^2 + \big(\bm{p} (\bm{x}_j) \bm{a}^3 - \bm{\theta}^{\mathrm{r}3}_j \big)^2 \bigg]. 
  \end{align}
  Furthermore, the resulting relative rotation vector field fulfills the interpolation property $\bm{\theta}^{\mathrm{r}i}(\bm{x}_j)=\bm{\theta}^{\mathrm{r}i}_j$. 
\end{remark}
\begin{remark}
  In our work, the moving least squares approaches are formulated in a local coordinate system with origin at $\bm{x}_p$, i.e., in the local system the coordinates of the interpolation scheme are given as $\bm{x}_p=\bm{0}$.
\end{remark}
\subsubsection{Quaternion interpolation}\label{sec-quaternion-vector-interpolation}

As second approach to approximate the rotation tensor at the point $\bm{x}_p$, the spherical weighted average (SWA) as proposed by Buss and Fillmore~\cite{buss2001spherical} shall be considered. Let $\hat{\bm{q}}^\mathrm{r}_1, \dots, \hat{\bm{q}}^\mathrm{r}_N \in \mathbb{H}^1$ be the relative unit
quaternions at $N$ given data points as defined above, and $\tilde{w}_1, \dots, \tilde{w}_N$ normalized weights (see~\eqref{exponential-weighting-function-SWA}) at these data points such that $\sum_j \tilde{w}_j =1$ and $\tilde{w}_j >0$. The spherical weighted average $\hat{\bm{q}}^\mathrm{r}_p$ at position $\bm{x}_p$ is the unit quaternion that minimizes the following weighted geodesic distance function
\begin{align}
  \hat{\bm{q}}^\mathrm{r}_p = \text{arg min}_{\hat{\bm{q}}^\mathrm{r}} f(\hat{\bm{q}}^\mathrm{r}) \quad \text{with} \quad 
  f(\hat{\bm{q}}^\mathrm{r}) = \frac{1}{2} \sum^N_{j=1} \tilde{w}_j \ \textrm{d}_{\mathbb{S}^3}(\hat{\bm{q}}^\mathrm{r},\hat{\bm{q}}^\mathrm{r}_j)^2 \quad \text{and} \quad \textrm{d}_{\mathbb{S}^3}(\hat{\bm{q}}^\mathrm{r},\hat{\bm{q}}^\mathrm{r}_j)= || \ln \left((\hat{\bm{q}}^\mathrm{r})^{-1} \hat{\bm{q}}^\mathrm{r}_j\right) ||, \label{SWA}
\end{align}
where $\textrm{d}_{\mathbb{S}^3}(\hat{\bm{q}}^\mathrm{r},\hat{\bm{q}}^\mathrm{r}_j)$ denotes the shortest geodesic distance between $\hat{\bm{q}}^\mathrm{r}$ and $\hat{\bm{q}}^\mathrm{r}_j$ on $\mathbb{S}^3$~\cite{buss2001spherical,angulo2014riemannian}. As demonstrated in Appendix~\ref{appendix:d_S3}, the shortest geodesic distance can be reformulated to give the following simple expression 
\begin{align}
  \textrm{d}_{\mathbb{S}^3}(\hat{\bm{q}}^\mathrm{r},\hat{\bm{q}}^\mathrm{r}_j)=\tilde{\theta}/2
  \quad \text{with} \quad \tilde{\theta}=||\tilde{\bm{\theta}}||, \quad
  \bm{R}(\tilde{\bm{\theta}} )
  =\bm{R}(\hat{\bm{q}}^\mathrm{r})^T \bm{R}(\hat{\bm{q}}^\mathrm{r}_j), \label{SWA-simplification}
\end{align}
where $\tilde{\theta}$ is the norm of the rotation vector associated with the relative rotation tensor between $\bm{R}(\hat{\bm{q}}^\mathrm{r})$ and $\bm{R}(\hat{\bm{q}}^\mathrm{r}_j)$. 
In~\cite{buss2001spherical} it is shown that the SWA approach fulfills partition of unity, i.e., $\hat{\bm{q}}^\mathrm{r}_p=\hat{\bm{q}}^\mathrm{r}_\text{const}$ if $\hat{\bm{q}}^\mathrm{r}_j = \hat{\bm{q}}^\mathrm{r}_\text{const} \,\, \forall j \in \{1,...,N\}$, and the interpolation property, i.e. $\hat{\bm{q}}^\mathrm{r}_p = \hat{\bm{q}}^\mathrm{r}_j$ if $\tilde{w}_j=1$ for one given $j$. It is important to note that the parameterization of quaternions is unique up to sign change ($\hat{\bm{q}}$ and $-\hat{\bm{q}}$ result in the same rotation tensor). As discussed in~\cite{buss2001spherical}, a unique solution $\hat{\bm{q}}^\mathrm{r}_p$ of~\eqref{SWA} can only be found if all quaternions $\hat{\bm{q}}^\mathrm{r}_j$ lie within the same hemisphere of $\mathbb{S}^3$. A procedure to enforce this prerequisite is presented in Section~\ref{sec:impose_uniqueness}. Eventually, Buss and Fillmore proposed two iterative algorithms to solve the optimization problem~\eqref{SWA}, one based on a steepest decent approach resulting in a linear convergence behavior, and one based on a Newton-type algorithm resulting in a quadratic convergence rate. In the present work, the latter approach has been employed. For algorithmic details and an in-depth mathematical analysis of the SWA approach, the interested reader is referred to~\cite{buss2001spherical}.

\begin{remark}
  For N=2, the SWA approach reduces to the well-known spherical linear interpolation (slerp), initially introduced by Ken Shoemake
\cite{shoemake1985animating} in the context of computer graphics. Based on two quaternions $\hat{\bm{q}}_1, \hat{\bm{q}}_2 \in \mathbb{H}^1$ and a weight $ w \in [0,1]$, the slerp scheme is defined according to
\begin{align}
  \hat{\bm{q}}(w)=\mathrm{slerp}(\hat{\bm{q}}_1, \hat{\bm{q}}_2,w) = (\hat{\bm{q}}_2 \hat{\bm{q}}^{-1}_1)^w \hat{\bm{q}}_1  \label{slerp-general},
\end{align}
which fulfills the interpolation property $\hat{\bm{q}}(w=0)=\hat{\bm{q}}_1$ and $\hat{\bm{q}}(w=1)=\hat{\bm{q}}_2$.
\end{remark}
\subsection{Eigenvalue interpolation}\label{sec:eigenvalue_interpolation}
In the following subsections, three different variants will be presented for interpolating the eigenvalues $\lambda^1, \lambda^2, \lambda^3$ of the stretch tensor $\bm{U}$ according to~\eqref{polar-decomposition} and~\eqref{spectral-decomposition} based on given eigenvalue data $\lambda^1_j, \lambda^2_j, \lambda^3_j$ at the points $\bm{x}_j$ with $j=1,...,N$.
\subsubsection{Logarithmic weighted average}\label{sec:eigenvalue_interpolation_log}
According to this variant, the eigenvalues are logarithmically transformed and individually interpolated as investigated in~\cite{yang2012feature} and~\cite{collard2014anisotropy}:
        \begin{align}
          \lambda^i_p = \textrm{GM}\{\lambda^i_1,...,\lambda^i_N\} :=
          \exp \Bigg(  \sum^N_{j=1}  \tilde{w}(\bm{x}_j) \ \ln (\lambda^i_j) \Bigg) \quad \text{for} \quad i=1,2,3, \label{GM}
        \end{align}
        where $\tilde{w}_j$ are the weights according to~\eqref{exponential-weighting-function-SWA} and fulfilling $\sum_j \tilde{w}_j =1$ as well as $\tilde{w}_j >0$. This averaging procedure can be identified as the weighted geometric mean (GM) of a general data $y_j>0$ fulfilling the essential property (see~\cite{yang2012feature})
        \begin{align}
          \textrm{min} \{y_1,...,y_N\} \leq \textrm{GM} \{y_1,...,y_N\} \leq \textrm{max} \{y_1,...,y_N\}, \label{GM-property}
        \end{align}
 i.e., it can be identified as a \textit{monotonic} interpolation scheme resulting in an interpolated eigenvalue that is always larger than (or equal to) the smallest eigenvalue and smaller than (or equal to) the largest eigenvalue in the data set. Among others, this characteristic evades the so-called swelling effect, i.e., the determinant of the interpolated tensor is more than the determinant of the original tensors (see~\cite{arsigny2006log}). Moreover, swelling leads to a decrease in fractional anisotropy (FA) and trace.
 Furthermore, when interpolating positive definite tensors, property~\eqref{GM-property} ensures that the interpolated eigenvalues remain positive, and, thus that also the interpolated tensor is positive definite, which is an important property for many physical applications.
 
\begin{remark}
Exemplarily, the upper bound of the monotonic interpolation property according to~\eqref{GM-property} shall be briefly verified in the following. For simplicity, let us assume that we consider general data $y_j>0$, and the data points are numbered in a manner such that $y_1$ represents the maximal value of the given data, i.e., $y_1=\mathrm{max} \{y_1,...,y_N\}$. Using the partition of unit property of the weights, i.e., $\sum_j \tilde{w}_j =1$, the interpolation scheme~\eqref{GM} can be rewritten as
\begin{align}
          y_p =
          \exp \Bigg(  \sum^N_{j=1}  \tilde{w}_j \ \ln (y_j) \Bigg) =
          \prod^N_{j=1} y_j^{\tilde{w}_j} =
          y_1^{\tilde{w}_1} \prod^N_{j=2} y_j^{\tilde{w}_j} =
          y_1^{\left(1-\sum^N_{j=2}\tilde{w}_j\right)} \prod^N_{j=2} y_j^{\tilde{w}_j} =
          y_1 \prod^N_{j=2} \left(\frac{y_j}{y_1}\right)^{\tilde{w}_j}.
\end{align}
If the weights are positive, i.e., $\tilde{w}_j>0$, this result can be used to state the following inequality, which concludes the proof of (the upper bound of) the monotonic interpolation property according to~\eqref{GM-property}:
\begin{align}
          y_p =
          y_1 \prod^N_{j=2} \left(\frac{y_j}{y_1}\right)^{\tilde{w}_j} \leq 
          y_1 \quad \text{if} \quad y_1\geq y_j, \,\, \tilde{w}_j>0 \quad \text{for} \quad j=1,...,N \quad \square \label{GM-property-proof}
\end{align}
\end{remark}
\subsubsection{Moving least squares eigenvalue approximation}\label{sec:eigenvalue_interpolation_mls}

The second variant for eigenvalue interpolation relies on a moving least squares approach as already presented in Section~\ref{sec-rotation-vector-interpolation}. Thereto, the eigenvalue fields $\lambda^i(\bm{x})$ are defined as polynomials according to
\begin{align}
  \lambda^i(\bm{x}) := \bm{p}(\bm{x}) \bm{a}^i,
  \label{LS-basis-interpolation_lambda}
\end{align}
The unknown coefficient vectors $\bm{a}^i$ are obtained by minimizing the weighted residual
\begin{align}
  r = \sum_{j=1}^N \tilde{w}(\bm{x}_j) \bigg[  \big(\bm{p} (\bm{x}_j) \bm{a}^1 - \lambda^{1}_j \big)^2 + \big(\bm{p} (\bm{x}_j) \bm{a}^2 - \lambda^{2}_j \big)^2 + \big(\bm{p} (\bm{x}_j) \bm{a}^3 - \lambda^{3}_j \big)^2 \bigg]. \label{LS-definition-lambda}
\end{align}
Minimization of the residual $r$ according to~\eqref{LS-definition-lambda} leads to the following result for the unknown coefficient vectors: 
 \begin{align}
  \bm{a}^i=\bm{P}^{-1} \bm{b}^i \quad \text{with} \quad \bm{P} = \sum_{j=1}^N \tilde{w}(\bm{x}_j)  \bm{p}(\bm{x}_j)^T \bm{p}(\bm{x}_j) \quad \textrm{and} \quad \bm{b}^i =  \sum_{j=1}^N \tilde{w}(\bm{x}_j) \bm{p}(\bm{x}_j)^T  {\lambda}^i_j.
\end{align}
The weighting functions $\tilde{w}(\bm{x}_j)$ are given by~\eqref{exponential-weighting-function-SWA}. This method is very beneficial when interpolating between negative and positive eigenvalues, for example, between negative-definite and positive-definite tensors. However, the nature of the interpolated field depends on the choice of polynomial basis and the data set. As a result, the method can not guarantee monotonic interpolation under all circumstances. 
\subsubsection{Logarithmic moving least squares eigenvalue approximation}\label{sec:eigenvalue_interpolation_logmls}

The third variant for eigenvalue interpolation represents a modification of the moving least squares approach presented in the last section. In particular, the final approximated fields $\lambda^i(\bm{x})$ are given as \textit{exponential} of a polynomial approximation according to
\begin{align}
  \lambda^i(\bm{x}) := \exp (\bm{p}(\bm{x}) \bm{a}^i),
  \label{LS-basis-interpolation-log}
\end{align}
Moreover, the unknown coefficient vectors $\bm{a}^i$ are obtained by minimizing the weighted residual
\begin{align}
  r = \sum_{j=1}^N \tilde{w}(\bm{x}_j) \bigg[  \big(\bm{p} (\bm{x}_j) \bm{a}^1 - \ln (\lambda^{1}_j) \big)^2 + \big(\bm{p} (\bm{x}_j) \bm{a}^2 - \ln (\lambda^{2}_j) \big)^2 + \big(\bm{p} (\bm{x}_j) \bm{a}^3 - \ln (\lambda^{3}_j) \big)^2 \bigg], \label{LS-definition-lambda-log}
\end{align}
including the deviation between the approximation function and the \textit{logarithm} of the data. Minimization of the residual~$r$ according to~\eqref{LS-definition-lambda-log} leads to the following result for the unknown coefficient vectors: 
\begin{align}
  \bm{a}^i=\bm{P}^{-1} \bm{b}^i \quad \text{with} \quad \bm{P} = \sum_{j=1}^N \tilde{w}(\bm{x}_j)  \bm{p}(\bm{x}_j)^T \bm{p}(\bm{x}_j) \quad \textrm{and} \quad \bm{b}^i =  \sum_{j=1}^N \tilde{w}(\bm{x}_j) \bm{p}(\bm{x}_j)^T  \ln ({\lambda}^i_j).
\end{align}
The weighting functions $\tilde{w}(\bm{x}_j)$ are again given by~\eqref{exponential-weighting-function-SWA}. Unlike the logarithmic weighted average in Section~\ref{sec:eigenvalue_interpolation_log}, the logarithmic moving least squares scheme does not result in a monotonic approximation (or interpolation for $N=m$). But, in contrast to the standard moving least squares approach in Section~\ref{sec:eigenvalue_interpolation_mls}, this logarithmic moving least squares approach ensures a non-negative approximation (interpolation). This is already very beneficial for the interpolation of symmetric, positive definite tensors since negative eigenvalues/determinants as well as singularities (i.e., zero eigenvalues/determinants) can be avoided. In contrast to the logarithmic weighted average in Section~\ref{sec:eigenvalue_interpolation_log}, this scheme can be extended to arbitrary polynomial orders.

\begin{remark}
  Again, when the order of the polynomial basis equals the number of data points ($m=N$), the logarithmic moving least squares approximation simplifies to a logarithmic interpolation according to 
  \begin{align}
      \lambda^i(\bm{x}) := \exp \left(\sum_{j=1}^m \tilde{p}_j(\bm{x}) \ln ({\lambda}^i_j) \right), \label{logarithmic-mls-interpolation}
  \end{align}
  where the interpolation functions $\tilde{p}_j(\bm{x})$ fulfill the interpolation property $\tilde{p}_j(\bm{x}_k)=\delta_{jk}$, with the Kronecker delta $\delta_{jk}$. Taking the exponential of the logarithmic interpolation according to~\eqref{logarithmic-mls-interpolation} ensures that the interpolation property is preserved for the final interpolated eigenvalues, i.e., $\lambda^i(\bm{x}_k)=\exp (\sum_{j=1}^m \delta_{jk} \ln ({\lambda}^i_j))=\lambda^i_k$.
\end{remark}
\subsubsection{Comparison of eigenvalue interpolation methods}\label{sec:eigenvalue_interpolation_comparison}

In this section, the properties of the three different eigenvalue interpolation schemes shall be briefly verified and compared using a one-dimensional data set. Thereto, data points $x_j $ and corresponding data values $y_j>0$ with $j=1,...,N$ are considered. In Section~\ref{sec:eigenvalue_interpolation_log} it has been demonstrated that the logarithmic weighted average (also denoted as weighed geometric mean) represents a positive and monotonic interpolation according to~\eqref{GM-property}, i.e., for given positive data values, the interpolated value also remains positive and lies between the minimum and maximum value of the given data. Clearly, due to the exponential mapping in~\eqref{LS-basis-interpolation-log}, also the proposed logarithmic moving least squares approach of Section~\ref{sec:eigenvalue_interpolation_logmls} leads to a positive interpolated value. In contrast to the weights $\tilde{w}(\bm{x}_j)>0$ used in the geometric mean, the polynomial shape functions $\bm{p}(\bm{x}_j)$ employed in the logarithmic moving least squares approach can also take on negative values. Therefore, the proof according to~\eqref{GM-property-proof} as well as the resulting monotonic interpolation property~\eqref{GM-property} is not valid for these shape functions. This means, for given positive data values the logarithmic moving least squares approach results in an interpolated value that also remains positive, but it does not necessarily lie between the minimum and maximum value of the given data. Eventually, the standard moving least squares approximation according to Section~\ref{sec:eigenvalue_interpolation_mls} fulfills neither of these properties, i.e., for given positive data values, the interpolated value can be negative, and, thus does not necessarily lie between the minimum and maximum value of the given data.

\begin{figure}
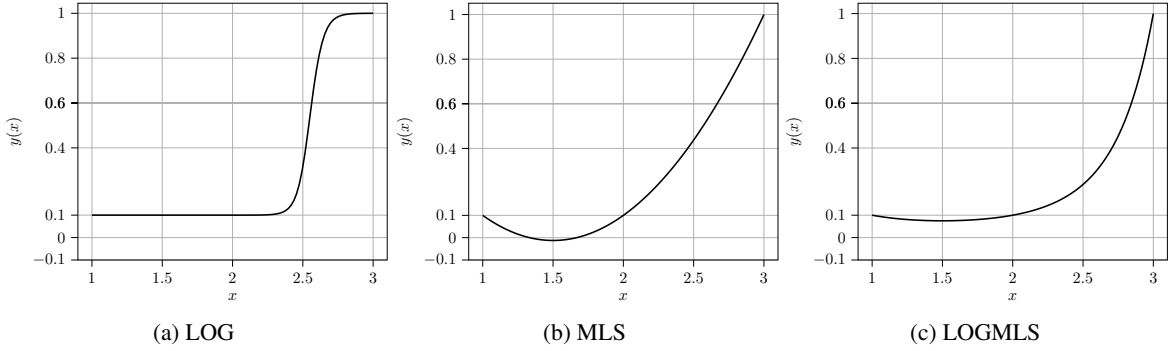

  \begin{subfigure}[b]{.32\textwidth}
    \centering
    \scalebox{0.6} { \input{pictures/Eigval_interpolation_comp_Log.tikz}}
    \caption{LOG} \label{fig:eigval_int_log} \end{subfigure}%
  \hspace*{-5pt}
  \begin{subfigure}[b]{.32\textwidth}
    \centering
    \scalebox{0.6} { \input{pictures/Eigval_interpolation_comp_MLS.tikz}}
    \caption{MLS} \label{fig:eigval_int_mls} \end{subfigure}%
  \centering
  \hspace*{-5pt}
  \begin{subfigure}[b]{.32\textwidth}
    \scalebox{0.6} { \input{pictures/Eigval_interpolation_comp_LogLS.tikz}}
    \caption{LOGMLS} \label{fig:eigval_int_logmls} \end{subfigure}%
  \caption{\small Comparison of eigenvalue interpolation methods based on the data set $(x_j,\ y_j(x)) = \{(1, 0.1),\
      (2,0.1)\, (3,1.0)\}$. Data is interpolated in the domain $x \in [1,\ 3]$ by weighted
    logarithmic mean (LOG), by moving least squares (MLS) method with a quadratic basis, and logarithmic moving least squares (LOGMLS) with a quadratic basis and depicted in (a), (b), and (c), respectively.}
  \label{fig:eigval_int_comp}
\end{figure}

To verify the aforementioned properties, a one-dimensional numerical test is performed. Consider the data set $(x_j,\ y_j(x)) = \{(1, 0.1),\ (2,0.1),\ (3,1.0)\}$. Now the interpolation is performed within the domain $x\in [1, \ 3]$ using (a) the logarithmic weighted average (LOG) according to Section~\ref{sec:eigenvalue_interpolation_log}, (b) moving least squares (MLS) with quadratic basis according to Section~\ref{sec:eigenvalue_interpolation_mls}, and (c) logarithmic moving least squares (LOGMLS) with quadratic basis according to Section~\ref{sec:eigenvalue_interpolation_logmls}. The interpolation results are visualized in Figure~\ref{fig:eigval_int_comp}, which confirms the aforementioned properties. Accordingly, the MLS scheme fails to preserve the monotonicity and positiveness of the input data in the interval $x \in [1,2]$. In contrast, the LOG scheme preserves both, monotonicity and positiveness, while resulting in a rather non-uniform interpolation, i.e., showing significant variations in the second derivative (curvature). Finally, the LOGMLS approach combines the properties of the other two by resulting in a uniform and strictly positive interpolation curve. Even though the LOGMLS interpolation is not monotonic, it is preferable in many practical applications. In practice, positiveness is typically more important than monotonicity since zero or negative eigenvalues might lead to a non-physical system behavior and to a singular tensor. Moreover, in contrast to the LOG scheme, the proposed LOGMLS approach can be extended to arbitrary interpolation orders.
\subsection{Overall tensor interpolation procedure}\label{sec:tensor_interpolation} Let $\bm{T}_1, \dots, \bm{T}_N \in
  \mathbb{R}^{3\times3}$ be invertible non-symmetric tensors defined at spatial positions $\bm{x}_1, \dots, \bm{x}_N \in
  \Omega$. The individual steps of the proposed interpolation procedure to find
$\bm{T}_p$ at position $\bm{x}_p$ are summarized in the following.
\begin{enumerate}[Step 1:]
  \item Perform polar decomposition $\bm{T}_j = \bm{R}_j \bm{U}_j$ of the tensors according to~\eqref{polar-decomposition}.  Thereto, solve the eigenvalue problem for the tensor $\bm{U}^2_j = \bm{T}^T_j\bm{T}_j$, resulting in the eigenvectors $\hat{\bm{n}}^1_j, \hat{\bm{n}}^2_j, \hat{\bm{n}}^3_j$ and the squared eigenvalues $(\lambda^{1}_j)^2, (\lambda^{2}_j)^2, (\lambda^{3}_j)^2$. The symmetric tensor $\bm{U}_j$ can then be represented as $\bm{U}_j = \sum_i \lambda^{i}_j \ \hat{\bm{n}}^i_j \otimes \hat{\bm{n}}^i_j $. Afterwards, the rotation tensor is recovered from $\bm{R}_j = \bm{T}_j \bm{U}^{-1}_j$. Be aware that the index $j=1,...,N$ refers to the given data points, and a repeated index $j$ does \textit{not} imply summation over this index.
  \item Compute the eigenvector tensors $\bm{Q}_j$ according to~\eqref{spectral-decomposition}. The rows of $\bm{Q}_j$ are given by the normalized eigenvectors $\hat{\bm{n}}^1_j, \hat{\bm{n}}^2_j, \hat{\bm{n}}^3_j$. The tensor $\bm{Q}_j$ is not unique since the permutation of rows will result in rotation tensors that also satisfy~\eqref{spectral-decomposition}. To impose uniqueness, we arrange the rows of $\bm{Q}_j$ (and also the eigenvalues within the diagonal tensor $\bm{\Lambda}_j$) in decreasing order of the eigenvalues, i.e.,
        the first row of $\bm{Q}_j$ corresponds to the eigenvector $\hat{\bm{n}}^1_j$ of the largest eigenvalue $\lambda^{1}_j$. With this procedure,
        the rotation vector is unique up to sign change of the eigenvectors. Since the determinant of the rotation tensor
        has to be $+1$, there are four possible ways to construct $\bm{Q}_j$ out of the calculated eigenvectors or their
        reflections. Such a non-uniqueness has to be avoided since it will introduce discontinuities in the interpolation scheme. In order
        to achieve a unique definition of the rotation tensor $\bm{Q}_j$, we choose the sign of the eigenvectors such that the relative rotations between associated eigenvectors (e.g., between all $\hat{\bm{n}}^1_j$ for $j=1,...,N$) at positions $\bm{x}_j$ close to the interpolation point $\bm{x}_p$ is reasonably small. This procedure is outlined in detail in
        Section~\ref{sec:impose_uniqueness}.

        \begin{remark} \label{remark:eigenvector_reoriention}
          For the proposed interpolation strategy, the assignment of eigenvalues and eigenvectors from different data points $\bm{x}_j$ is a critical issue. As outlined above, the default approach to solve this problem is to arrange eigenvalues (and associated eigenvectors) at each data point in decreasing order, i.e., interpolation takes place between all largest eigenvalues, between all medium eigenvalues and between all smallest eigenvalues from the N data points. As long as the data points $\bm{x}_j$ are located close enough to the interpolation point $\bm{x}_p$, it can be assumed that the ranges of largest, medium and smallest eigenvalues are clearly separated, and this procedure is justified. However, in scenarios where an overlap occurs between the ranges of largest, medium and smallest eigenvalues, this procedure can be suboptimal. As demonstrated in Section~\ref{sec:application_cont_mechanics}, in some applications (e.g., interpolation of the deformation gradient in slender structures) an alternative strategy might be favorable, where eigenvectors (and associated eigenvalues) from different data points $\bm{x}_j$ are assigned by considering their orientation relative to the eigenvectors at a reference point (e.g., one of the data points $\bm{x}_j$).
        \end{remark}
  \item To ensure objectivity of the interpolation procedure, $\bm{R}_j $ and $\bm{Q}_j$ are transformed into relative rotation tensors based on a reference rotation tensor as outlined at the beginning of Section~\ref{sec:paramateirzed-interpolation}. Depending on the chosen rotation interpolation strategy, either the associated relative rotation vectors ${\bm{\theta}}^\mathrm{r}_j$ (rotation vector interpolation according to Section~\ref{sec-rotation-vector-interpolation}) or the relative quaternions $\bm{q}^\mathrm{r}_j$ (quaternion interpolation according to Section~\ref{sec-quaternion-vector-interpolation}) are extracted from the relative rotation tensors. After the rotation interpolation based on either relative rotation vectors or quaternions, the associated relative rotation tensors $\bm{R}^\mathrm{r}_p$ and $\bm{Q}^\mathrm{r}_p$ at the interpolation point $\bm{x}_p$ have to be calculated. It is emphasized that the calculation of a rotation tensor out of a rotation vector or quaternion is unique since the maps $\bm{R}(\bm{\theta}): \mathbb{R}^3 \mapsto \mathbb{SO}(3) $ and $\bm{R} (\hat{\bm{q}}):\mathbb{H}^1 \mapsto \mathbb{SO}(3)$ are isomorphic and $2:1$ homomorphic, respectively. Eventually, the relative rotation tensors $\bm{R}^\mathrm{r}_p$ and $\bm{Q}^\mathrm{r}_p$ are converted into the sought-after absolute rotation tensors $\bm{R}_p$ and $\bm{Q}_p$ according to~\eqref{final-interpolated-tensor}. 
  \item In a next step the eigenvalue interpolation is performed based on  Section~\ref{sec:eigenvalue_interpolation_log}, Section~\ref{sec:eigenvalue_interpolation_mls}, or Section~\ref{sec:eigenvalue_interpolation_logmls}. Within this work, we follow the logarithmic interpolation scheme according to Section~\ref{sec:eigenvalue_interpolation_log} if not specified differently. Once the interpolated eigenvalues $\lambda^{1}_p, \lambda^{2}_p, \lambda^{3}_p$ are found, the diagonal eigenvalue tensor $\bm{\Lambda}_p$ is constructed.
  \item Finally, the interpolated tensor at the interpolation point $\bm{x}_p$ is constructed from $\bm{T}_p=\bm{R}_p\{\bm{Q}_p^T\bm{\Lambda}_p \bm{Q}_p\}$.
\end{enumerate}
From the two different variants for rotation interpolation, i.e., based on either rotation vectors (R) according to Section~\ref{sec-rotation-vector-interpolation} or quaternions (Q) according to Section~\ref{sec-quaternion-vector-interpolation}, and the three different variants for eigenvalue interpolation, i.e., based on the logarithmic weighted average (LOG) according to Section~\ref{sec:eigenvalue_interpolation_log}, moving least squares (MLS) according to Section~\ref{sec:eigenvalue_interpolation_mls}, or logarithmic moving least squares (LOGMLS) according to Section~\ref{sec:eigenvalue_interpolation_logmls}, we get the six different interpolation variants R-LOG, Q-LOG, R-MLS , Q-MLS, R-LOGMLS, and Q-LOGMLS which will be investigated in the following examples. In the special case of symmetric tensors $\bm{T}_p$, only the tensors $\bm{Q}_j$ will require a rotation interpolation ($\bm{R}_j = \bm{I}$).
\subsubsection{Imposing uniqueness on eigenvectors and quaternions} \label{sec:impose_uniqueness}
\begin{figure}[ht]
  \centering
  \scalebox{0.75} {\input{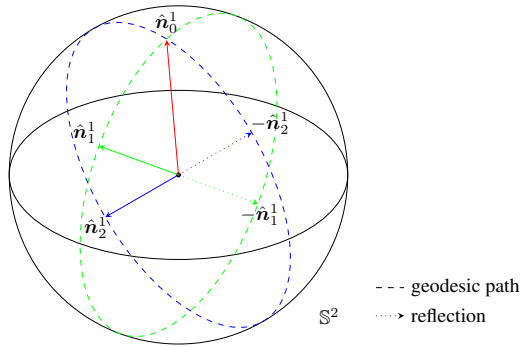}}
  \caption{\small Illustration of the eigenvector reorientation procedure. Consider the reference eigenvector
  $\hat{\bm{n}}^{1}_0$ as well as two neighbouring eigenvectors $\hat{\bm{n}}^1_1$ and $\hat{\bm{n}}^1_2$ together with their reflections $-\hat{\bm{n}}^1_1$ and $-\hat{\bm{n}}^1_2$, mapped onto
  $\mathbb{S}^2$. The green dashed line represents the geodesic path between $\hat{\bm{n}}^{1}_0$
  and $\hat{\bm{n}}^1_1$. Here vector $\hat{\bm{n}}^1_1$ is chosen for interpolation as
  $\textrm{d}_{\mathbb{S}^2}(\hat{\bm{n}}^{1}_0,\hat{\bm{n}}^1_1)<
    \textrm{d}_{\mathbb{S}^2}(\hat{\bm{n}}^{1}_0,-\hat{\bm{n}}^1_1)$. In the case of $\hat{\bm{n}}^1_2$, the reflected vector (blue dotted arrow) is chosen as
  $\textrm{d}_{\mathbb{S}^2}(\hat{\bm{n}}^{1}_0,-\hat{\bm{n}}^1_2)<
    \textrm{d}_{\mathbb{S}^2}(\hat{\bm{n}}^{1}_0,\hat{\bm{n}}^1_2)$.}
  \label{fig:eigenvector-reorientation}
\end{figure}

As discussed above, the eigenvectors $\hat{\bm{n}}^1_j, \hat{\bm{n}}^2_j, \hat{\bm{n}}^3_j$ of the tensors $\bm{U}_j$ are only defined up to their signs. The basic idea of the following re-orientation procedure is to define the orientation of the eigenvectors at all data points such that the angular distance between the eigenvectors is minimized. In a first step, the eigenvectors $\hat{\bm{n}}^1_j$ corresponding to the largest eigenvalues are considered. For the following procedure, a reference eigenvector $\hat{\bm{n}}^1_0$ is required, which can be conveniently chosen as an arbitrary eigenvector $\hat{\bm{n}}^1_j$ out of the data points $\bm{x}_j$. Within this work, the data point that is closest to the interpolation point $\bm{x}_p$ is chosen for this purpose as it has most influence on the interpolated values at $\bm{x}_p$. Now, for every eigenvector $\hat{\bm{n}}^1_j$ the geodesic distance with respect to $\hat{\bm{n}}^1_0$ is calculated according to $\textrm{d}_{\mathbb{S}^2}(\hat{\bm{n}}^1_0,\hat{\bm{n}}^1_j)= \arccos(\hat{\bm{n}}^1_0\cdot\hat{\bm{n}}^1_j)$ (note, that the unit eigenvectors $\hat{\bm{n}}^1_0$ and $\hat{\bm{n}}^1_j$ are elements of the unit-sphere $\mathbb{S}^2$, and the geodesic distance represents the enclosed angle). The sign of an eigenvector $\hat{\bm{n}}^1_j$ is inverted if $\textrm{d}_{\mathbb{S}^2}(\hat{\bm{n}}^1_0,-\hat{\bm{n}}^1_j)< \textrm{d}_{\mathbb{S}^2}(\hat{\bm{n}}^1_0,\hat{\bm{n}}^1_j)$ (see Figure~\ref{fig:eigenvector-reorientation}). The same procedure is applied to the eigenvectors $\hat{\bm{n}}^2_j$ associated with the median eigenvalues, and the third eigenvector $\hat{\bm{n}}^3_j$ at every data point is found by the vector product of the first and second eigenvector. Theoretically, there is an ambiguity if the angular separation is $\pm\pi/2$. However, for data points $\bm{x}_j$ that are reasonably close to the interpolation point $\bm{x}_p$, it is expected that the angle enclosed by associated eigenvectors is always smaller than $\pi/2$.

The same strategy is adopted to choose between quaternions $\hat{\bm{q}}$ and their antipodes $-\hat{\bm{q}}$ in the context of the quaternion-based interpolation strategy to ensure that all quaternions lie in the same hemisphere as discussed in Section~\ref{sec-quaternion-vector-interpolation}. An arbitrary quaternion $\hat{\bm{q}}^\mathrm{r}_j$ has to be chosen as reference quaternion
$\hat{\bm{q}}^\mathrm{r}_0$ for this procedure. Within this paper, again the data point closest to the interpolation point is chosen for this purpose. For all data points $j=1,...,N$, the geodesic distance between $\hat{\bm{q}}^\mathrm{r}_0$ and $\hat{\bm{q}}^\mathrm{r}_j$ is computed as $\textrm{d}_{\mathbb{S}^3}(\hat{\bm{q}}^\mathrm{r}_0,\hat{\bm{q}}^\mathrm{r}_j) = \arccos({\hat{\bm{q}}}^\mathrm{r}_0 \cdot {\hat{\bm{q}}}^\mathrm{r}_j)$ (see Appendix~\ref{appendix:d_S3}). The sign of a quaternion ${\hat{\bm{q}}}^\mathrm{r}_j$ is inverted if
$\textrm{d}_{\mathbb{S}^3}({\hat{\bm{q}}}^\mathrm{r}_0,-{\hat{\bm{q}}}^\mathrm{r}_j)<
  \textrm{d}_{\mathbb{S}^3}({\hat{\bm{q}}}^\mathrm{r}_0,{\hat{\bm{q}}}^\mathrm{r}_j)$.
\begin{figure}[ht]
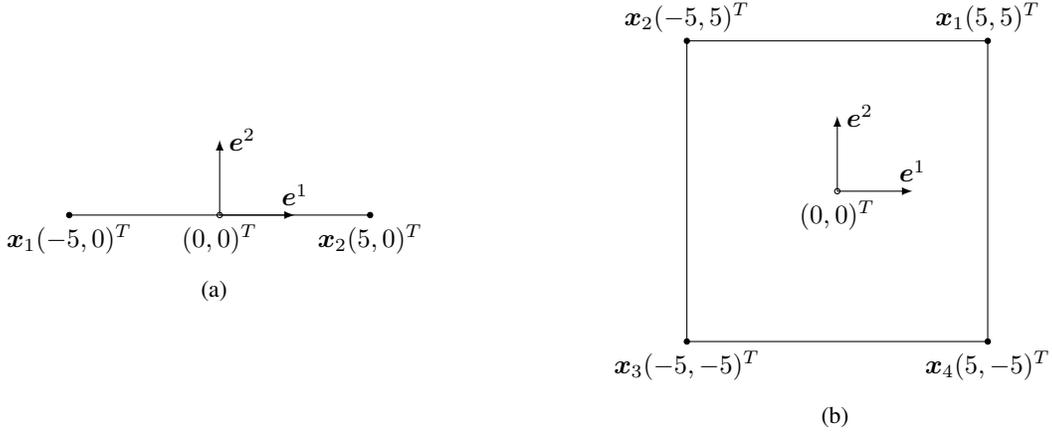

  \begin{subfigure}[h]{.5\textwidth}  \centering
    \input{pictures/problemsetup_1d.tikz}
    \caption{} \label{fig:problem_setup_1d} \end{subfigure}%
  \begin{subfigure}[h]{.5\textwidth}  \centering
    \input{pictures/problemsetup_2d.tikz}
    \caption{} \label{fig:problem_setup_2d} \end{subfigure}%
  \caption{\small Problem setup for (a) 1-dimensional examples with two data points at $\bm{x}_1=(-5,0)^T$ and
    $\bm{x}_2=(5,0)^T$ and (b) 2-dimensional examples with four data points at $\bm{x}_1=(5,5)^T$, $\bm{x}_2=(-5,5)^T$, $\bm{x}_3=(-5,-5)^T$, and $\bm{x}_4=(5,-5)^T$.}
  \label{fig:problem_steup}
\end{figure}

\begin{figure}[ht]
  \centering
  \begin{subfigure}[b]{.3\linewidth}
    \centering
    \includegraphics[trim=0cm 12cm 0cm 0cm, clip,width=\textwidth]{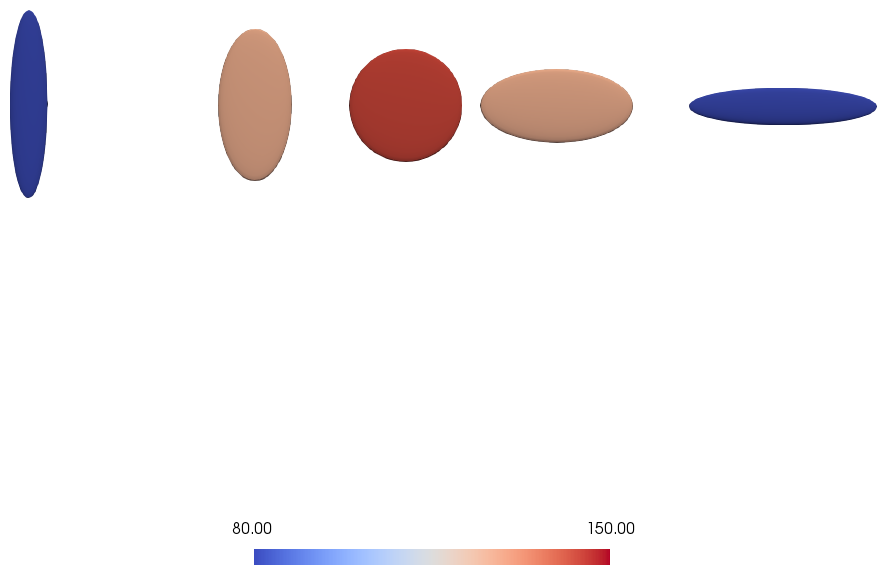}
    \caption{E: Case 1 and 2} \label{fig:Sym_1D_AMB_Euclidean} \end{subfigure}%
   \\%
   \centering
   \begin{subfigure}[b]{.3\linewidth}
     \centering
     \includegraphics[trim=0cm 12cm 0cm 0cm, clip,width=\textwidth]{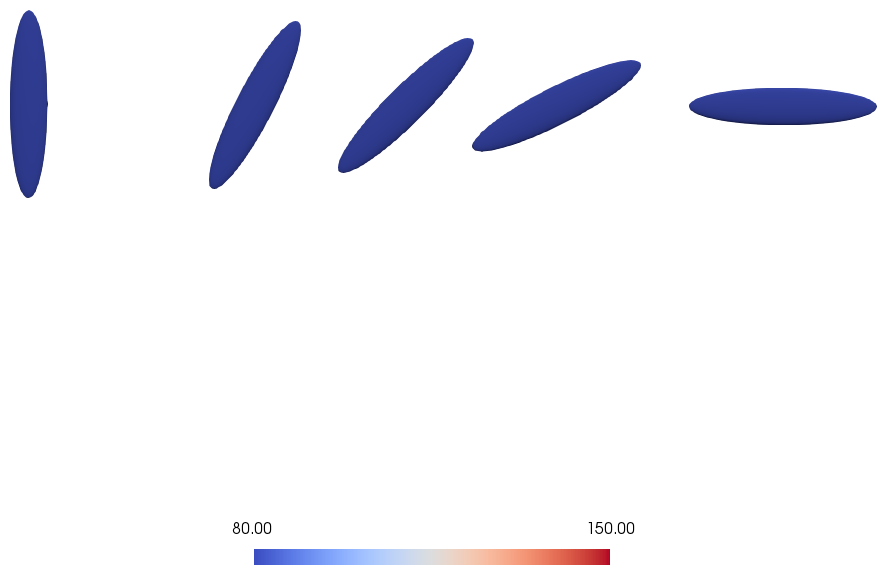}
     \caption{R-MLS: Case 1} \label{fig:Sym_1D_R_MLS_AMB_0} \end{subfigure}%
    \\%
    \centering
    \begin{subfigure}[b]{.3\linewidth}
      \centering
      \includegraphics[trim=0cm 12cm 0cm 0cm, clip,width=\textwidth]{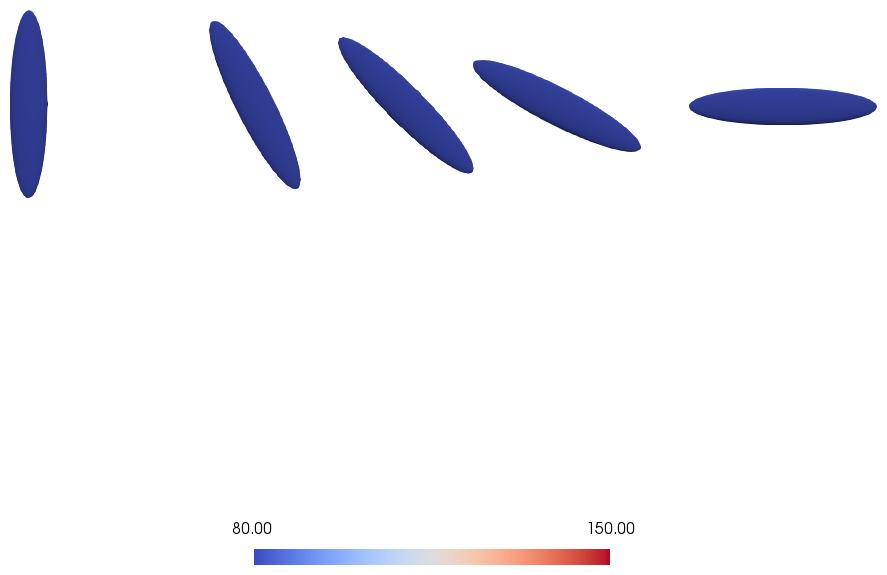}
      \caption{R-MLS: Case 2} \label{fig:Sym_1D_R_MLS_AMB_180} \end{subfigure}%
      \\
      \begin{subfigure}[b]{\linewidth}
        \centering
        \includegraphics[trim=0cm 0cm 0cm 18.0cm,clip,width=0.5\textwidth]{pictures/euclidean_angle_179.9999_1d_symmetric_exp_amb.png}
      \end{subfigure}%
      \\  \vspace{-50pt} \hspace{0pt}
      \begin{subfigure}[b]{\linewidth}
        \begin{tikzpicture}
          \coordinate [label = above:{$\bm{e}^1$}](x) at (1,0); \coordinate [label =
          right:{$\bm{e}^2$}](y) at (0,1); \coordinate [](C) at (0, 0); \draw[-latex] (C)--(x);
          \draw[-latex] (C)--(y);
        \end{tikzpicture}
      \end{subfigure}
  \caption{\small Comparison of rotation vector-based (R-MLS) and Euclidean (E) interpolation for interpolation between two tensors based on an eigenvector reorientation procedure. The two tensors $\bm{T}_1$ and $\bm{T}_2$ are placed at  $\bm{x}_1$  and $\bm{x}_2$, respectively (see Figure~\ref{fig:problem_setup_1d}). Tensor $\bm{T}_1$ is defined by the eigenvalues $\{\lambda^1_1, \lambda^2_1, \lambda^3_1\}=\{20, 4.0, 1.0\}$ whose  primary eigenvector is oriented at $\pi/2$ with respect to coordinate axis $\bm{e}^1$ (i.e., $\sphericalangle (\bm{e}^1,\hat{\bm{n}}^1_1) = \frac{\pi}{2})$. Tensor $\bm{T}_2$ has the  same eigenvalues as  $\bm{T}_1$. Two cases are considered based on the relative orientation of the primary eigenvector of $\bm{T}_2$. Case 1: $\sphericalangle (\bm{e}^1,\hat{\bm{n}}^1_2) \approx 0$ and case 2: $\sphericalangle (\bm{e}^1,\hat{\bm{n}}^1_2) \approx \pi$. The ellipsoidal representation of the interpolated tensors is illustrated for: (a) E (cases 1 and 2 identical), (b) R-MLS of case 1, and  (c) R-MLS of case 2. The ellipsoid color represents the tensor determinant.}
  \label{fig:reorienation_strategy}
\end{figure}

Finally, the eigenvector reorientation method presented above shall be illustrated by a numerical test case, considering two symmetric anisotropic tensors $\bm{T}_1$ and $\bm{T}_2$ at $\bm{x}_1$ and $\bm{x}_2$, respectively (see Figure~\ref{fig:problem_setup_1d} for the problem setup). The eigenvalues of both tensors are $\{\lambda^1_{1,2}, \lambda^2_{1,2}, \lambda^3_{1,2}\}=\{20, 4.0, 1.0\}$. The primary eigenvector $\hat{\bm{n}}^1_2$, i.e., the eigenvector associated with the largest eigenvalue, of tensor $\bm{T}_2$ is oriented at $\pi/2$ with respect to the coordinate axis $\bm{e}^1$, i.e., $\sphericalangle (\bm{e}^1,\hat{\bm{n}}^1_2)=\pi/2$. Two cases are considered based on the eigenvector orientation of $\bm{T}_1$. Case 1: $\sphericalangle (\bm{e}^1,\hat{\bm{n}}^1_1) \approx 0$ and case 2: $\sphericalangle (\bm{e}^1,\hat{\bm{n}}^1_1) \approx \pi$. Note that the respective eigenvectors resulting from these two cases are antiparallel, i.e., scaled by $-1$. Thus, both cases result in the same tensor. The interpolation is carried out by two different approaches, either an Euclidean (E), i.e., component-wise, tensor interpolation (see Section~\ref{sec:Results} for more details) and the R-MLS scheme using a rotation vector-based rotation interpolation. Note that the scheme used for eigenvalue interpolation (LOG, MLS or LOGMLS) does not influence the results for this example, since the eigenvalues of both tensors are identical. The results are illustrated in Figures~\ref{fig:Sym_1D_AMB_Euclidean} to~\ref{fig:Sym_1D_R_MLS_AMB_180}. Since the components of the final tensor $\bm{T}_1$ are identical for both aforementioned cases, the Euclidean interpolation yields the same result in both cases, which is illustrated in Figure~\ref{fig:Sym_1D_AMB_Euclidean}. Even though the two anisotropic tensors $\bm{T}_1$ and $\bm{T}_2$ only differ via a relative rotation while having identical eigenvalues, the Euclidean scheme leads to an interpolated tensor at the midpoint that is isotropic (the ellipsoidal representation results in a sphere). In other words, this simple interpolation scheme cannot preserve the anisotropy of the original tensor data, which is denoted as swelling. In contrast, the proposed approaches (e.g., R-MLS) result in the desired pure rotation interpolation in case of identical eigenvalues of both tensors, i.e., the anisotropy of the tensors is exactly preserved in this case. However, in the presented extreme case that the two tensors $\bm{T}_1$ and $\bm{T}_2$ differ by a relative rotation of $\pi/2$, the eigenvector interpolation becomes non-unique according to the two aforementioned cases of eigenvector orientation. The interpolation results for these two cases based on the R-MLS scheme are illustrated in Figures~\ref{fig:Sym_1D_R_MLS_AMB_0} and~\ref{fig:Sym_1D_R_MLS_AMB_180}. Finally, it shall be mentioned, that for practically relevant examples (i.e., when the tensor data points are reasonably close to the interpolation point) the relative orientation angle between the tensor eigenvector bases is typically smaller than $\pi/2$, thus the proposed interpolation schemes are unique. Moreover, in Section~\ref{sec:application_cont_mechanics}, an alternative scheme is presented to uniquely determine eigenvector orientations in the context of nonlinear continuum mechanics, which also works in case of very large relative rotations.

\subsubsection{Invariant properties of interpolation methods}
The tensor interpolation schemes proposed in the last sections are scaling invariant, i.e., under arbitrary scaling of all tensors $\bm{T}_j \in \mathbb{R}^{3 \times 3}, \,\, j=1,...,N$ with a positive scalar $\alpha \in \mathbb{R}^+$ the shape of the interpolated tensor $\bm{T}_p= \mathrm{Int}(\bm{T}_j ) \in \mathbb{R}^{3 \times 3}$ remains unchanged: 
\begin{align}
  \begin{split}
    \forall \alpha \in \mathbb{R}^+: \mathrm{Int}(\alpha \bm{T}_j )&= \mathrm{Int}(\bm{R}_j (\alpha \bm{U}_j) ) = \mathrm{Int}( \bm{R}_j  \bm{Q}_j^{T} (\alpha \bm{\Lambda}_j )\bm{Q}_j ) \\
    &= \mathrm{Int}(\bm{R}_j) \ \mathrm{Int}(\bm{Q}_j^{T}) \ \mathrm{Int}(\alpha \bm{\Lambda}_j)  \ \mathrm{Int}(\bm{Q}_j) = \alpha \  \mathrm{Int}(\bm{R}_j) \ \mathrm{Int}(\bm{Q}_j^{T}) \ \mathrm{Int}(\bm{\Lambda}_j)  \ \mathrm{Int}(\bm{Q}_j)\\
    &= \alpha \ \mathrm{Int}(\bm{T}_j ),
  \end{split}
\end{align}
where the operator $\mathrm{Int}(\cdot)$ represents the interpolation. In the context of the eigenvalue interpolation, the scaling invariance $\mathrm{Int}(\alpha \bm{\Lambda}_j)=\alpha \mathrm{Int}(\bm{\Lambda}_j)$ is well-known and can be verified in a straight-forward manner for all the three employed schemes, i.e., the logarithmic weithed average (LOG) according to Section~\ref{sec:eigenvalue_interpolation_log}, the moving least squares (MLS) approach according to Section~\ref{sec:eigenvalue_interpolation_mls}, and the logarithmic moving least squares (LOGMLS) approach according to Section~\ref{sec:eigenvalue_interpolation_logmls}.  Among others, scaling invariance makes the methods independent of the choice of physical units. In addition to scaling invariance, the rotation
vector-based and quaternion-based rotation interpolation schemes according to Section~\ref{sec-rotation-vector-interpolation} and~\ref{sec-quaternion-vector-interpolation} are rotational invariant. This means, if all rotation tensors $\bm{R}_j, \,\, j=1,...,N$ are rotated by a constant rotation tensor $\bm{M} \in \mathbb{SO}(3)$, also the interpolated rotation tensor $\bm{R}_p=\mathrm{Int}(\bm{R}_j) \in \mathbb{SO}(3)$ is rotated by $\bm{M}$:
\begin{align}
  \forall \bm{M} \in \mathbb{SO}(3):\quad   \mathrm{Int}(\bm{M} \bm{R}_j(\bm{\theta}))= \bm{M}\mathrm{Int}(\bm{R}_j(\bm{\theta})) \quad \textrm{and} \quad  \mathrm{Int}(\bm{M} \bm{R}_j(\bm{q}) )= \bm{M}\mathrm{Int}(\bm{R}_j(\bm{q})).
\end{align}
This property is fulfilled since relative rotation vectors (and quaternions) are used for interpolation, as demonstrated in~\cite{Crisfield1999}. Thus, due to $\mathrm{Int}(\bm{M} \bm{R}_j)=\bm{M} \mathrm{Int}(\bm{R}_j)$ and $\mathrm{Int}(\bm{Q}_j \bm{M}^T)=\mathrm{Int}(\bm{Q}_j) \bm{M}^T$, also the following rotation invariance is fulfilled for the tensor $\bm{T}_p=\mathrm{Int}(\bm{T}_j)=\mathrm{Int}(\bm{R}_j) \ \mathrm{Int}(\bm{Q}_j^{T}) \ \mathrm{Int}(\bm{\Lambda}_j)  \ \mathrm{Int}(\bm{Q}_j)$:
\begin{align}
  \forall \bm{M} \in \mathbb{SO}(3):\quad   \mathrm{Int}(\bm{M} \bm{T}_j \bm{M}^T)= \bm{M}  \mathrm{Int}(\bm{T}_j) \bm{M}^T. \label{eqn:rotationinvariance}
\end{align}
This property makes the interpolation methods objective, i.e., invariant under arbitrary rigid body rotations. Finally, the combined scaling and rotation invariance can be stated as:
\begin{align}
  \forall \alpha \in \mathbb{R}^+ \quad \textrm{and} \quad \forall \bm{M} \in \mathbb{SO}(3):\quad   \mathrm{Int}(\bm{M}\alpha  \bm{T}_j \bm{M}^T)=\alpha (\bm{M}  \mathrm{Int}(\bm{T}_j) \bm{M}^T).
\end{align}
The scaling and rotation invariance is highly desirable in applications such as DTI processing and problems of nonlinear continuum mechanics (solved, e.g., by the finite element method).

\begin{figure}
   \centering
  \begin{subfigure}[b]{.33\linewidth}\hspace{10pt}
    \includegraphics[trim=0cm 7cm 0cm 0cm,clip,width=0.80\textwidth]{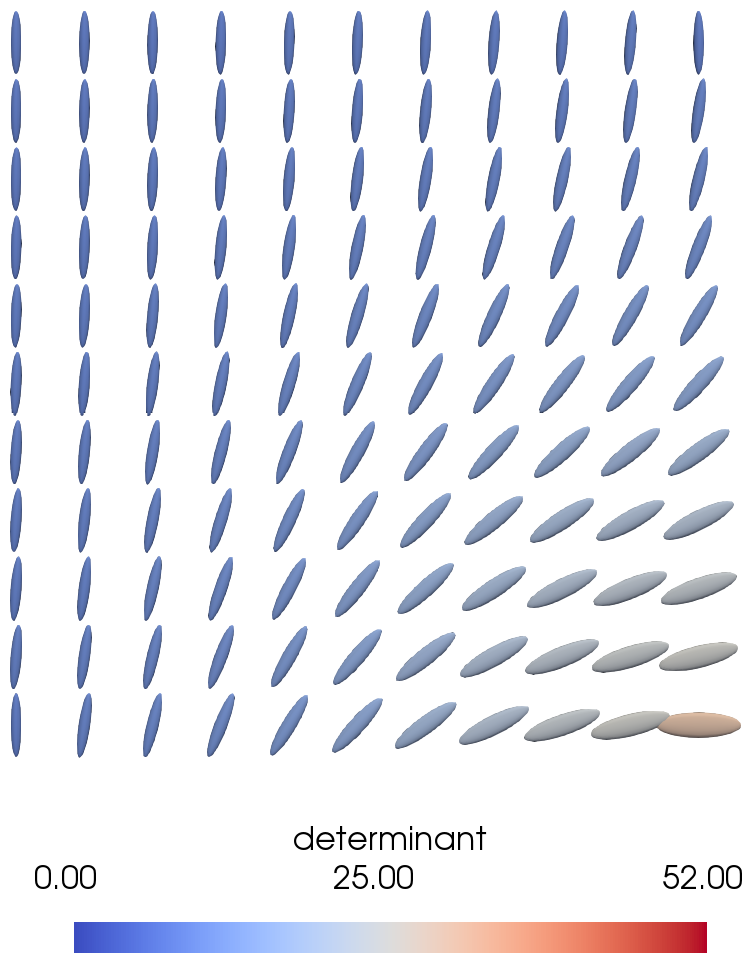}
    \caption{Quaternion: $  \Int(\bm{T}_j)$ }  \label{fig:obj_2d_sym_quaternion_1}
  \end{subfigure}
  \begin{subfigure}[b]{.33\linewidth}\hspace{10pt}
    \includegraphics[trim=0cm 7cm 0cm 0cm,clip,width=0.80\textwidth]{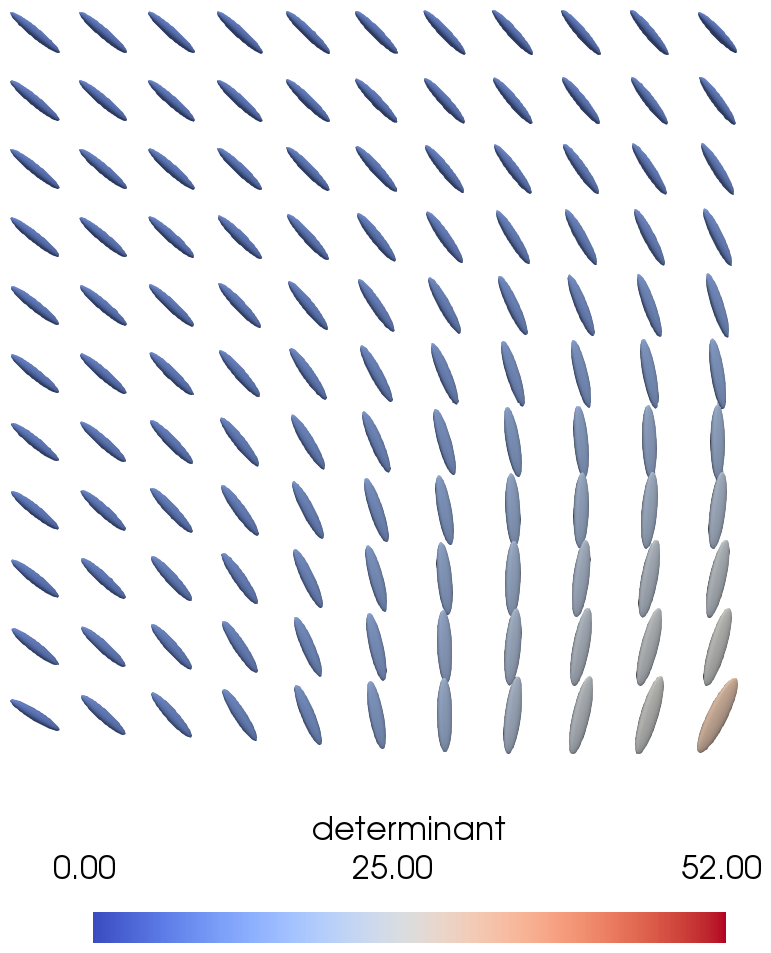}
    \caption{Quaternion: $ \Int(\bm{M} \bm{T}_j \bm{M}^T) $}  \label{fig:obj_2d_sym_quaternion_2}
  \end{subfigure}
  \begin{subfigure}[b]{.33\linewidth}\hspace{10pt}
    \includegraphics[trim=0cm 7cm 0cm 0cm,clip,width=0.80\textwidth]{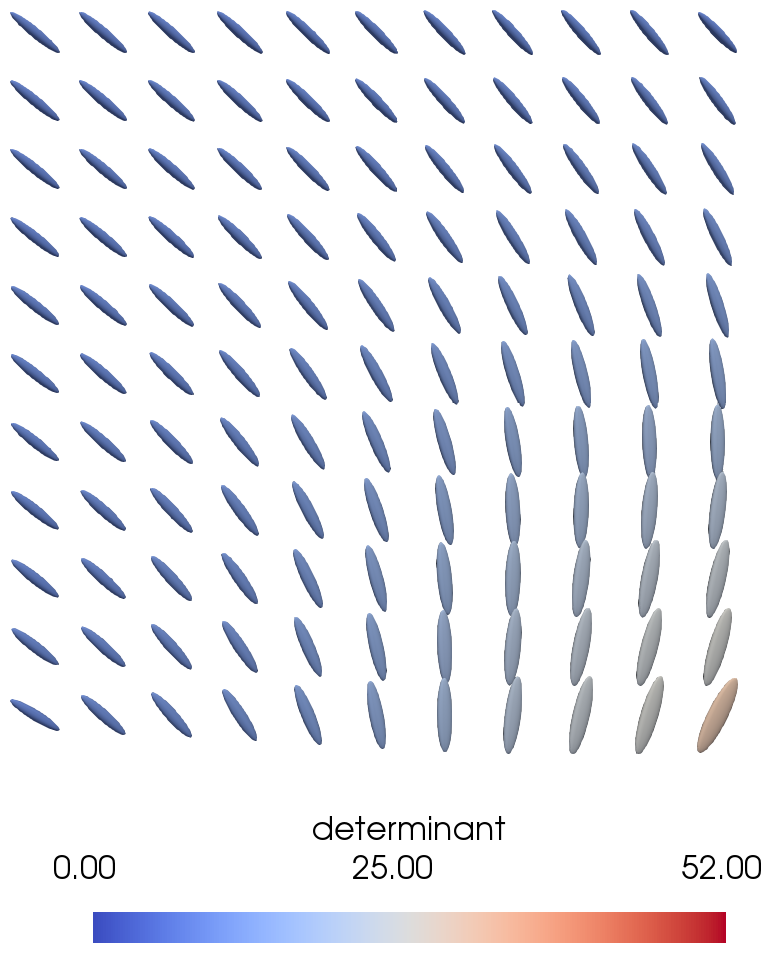}
    \caption{Quaternion: $ \bm{M} \Int(\bm{T}_j) \bm{M}^T$} \label{fig:obj_2d_sym_quaternion_3}
  \end{subfigure} \\ \vspace{4pt}
  \centering
  \begin{subfigure}[b]{.33\linewidth}\hspace{10pt}
    \includegraphics[trim=0cm 7cm 0cm 0cm,clip,width=0.80\textwidth]{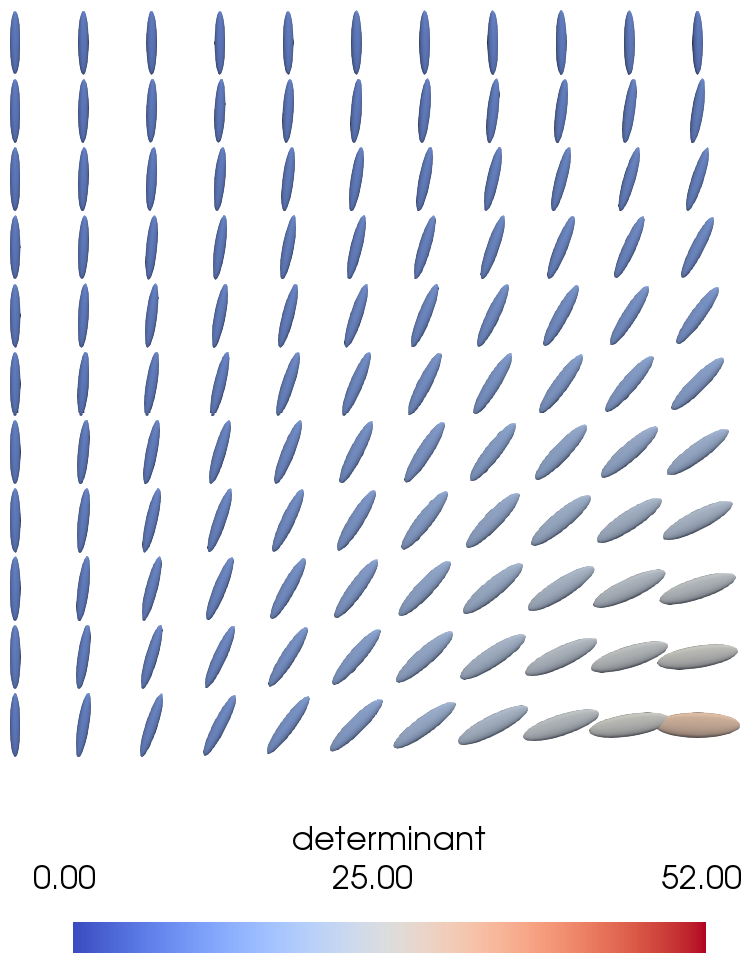}
    \caption{Rotation vector: $  \Int(\bm{T}_j)$ }   \label{fig:obj_2d_sym_rotationvector_1}
  \end{subfigure}
  \begin{subfigure}[b]{.33\linewidth}\hspace{10pt}
    \includegraphics[trim=0cm 7cm 0cm 0cm,clip,width=0.80\textwidth]{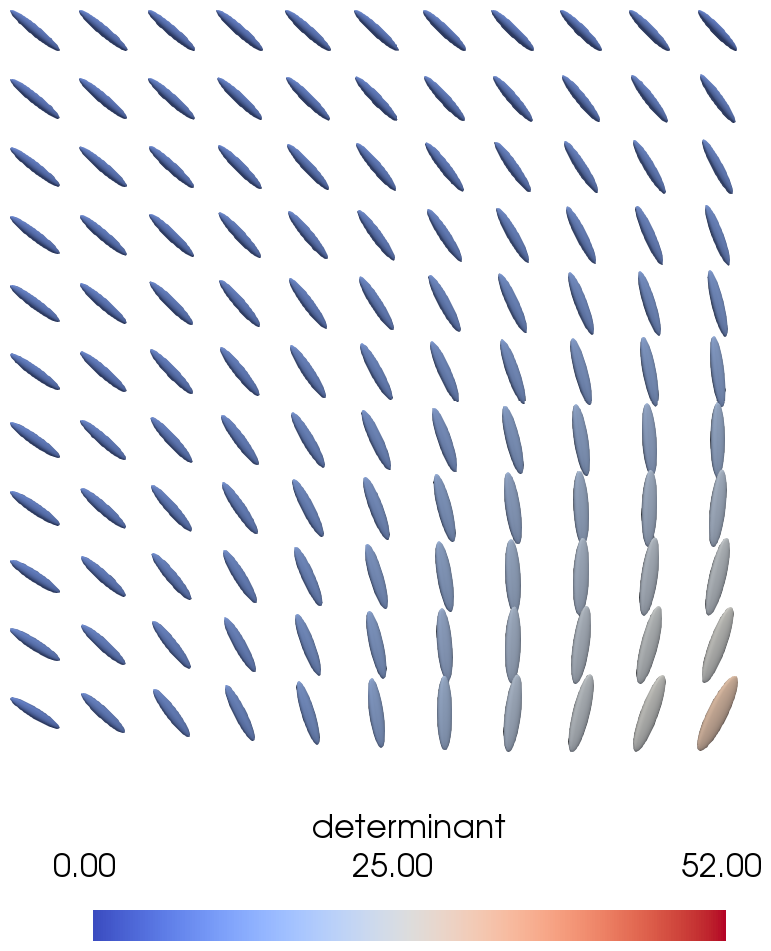}
    \caption{Rotation vector: $ \Int(\bm{M} \bm{T}_j \bm{M}^T) $}
    \label{fig:obj_2d_sym_rotationvector_2}
  \end{subfigure}
  \begin{subfigure}[b]{.33\linewidth}\hspace{10pt}
    \includegraphics[trim=0cm 7cm 0cm 0cm,clip,width=0.80\textwidth]{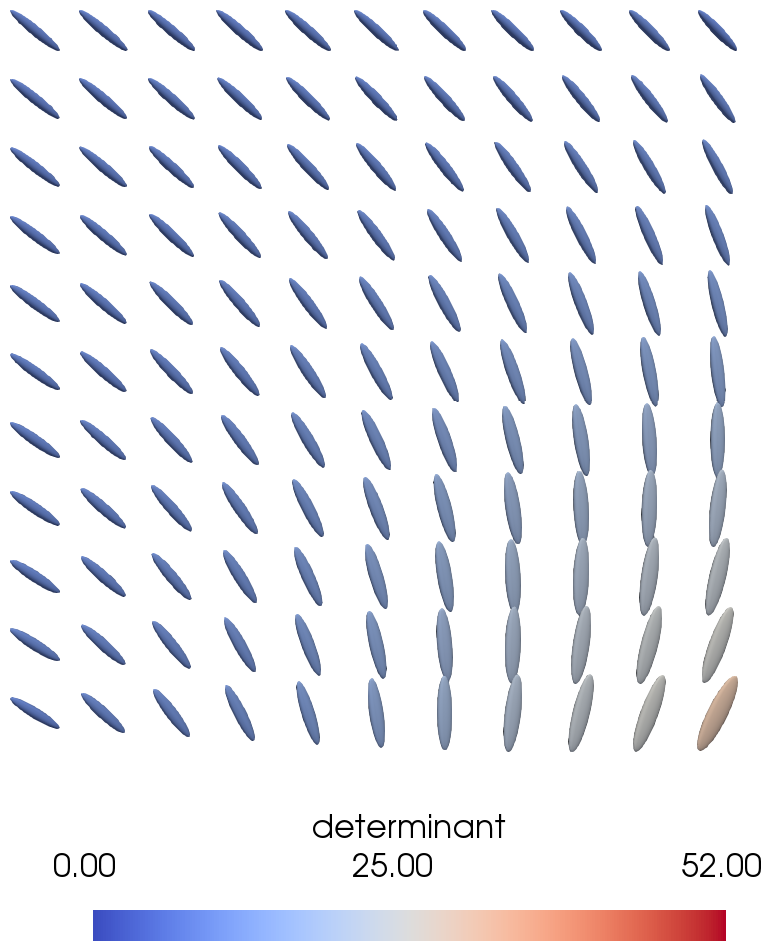}
    \caption{Rotation vector: $ \bm{M} \Int(\bm{T}_j) \bm{M}^T$}
    \label{fig:obj_2d_sym_rotationvector_3}
  \end{subfigure} \\  \vspace{-32pt} \hspace{0pt}
  \begin{subfigure}[b]{\linewidth}
    \begin{tikzpicture}
      \coordinate [label = above:{$\bm{e}^1$}](x) at (1,0);
      \coordinate [label = right:{$\bm{e}^2$}](y) at (0,1);
      \coordinate [](C) at (0, 0);
      \draw[-latex] (C)--(x);
      \draw[-latex] (C)--(y);
    \end{tikzpicture}
  \end{subfigure}
  \caption{\small Verification of rotation invariance for interpolation of four symmetric tenors. Subfigures (a) and (d): Interpolation results for original tensors for quaternion- and rotation vector-based rotation interpolations methods. Subfigures (b)
    and (e): Interpolation results when a rigid body rotation $\bm{M} \bm{T}_j \bm{M}^T $ with
    $\bm{M}=\bm{M}_{\bm{e}^3}(\frac{\pi}{3})\bm{M}_{\bm{e}^2}(\frac{\pi}{6})\bm{M}_{\bm{e}^1}(\frac{\pi}{12})$ is pre-multiplied to the input data. Subfigures (c) and (f): Post-multiplication of the original interpolated tensor field in (a) and (d) according to $\bm{M}
      \mathrm{Int}(\bm{T}_j)\ \bm{M}^T$.}
  \label{fig:Symmetric_Tensor_interpolation_2D_Objectivity}
\end{figure}
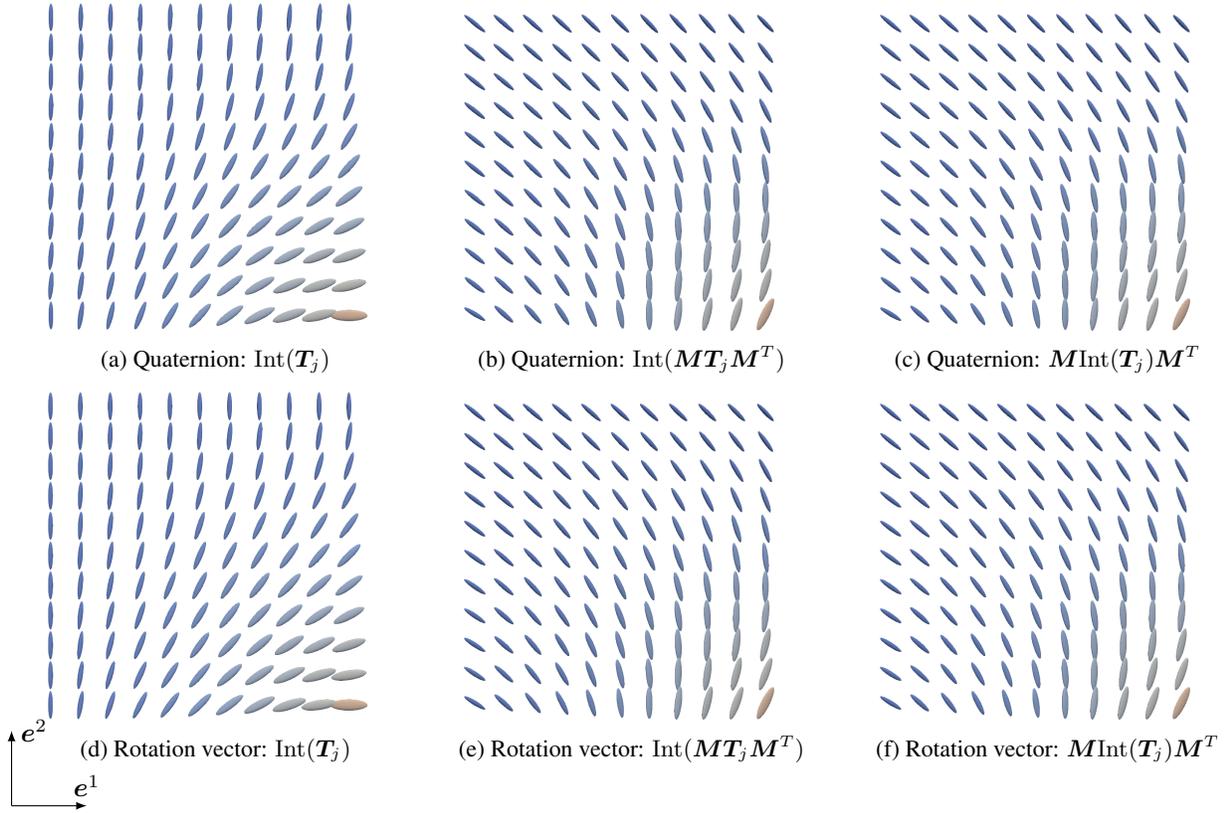
\begin{figure}
  \centering
    \input{pictures/2d_sym_obj_comparison_midpoint.tikz}
  \caption{\small Verification of rotation invariance for interpolation of four symmetric tenors. Deviation in eigenvector orientation at the interpolation point $\bm{x}_p=(0,0)^T$ according to $\parallel \theta_{\bm{n}} \parallel = \Big(\sum^{3}_{i=1} \big[\arccos (\bm{n}^i_{p, {\mathrm{Int}}
    \left( \bm{M} \bm{T}_{j} \bm{M}^{T} \right)} \cdot \bm{n}^i_{p, \bm{M} {\mathrm{Int}} \left(
    \bm{T}_{j} \right) \bm{M}^{T}})\big]^{2}\Big)^{1/2}$ for the proposed methods R-LOG and Q-LOG as well as for four methods from the literature: Euclidean interpolation (E), Log-Euclidean interpolation (LOG-E), Cholesky interpolation (C) and Log-Cholesky interpolation (LOG-C).}
      \label{fig:obj_2d_sym_comparison}
\end{figure}

In the following, the rotation invariance shall be verified numerically using four tensors. Three identical tensors with primary eigenvector orientation of $\sphericalangle
  (\bm{e}^1,\hat{\bm{n}}^1_{1,2,3}) \approx \pi/2$ with eigenvalues $\{\lambda^1_{1,2,3},
  \lambda^2_{1,2,3}, \lambda^3_{1,2,3}\}=\{7.5,1.25, 1.0\}$ are placed at $\bm{x}_1=(5,5)^T$,
$\bm{x}_2=(-5,5)^T$, and $\bm{x}_3=(-5,-5)^T$, see Figure~\ref{fig:problem_setup_2d}. The fourth tensor is positioned at $\bm{x}_4=(5,-5)^T$ with a primary eigenvector orientation of $\sphericalangle (\bm{e}^1,\hat{\bm{n}}^1_{4})=0$ with eigenvalues $\{\lambda^1_{4}, \lambda^2_{4},
  \lambda^3_{4}\}=\{10, 3, 1.0 \}$. The tensors are rotated by a constant rotation tensor $\bm{M}=\bm{M}_{\bm{e}^3}(\frac{\pi}{3})\bm{M}_{\bm{e}^2}(\frac{\pi}{6})\bm{M}_{\bm{e}^1}(\frac{\pi}{12})$ (see~\cite{jog2015continuum}, Equation 1.108), where $\bm{M}_{\bm{e}^i}(\theta^j)$ is the rotation about the coordinate axis ${\bm{e}^i}$ by an angle $\theta^j$,   resulting in a general 3D rotation. The results are visualized in Figures~\ref{fig:obj_2d_sym_quaternion_1} to~\ref{fig:obj_2d_sym_quaternion_3} for the proposed Q-LOG scheme (quaternion-based interpolation or rotations), and in Figures~\ref{fig:obj_2d_sym_rotationvector_1} to~\ref{fig:obj_2d_sym_rotationvector_3} for the proposed R-LOG scheme (rotation vector-based interpolation of rotations). Specifically, Figures~\ref{fig:obj_2d_sym_quaternion_1} and~\ref{fig:obj_2d_sym_rotationvector_1} represent the non-rotated, original tensor field resulting from the interpolation $\mathrm{Int}(\bm{T}_j), \,\, j=1,2,3,4$; Figures~\ref{fig:obj_2d_sym_quaternion_2} and~\ref{fig:obj_2d_sym_rotationvector_2} represent the tensor field resulting from interpolation of the rotated tensors $\bm{M} \bm{T}_j \bm{M}^T, \,\, j=1,2,3,4$, and  Figures~\ref{fig:obj_2d_sym_quaternion_3} and~\ref{fig:obj_2d_sym_rotationvector_3} represent the original, interpolated tensor field subsequently rotated according to $ \bm{M} \mathrm{Int}(\bm{T}_j) \bm{M}^T$. Visually, no difference can be seen between the variants $ \mathrm{Int}(\bm{M} \bm{T}_j \bm{M}^T) $ and $ \bm{M} \mathrm{Int}(\bm{T}_j) \bm{M}^T$ for both interpolation schemes, which confirms rotation invariance according to~\eqref{eqn:rotationinvariance}. The same procedure is carried out for four well-established tensor interpolation schemes from literature, i.e., for the so-called Euclidean (E), Log-Euclidean (LOG-E), Cholesky (C), and Log-Cholesky (LOG-C) interpolation as introduced in Section~\ref{sec:Results}. In order to quantify the deviation in eigenvector orientation at the interpolation point $\bm{x}_p=(0,0)^T$ due to non-objectivity, the following metric is introduced
\begin{align}
   \parallel \theta_{\bm{n}} \parallel = \Big(\sum^{3}_{i=1} \big[\arccos (\bm{n}^i_{p, {\mathrm{Int}}
    \left( \bm{M} \bm{T}_{j} \bm{M}^{T} \right)} \cdot \bm{n}^i_{p, \bm{M} {\mathrm{Int}} \left(
    \bm{T}_{j} \right) \bm{M}^{T}})\big]^{2}\Big)^{1/2}, 
\end{align}
which is plotted for the two proposed interpolation schemes and the four schemes from the literature in Figure~\ref{fig:obj_2d_sym_comparison}. It can be seen that, apart from the proposed methods, only the Euclidean interpolation scheme is objective, owing to the distributive property of the tensor multiplication. The other three interpolation schemes from literature (LOG-E, C, and LOG-C) suffer from severe non-objectivity.
\section{Results} \label{sec:Results} This section presents numerical examples to verify the proposed tensor interpolation schemes using synthetic data for non-symmetric and symmetric tensors. For the case of symmetric tensors, also existing interpolation schemes from literature can be employed. For comparison, we consider the following four schemes from literature:
\begin{enumerate}
    \item Euclidean interpolation (E): This scheme represents a component-wise
weighted average $\bm{T}_p= \sum^N_{j=1} w_j \bm{T}_j$.
    \item Cholesky interpolation (C)~\cite{wang2004constrained}: First, the tensor is decomposed into lower triangular matrices according to $\bm{T}_j= \bm{L}_j \bm{L}^T_j $, followed by a component-wise weighted averaging of $\bm{L}_j$ according to $\bm{L}_p= \sum^N_{j=1} w_j \bm{L}_j$. Finally, the interpolated tensor reads  $\bm{T}_p = \bm{L}_p \bm{L}_p^T$.
    \item Log-Euclidean interpolation (LOG-E)~\cite{arsigny2006log}: First, perform the eigendecompostion (see~\eqref{spectral-decomposition}) such that  $\bm{T}_j := \bm{Q}_j^T  \bm{\Lambda}_j \bm{Q}_j$ then, the tensor logarithms are computed based on the definition $\ln(\bm{T}_j) := \bm{Q}_j^T \ln(\bm{\Lambda}_j) \bm{Q}_j$. Finally, the interpolated tensor is determined as the exponential of the weighted average of the tensor logarithms according to $\bm{T}_p= \exp(\sum^N_{j=1} w_j \ \ln(\bm{T}_j))$. 
  \item Log-Cholesky (LOG-C)~\cite{lin2019riemannian}: First,
the tensor is decomposed into lower triangular tensors according to $\bm{T}_j= \bm{L}_j \bm{L}^T_j$ such that $
  \bm{L}_j = \left\lfloor  \bm{L}_j \right\rfloor + \left\lceil \bm{L}_j \right\rceil $, where
$\left\lfloor \cdot \right\rfloor $ is the strictly positive diagonal part of $\bm{L}_j$ and
$\left\lceil  \cdot \right\rceil $ is the remaining unrestricted part of $\bm{L}_j$. Then, the lower triangular matrix is
interpolated according to $ \bm{L}_p = \exp(\sum^N_{j=1} w_j \ \ln(\left\lfloor\bm{L}_j \right\rfloor)) +
  \sum^N_{j=1} w_j \left\lceil \bm{L}_j \right\rceil $. Finally, the interpolated tensor is reconstructed as $\bm{T}_p= \bm{L}_p \bm{L}^T_p$.
\end{enumerate}

For a quantitative assessment of the shape and size of the interpolated tensors we consider the following metrics: determinant, trace, fractional anisotropy
(FA) and Hilbert anisotropy (HA)~\cite{birkhoff1957extensions}. FA is a rotation-invariant
dimensionless parameter defined as 
\begin{align}
\textrm{FA}= \sum^3_{i=1} 3(\lambda_i - \bar{\lambda})^2/
  \sum^3_{i=1} 2\lambda^2_i \in [0,\ 1],
\end{align}
where $\bar{\lambda}=(\lambda_1+\lambda_2+\lambda_3)/3$ is the arithmetic mean of the eigenvalues. An
isotropic tensor results in $\text{FA}=0$, and for a highly anisotropic tensor $\text{FA} \rightarrow 1$. The rotation invariant parameter HA
is defined as 
\begin{align*}
\textrm{HA}=\ln(\lambda_\text{max}/\lambda_\text{min}),
\end{align*}
where $\lambda_\text{max}$ and $\lambda_\text{min}$ are the maximal and minimal eigenvalue, respectively. An
isotropic tensor results in $\text{HA}=0$, and for a highly anisotropic tensor $\text{HA} \rightarrow \infty$. Note that we always have positive eigenvalues for the considered positive definite tensors. The dimensionless parameters FA and HA are scaling invariant, i.e., they only depend on the shape of the tensor and remain unchanged when scaling the tensor by a scalar factor. Moreover, we quantify the relative orientation between two symmetric tensors using the cosine of the angle included by the primary eigenvectors (i.e., the eigenvectors associated with the maximal eigenvalue) according to
\begin{align}
 \cos[\sphericalangle(\hat{\bm{n}}^1_1,\hat{\bm{n}}^1_2)] =\hat{\bm{n}}^1_1 \cdot \hat{\bm{n}}^1_2.
\end{align}
\subsection{Interpolation of symmetric, positive-definite tensors}

In the first two numerical examples, symmetric tensors are considered. This scenario allows for comparison with well-known tensor interpolation schemes from the literature such as Euclidean, Log-Euclidean, Cholesky, and Log-Cholesky interpolation, which have been designed for symmetric tensors only.
\subsubsection{Interpolation of two symmetric tensors} \label{sec:interpolation_two_tensor_pd}
\begin{figure}[ht]
  \centering
  \begin{subfigure}[b]{.3\linewidth}
    \centering
    \includegraphics[trim=0cm 12cm 0cm 0cm, clip,width=0.95\textwidth]{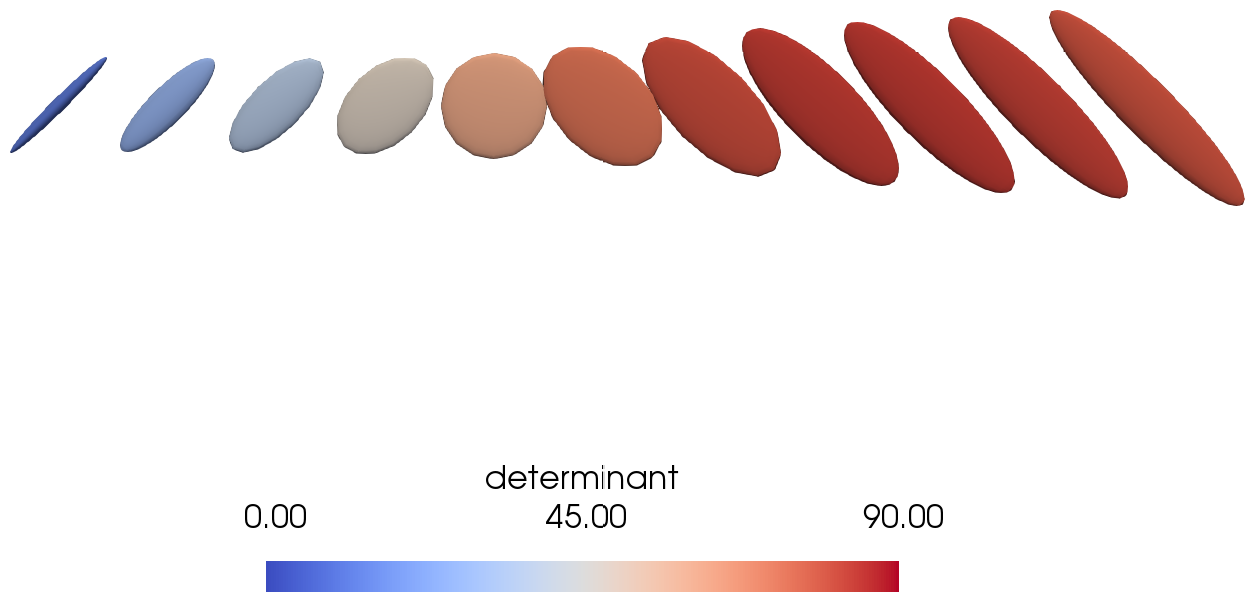}
    \caption{Euclidean} \label{fig:Sym_1D_Euclidean} \end{subfigure}%
  \begin{subfigure}[b]{.3\linewidth}
    \centering
    \includegraphics[trim=0cm 12cm 0cm 0cm,clip,width=0.95\textwidth]{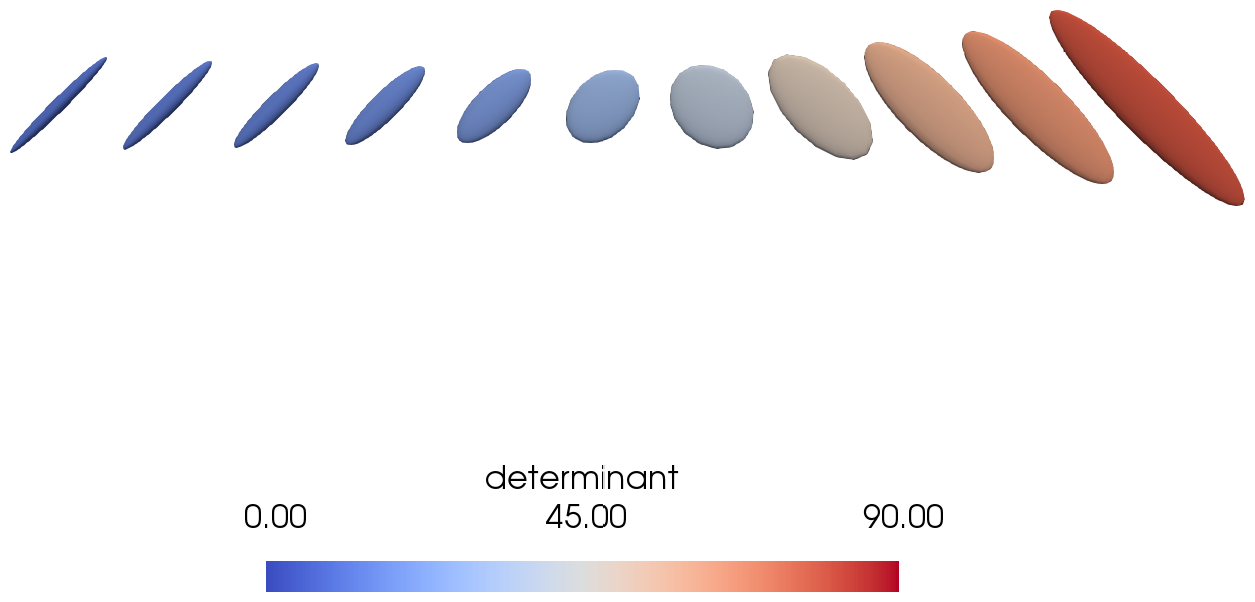}
    \caption{Log-Euclidean} \label{fig:Sym_1D_logEuclidean}
  \end{subfigure}
  \begin{subfigure}[b]{.3\linewidth}
    \centering
    \includegraphics[trim=0cm 12cm 0cm 0cm,clip,width=0.95\textwidth]{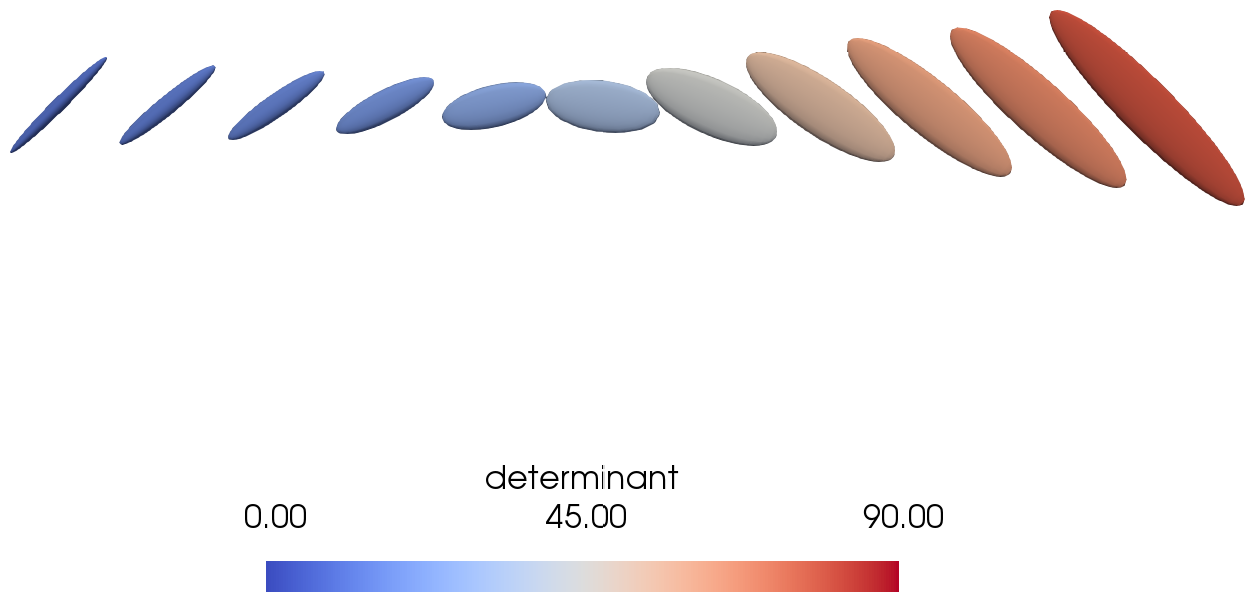}
    \caption{Cholesky} \label{fig:Sym_1D_Cholesky} \end{subfigure}\\%
  \vspace{0cm}
  \begin{subfigure}[b]{.3\linewidth} \vspace{8pt}
    \centering
    \includegraphics[trim=0cm 12cm 0cm 0cm,clip,width=0.95\textwidth]{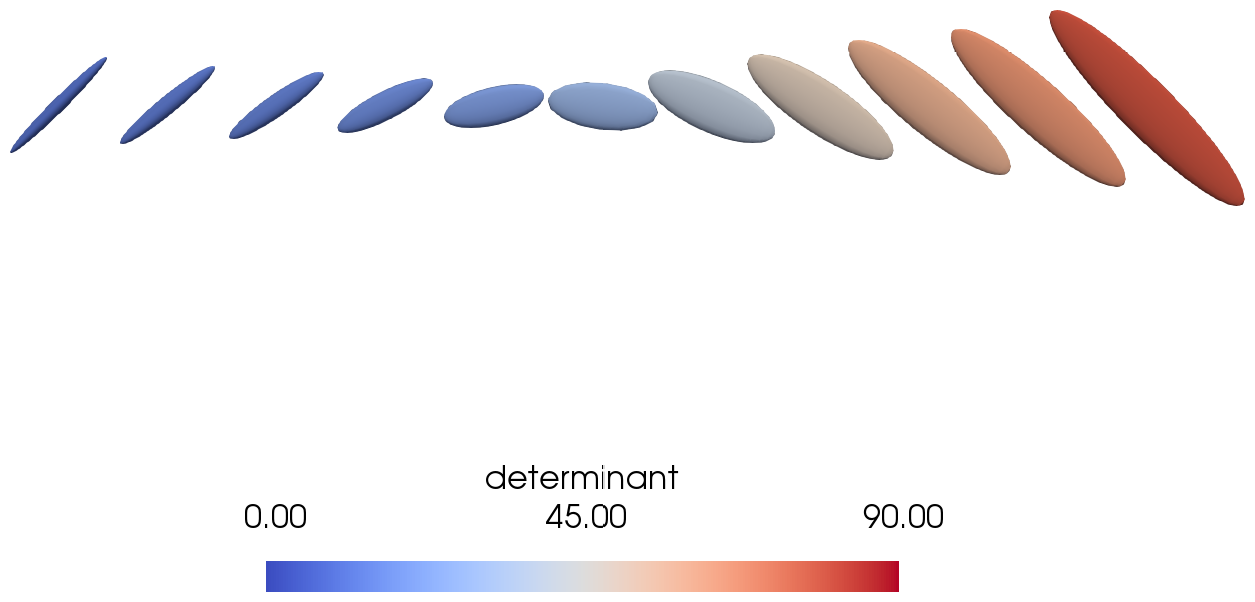}
    \caption{Log-Cholesky} \label{fig:Sym_1D_logCholesky} \end{subfigure}%
  \begin{subfigure}[b]{.3\linewidth} \vspace{8pt}
    \centering
    \includegraphics[trim=0cm 12cm 0cm 0cm,clip,width=0.95\textwidth]{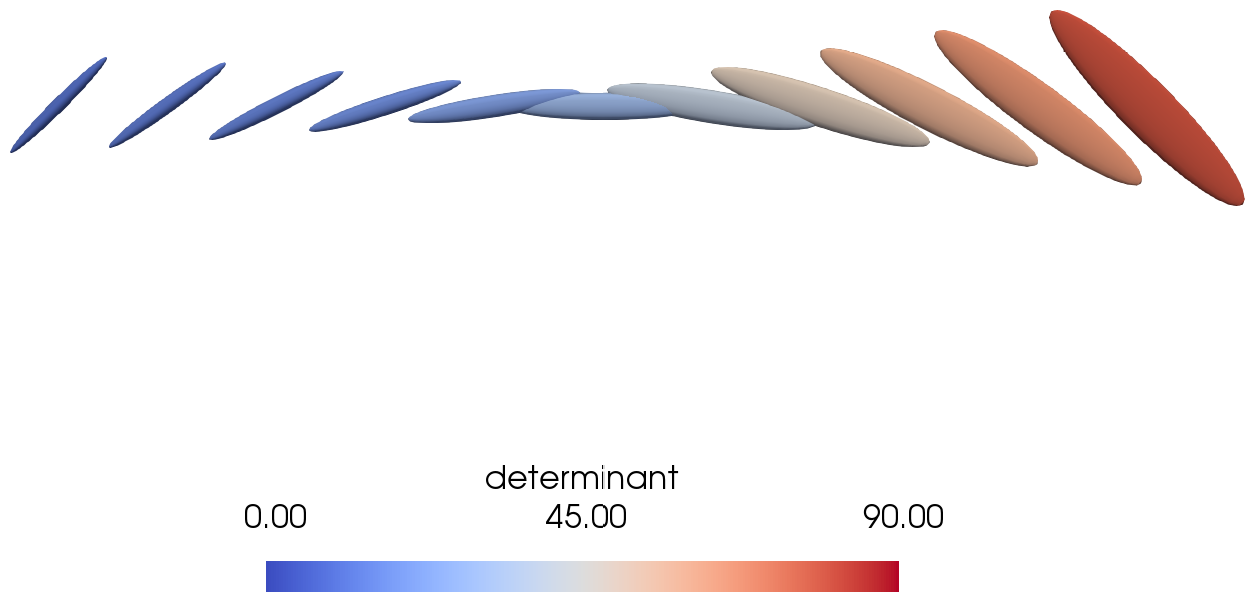}
    \caption{R-LOG} \label{fig:Sym_1D_RV} \end{subfigure}%
  \begin{subfigure}[b]{.3\linewidth} \vspace{8pt}
    \centering
    \includegraphics[trim=0cm 12cm 0cm 0cm,clip,width=0.95\textwidth]{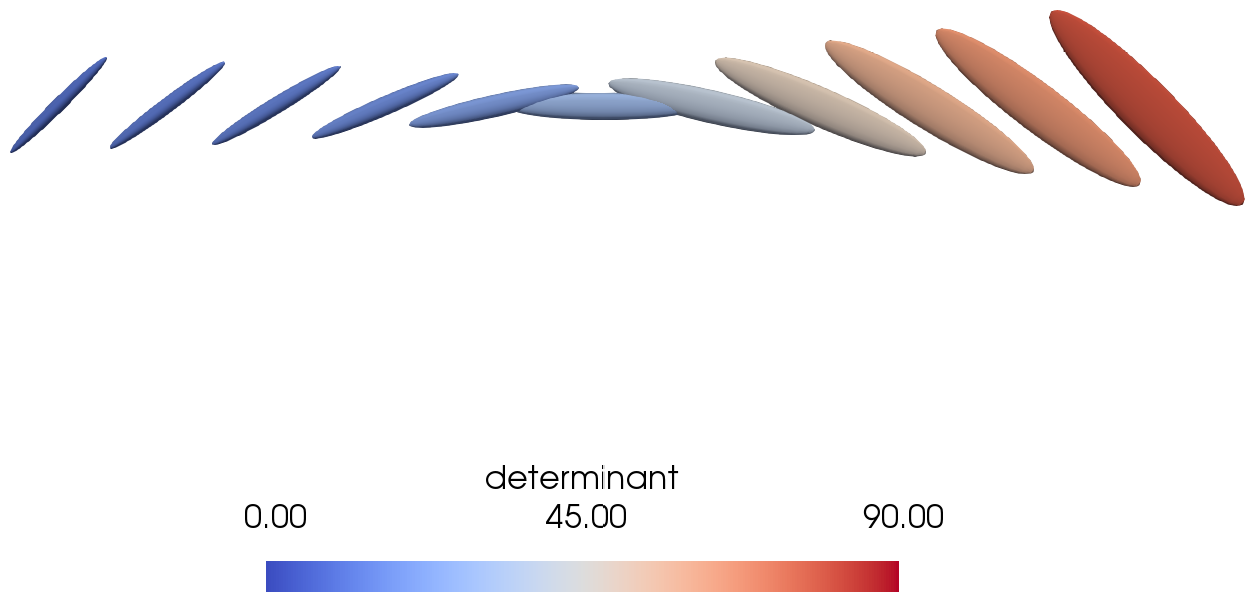}
    \caption{Q-LOG } \label{fig:Sym_1D_Quad} \end{subfigure}\\
    \begin{subfigure}[b]{.3\linewidth}\vspace{8pt}
      \centering
      \includegraphics[trim=0cm 12cm 0cm 0cm,clip,width=0.95\textwidth]{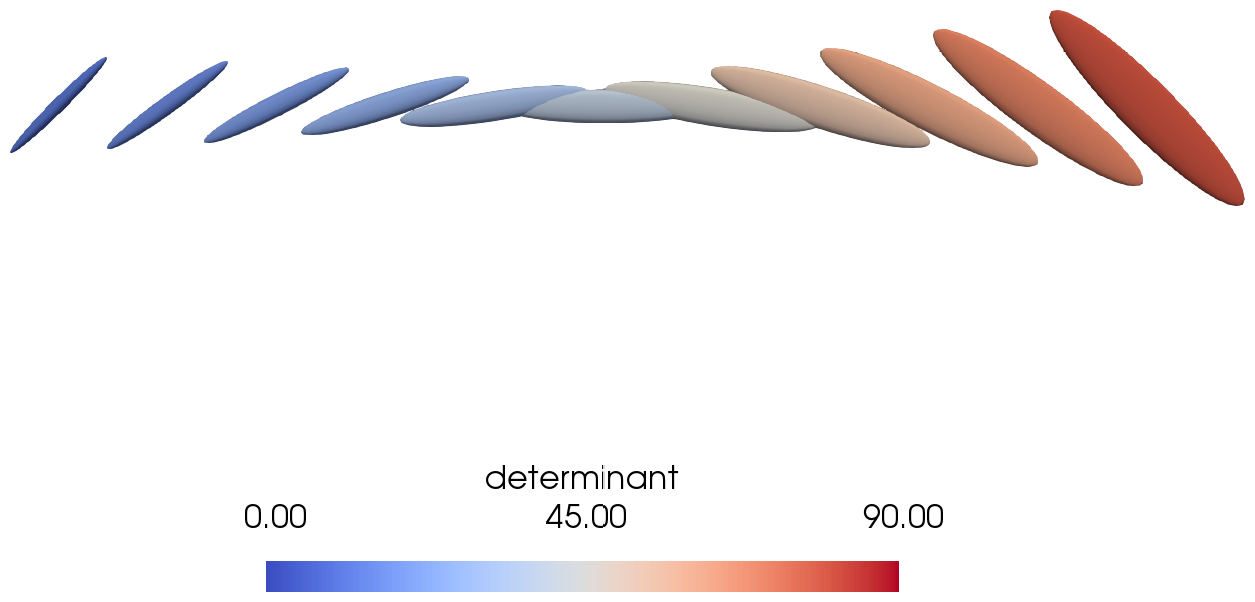}
      \caption{R-MLS } \label{fig:Sym_1D_RV_MLS} \end{subfigure} 
     \begin{subfigure}[b]{.3\linewidth}\vspace{8pt}
      \centering
      \includegraphics[trim=0cm 12cm 0cm 0cm,clip,width=0.95\textwidth]{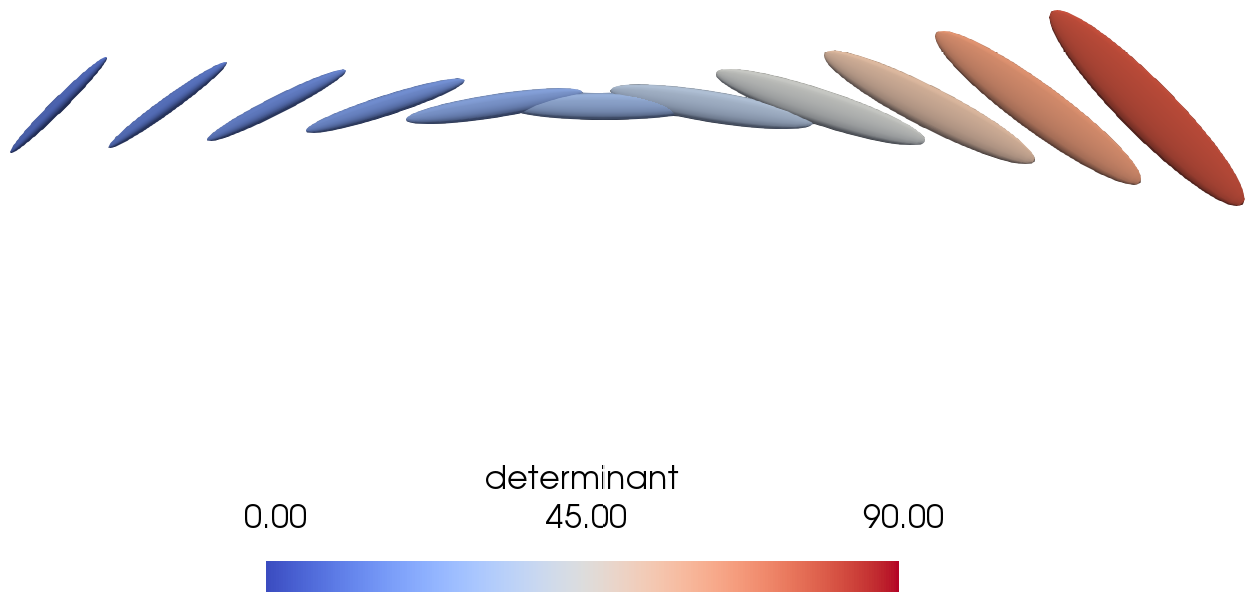}
      \caption{R-LOGMLS } \label{fig:Sym_1D_RV_LogMLS} \end{subfigure} \\%
  \begin{subfigure}[b]{\linewidth}
    \centering
    \includegraphics[trim=0cm 0cm 0cm 17.5cm,clip,width=0.4\textwidth]{pictures/euclidean_1d_symmetric_exp.png}
  \end{subfigure}%
   \\  \vspace{-50pt} \hspace{0pt}
  \begin{subfigure}[b]{\linewidth}
    \begin{tikzpicture}
      \coordinate [label = above:{$\bm{e}^1$}](x) at (1,0); \coordinate [label =
      right:{$\bm{e}^2$}](y) at (0,1); \coordinate [](C) at (0, 0); \draw[-latex] (C)--(x);
      \draw[-latex] (C)--(y);
    \end{tikzpicture}
  \end{subfigure}
  \caption{\small Ellipsoidal representation of tensors showcasing the shape and orientation for interpolation between two symmetric tensors using different interpolation schemes. Here the color of the ellipsoid is its determinant. Two tensors $\bm{T}_1$ and $\bm{T}_2$ are placed at $\bm{x}_1=(-5,0)^T$ and $\bm{x}_2=(5,0)^T$, respectively  (see Figure~\ref{fig:problem_setup_1d}).   Tensor $\bm{T}_1$ is defined by eigenvalues $\{\lambda^1_1, \lambda^2_1, \lambda^3_1\}=\{10, 1.0, 1.0\}$  and primary eigenvector orientation $\sphericalangle(\bm{e}^1,\hat{\bm{n}}^1_{1})=\pi/4$. Tensor $\bm{T}_2$  is constituted by eigenvalues  $\{\lambda^1_2, \lambda^2_2, \lambda^3_2\}=\{20, 4.0, 1.0\}$ and primary eigenvector orientation $\sphericalangle(\bm{e}^1,\hat{\bm{n}}^1_{2}) \approx -\pi/4$.}
  \label{fig:Symmetric_Tensor_interpolation_1D}
\end{figure}

To verify the robustness of the proposed interpolation schemes, the extreme case of interpolating between two anisotropic tensors that are nearly orthogonal (i.e., relative angle between the primary eigenvectors of the tensors $\sphericalangle (\hat{\bm{n}}^1_1,\hat{\bm{n}}^1_2) \approx\pi/2$) is considered as illustrated in Figure~\ref{fig:Symmetric_Tensor_interpolation_1D}. The
first tensor $\bm{T}_1$ is located at $\bm{x}_1=(-5,0)^T$ (see Figure~\ref{fig:problem_setup_1d}) with eigenvalues $\{\lambda^1_1, \lambda^2_1, \lambda^3_1\}=\{10, 1.0, 1.0\}$ and primary eigenvector orientation (i.e., angle between the global Cartesian base vector $\bm{e}^1$ and the eigenvector $\hat{\bm{n}}^1$ associated with the maximal eigenvalue) of $\sphericalangle (\bm{e}^1,\hat{\bm{n}}^1_1)=\pi/4$. The second tensor $\bm{T}_2$ is located at $\bm{x}_2=(5,0)^T$ with eigenvalues $\{\lambda^1_2, \lambda^2_2, \lambda^3_2\}=\{20, 4.0, 1.0\}$ and primary eigenvector orientation of $\sphericalangle (\bm{e}^1,\hat{\bm{n}}^1_2)= 0.99 \cdot (-\pi/4) \approx -\pi/4$. Note that the primary eigenvector orientation of the second tensor has been chosen such that the relative orientation between the tensors $\sphericalangle (\hat{\bm{n}}^1_1,\hat{\bm{n}}^1_2)$ is slightly smaller than $\pi/2$, which represents the most challenging scenario. For a relative angle of exactly $\pi/2$, the proposed eigenvector re-orientation scheme (see Section~\ref{sec:impose_uniqueness}) results in non-unique solutions since $\sphericalangle (\hat{\bm{n}}^1_1,\hat{\bm{n}}^1_2)=\sphericalangle (\hat{\bm{n}}^1_1,-\hat{\bm{n}}^1_2)$. However, in practical applications the data points should be reasonably close to the interpolation point. Thus, the relative orientation between the tensors should be significantly smaller than $\pi/2$.

When looking at the results for the Euclidean and Log-Euclidean interpolation schemes (Figures~\ref{fig:Sym_1D_Euclidean} and~\ref{fig:Sym_1D_logEuclidean}), the ellipsoidal eigenvalue/eigenvector representation of the tensor at positions close to $\bm{x}_p=(0,0)^T$ shows a disk-like shape, which demonstrates that the anisotropic shape of the tensor is not preserved by the interpolation scheme. This effect is less pronounced but still present for the Cholesky and
Log-Cholesky interpolation schemes (Figures~\ref{fig:Sym_1D_Cholesky} and~\ref{fig:Sym_1D_logCholesky}). In contrast, the
proposed interpolation schemes Q-LOG, R-LOG, R-MLS, and R-LOGMLS  (Q-MLS (Q-LOGMLS) has not been plotted since the results are very similar to R-MLS (R-LOGMLS)) preserve the anisotropic shape of the tensors very well and lead to a smooth transition in tensor orientation and shape as depicted in Figures~\ref{fig:Sym_1D_RV},~\ref{fig:Sym_1D_Quad},~\ref{fig:Sym_1D_RV_MLS}, and~\ref{fig:Sym_1D_RV_LogMLS}. Remarkably, the orientation of the interpolated tensor at $\bm{x}_p=(0,0)^T$ is the arithmetic mean of the orientations of the tensors $\bm{T}_1$ and $\bm{T}_2$, i.e., $\sphericalangle (\bm{e}^1,\hat{\bm{n}}^1_p)=[\sphericalangle (\bm{e}^1,\hat{\bm{n}}^1_1)+\sphericalangle (\bm{e}^1,\hat{\bm{n}}^1_2)]/2\approx 0$.

\begin{figure}[ht]
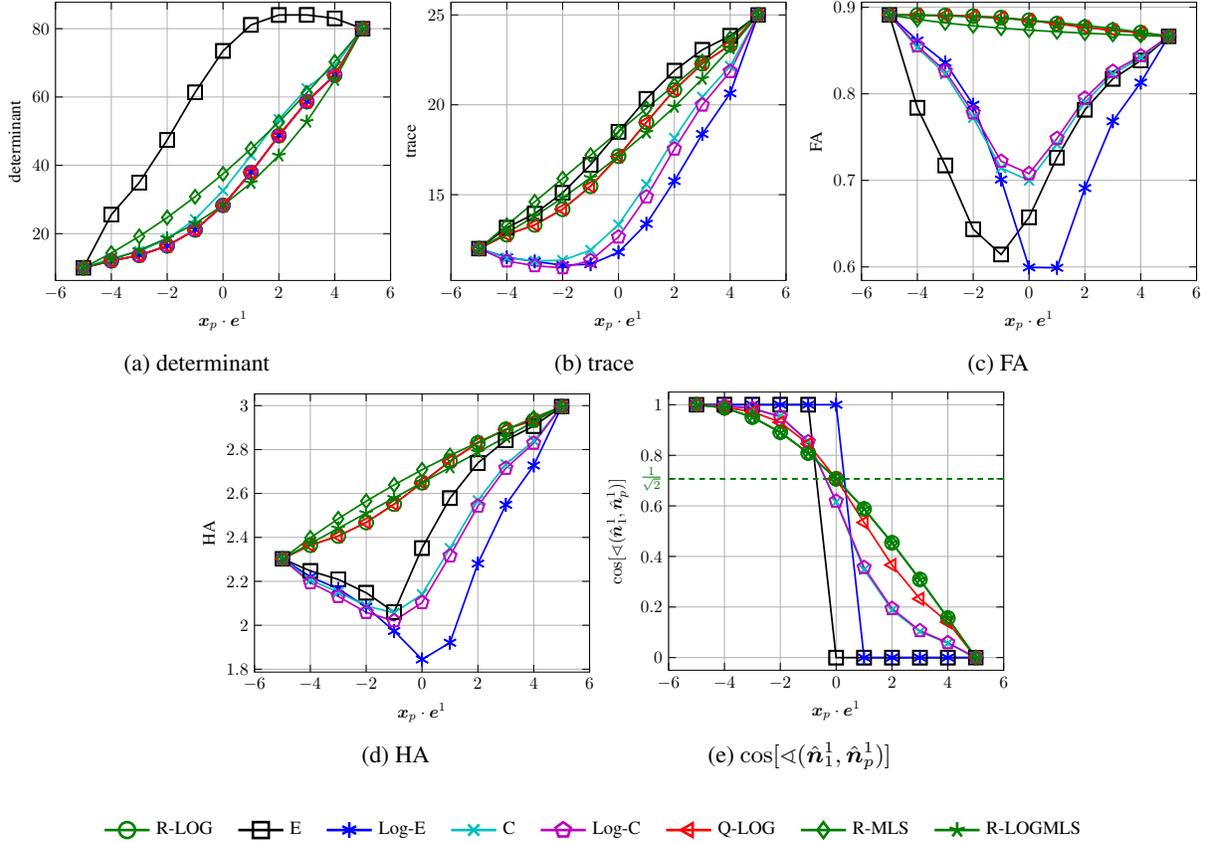

  \centering
  \begin{subfigure}[b]{.32\linewidth}
    \scalebox{0.65}{ \input{pictures/invariants__1d_symmetric_exp_determinant.tikz}}
    \caption{determinant} \label{fig:Sym_1D_determinant} \end{subfigure}%
  \begin{subfigure}[b]{.32\linewidth}
    \scalebox{0.65}{ \input{pictures/invariants__1d_symmetric_exp_trace.tikz}}
    \caption{trace} \label{fig:Sym_1D_trace}
  \end{subfigure}
  \begin{subfigure}[b]{.32\linewidth}
    \scalebox{0.65}{   \input{pictures/invariants__1d_symmetric_exp_FA.tikz}}
    \caption{FA}  \label{fig:Sym_1D_FA} \end{subfigure} \\ \vspace*{4pt}
  \begin{subfigure}[b]{.32\linewidth}
    \centering
    \scalebox{0.65}{\input{pictures/invariants__1d_symmetric_exp_HA.tikz}}
    \caption{HA} \label{fig:Sym_1D_HA}
  \end{subfigure}
  \begin{subfigure}[b]{.32\linewidth}
    \centering
    \scalebox{0.65}{\input{pictures/invariants__1d_symmetric_exp_IA.tikz}}
    \caption{\cosIA} \label{fig:Sym_1D_cosIA} \end{subfigure}\\
  \vspace{-160pt}
  \begin{subfigure}[b]{1\linewidth}
    \centering
    \input{pictures/invariants__1d_symmetric_exp_legend_final.tikz}
  \end{subfigure}
  \caption{\small Plot of different tensor metrics for interpolation between two symmetric tensors
    as displayed in Figure \ref{fig:Symmetric_Tensor_interpolation_1D}.}
  \label{fig:Invariants_Symmetric_Tensor_interpolation_1D}
\end{figure}

The aforementioned trends are confirmed, and become even clearer, when looking at Figure~\ref{fig:Invariants_Symmetric_Tensor_interpolation_1D}, illustrating the evolution of the aforementioned metrics (determinant, trace, FA, HA, \cosIA) of the interpolated tensor for the different interpolation schemes. As expected, all methods except the Euclidean interpolation
show a monotonic increase of the determinant (Figure~\ref{fig:Sym_1D_determinant}). A slight non-monotonic decrease in
the interpolated trace is observed for the Log-Euclidean, Cholesky, and Log-Cholesky schemes (see Figure 
\ref{fig:Sym_1D_trace}). The anisotropy metrics FA and HA are plotted in Figures~\ref{fig:Sym_1D_FA} and~\ref{fig:Sym_1D_HA}. It is striking that all methods from literature (E, LOG-E, C, and LOG-C) lead to a strongly non-monotonic interpolation in both anisotropy metrics FA and HA, which is not present for the proposed schemes (Q-LOG, R-LOG, R-MLS, and R-LOGMLS).

The cosine of the angle included between the
primary eigenvector of $\bm{T}_1$ and the primary eigenvector of the interpolated tensor $\bm{T}_p$, i.e., $\cos [\sphericalangle (\hat{\bm{n}}^1_1,\hat{\bm{n}}^1_p)]$ is plotted in Figure~\ref{fig:Sym_1D_cosIA}. From this plot, it is evident that the tensor orientation is not considered in the Euclidean and Log-Euclidean interpolation scheme, which leads to a jump in the eigenvector orientation between two neighboring interpolation points. For the Cholesky and Log-Cholesky schemes, the interpolated tensor
close to the position $\bm{x}=(0,0)^T$ is rather orientated towards $\bm{T}_2$. For the proposed schemes, the orientation changes gradually such that the orientation of the interpolated tensor at $\bm{x}_p=(0,0)^T$ is the arithmetic mean of the orientations of the tensors $\bm{T}_1$ and $\bm{T}_2$ (dashed green line at \cosIA $\approx 1/ \sqrt{2}$).

To sum up, it can be concluded that the proposed schemes (Q-LOG, R-LOG, R-MLS, R-LOGMLS, Q-MLS$^*$, and Q-LOGMLS\footnote[1]{not plotted}) lead to a smooth and monotonic interpolation in all considered metrics. This is not the case for the interpolation methods from literature (E, LOG-E, C, and LOG-C).
\subsubsection{Interpolation of four symmetric tensors} \label{example-four-symmetric-tensors}
The second numerical test case is carried out for four anisotropic tensors and associated data points placed at the four corners of a square with $x,y \in [-5;5]$ (see Figure~\ref{fig:problem_setup_2d}). Three of these four tensors are chosen identical, which are located at $\bm{x}_1=(5,5)^T$, $\bm{x}_2=(-5,5)^T$, and $\bm{x}_3=(-5,-5)^T$ and exhibit a primary eigenvector orientation of $\sphericalangle (\bm{e}^1,\hat{\bm{n}}^1_{1,2,3}) \approx \pi/2$ with eigenvalues $\{\lambda^1_{1,2,3}, \lambda^2_{1,2,3}, \lambda^3_{1,2,3}\}=\{7.5,1.25, 1.0\}$. The fourth tensor is located at $\bm{x}_4=(5,-5)^T$ with a primary eigenvector orientation of $\sphericalangle (\bm{e}^1,\hat{\bm{n}}^1_{4})=0$ with eigenvalues $\{\lambda^1_{4}, \lambda^2_{4}, \lambda^3_{4}\}=\{10, 3, 1.0 \}$.

\newcounter{mycounter}
\setcounter{mycounter}{1}
\renewcommand{\thefigure}{\arabic{figure}.\Alph{mycounter}}
\begin{figure}[ht]
  \centering
  \begin{subfigure}[b]{.30\linewidth} \hspace{15pt}
    \includegraphics[trim=0cm 7cm 0cm 0cm,clip,width=0.85\textwidth]{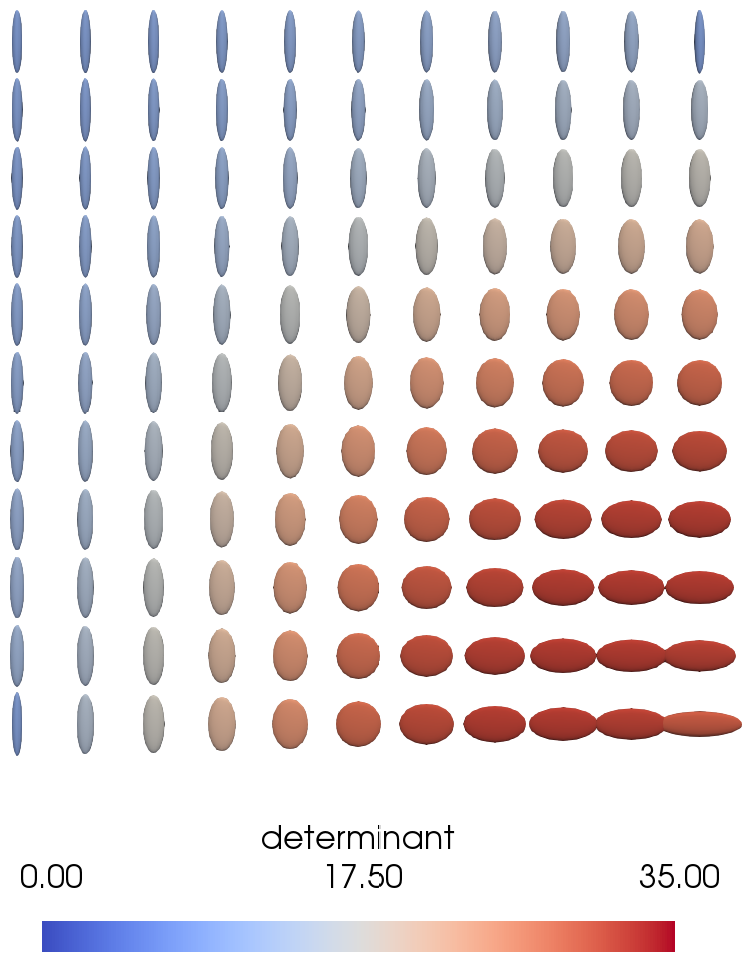}
    \caption{Euclidean} \label{fig:Sym_2D_determinant}
  \end{subfigure}
  \begin{subfigure}[b]{.34\linewidth} \hspace{-5pt}
    \includegraphics[trim=0cm 0cm 0cm 0cm,clip,width=1\textwidth]{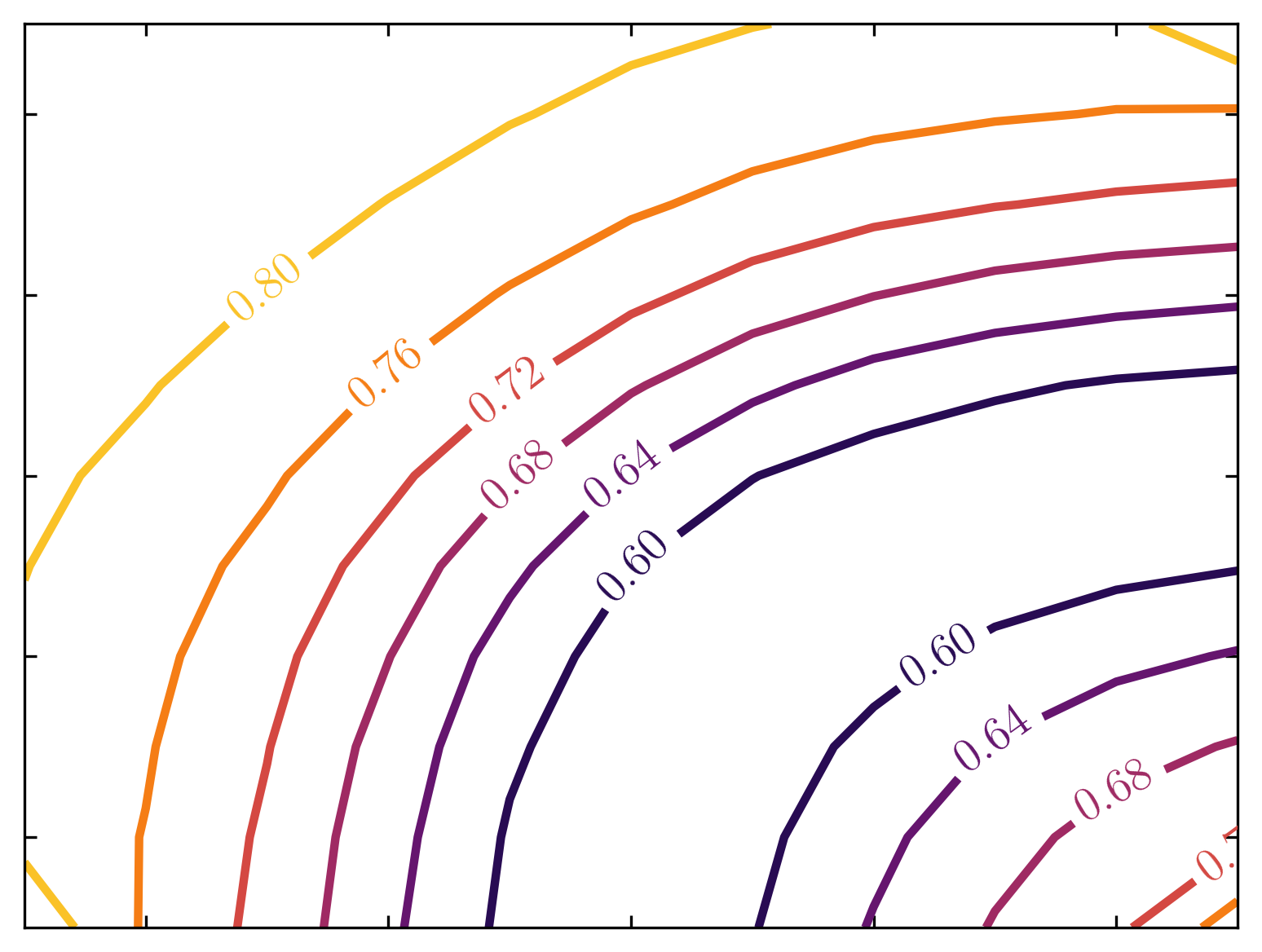}
    \caption{Euclidean: FA} \label{fig:Sym_2D_Euclidean_FA}
  \end{subfigure}
  \begin{subfigure}[b]{.34\linewidth}
    \includegraphics[trim=0cm 0cm 0cm 0cm,clip,width=1\textwidth]{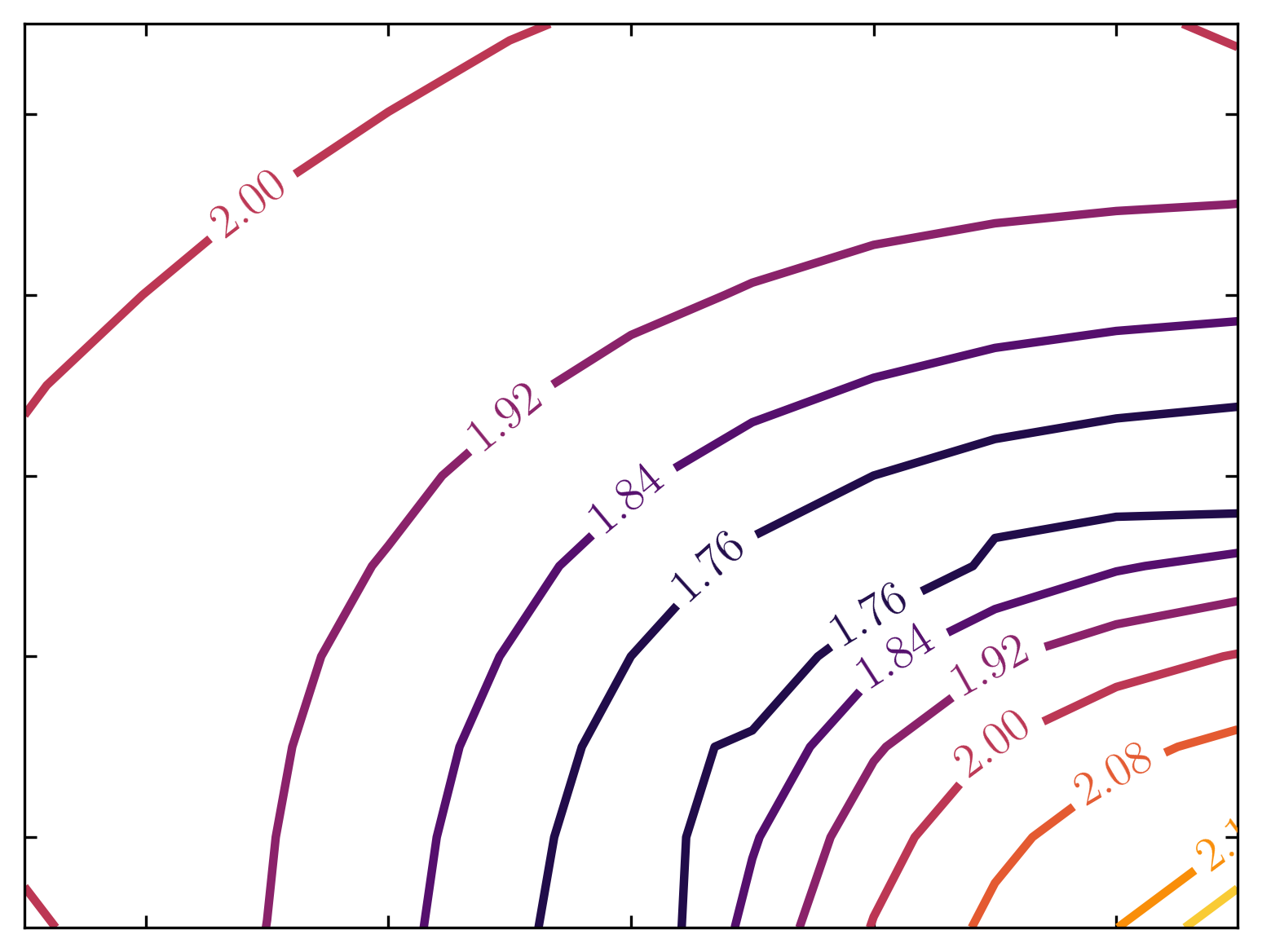}
    \caption{Euclidean: HA}  \label{fig:Sym_2D_Euclidean_HA} \end{subfigure} \\ \vspace{4pt}
  \centering
  \begin{subfigure}[b]{.30\linewidth} \hspace{15pt}
    \includegraphics[trim=0cm 7cm 0cm 0cm,clip,width=0.85\textwidth]{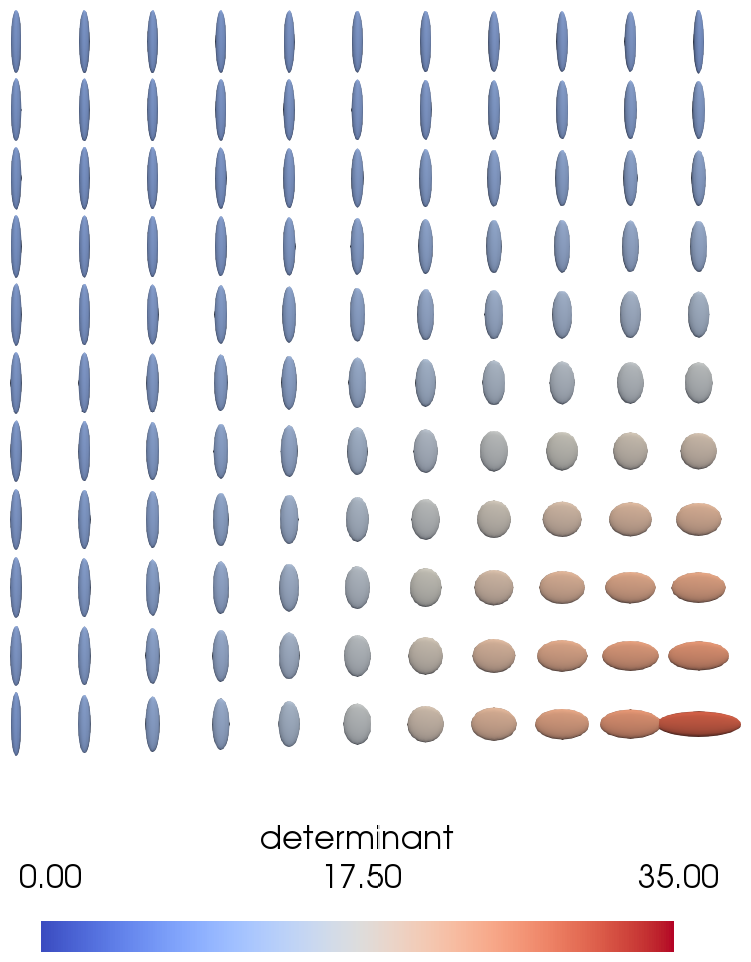}
    \caption{Log-Euclidean} \label{fig:Sym_2D_LogEuclidean}
  \end{subfigure}
  \begin{subfigure}[b]{.34\linewidth} \hspace{-5pt}
    \includegraphics[trim=0cm 0cm 0cm 0cm,clip,width=1\textwidth]{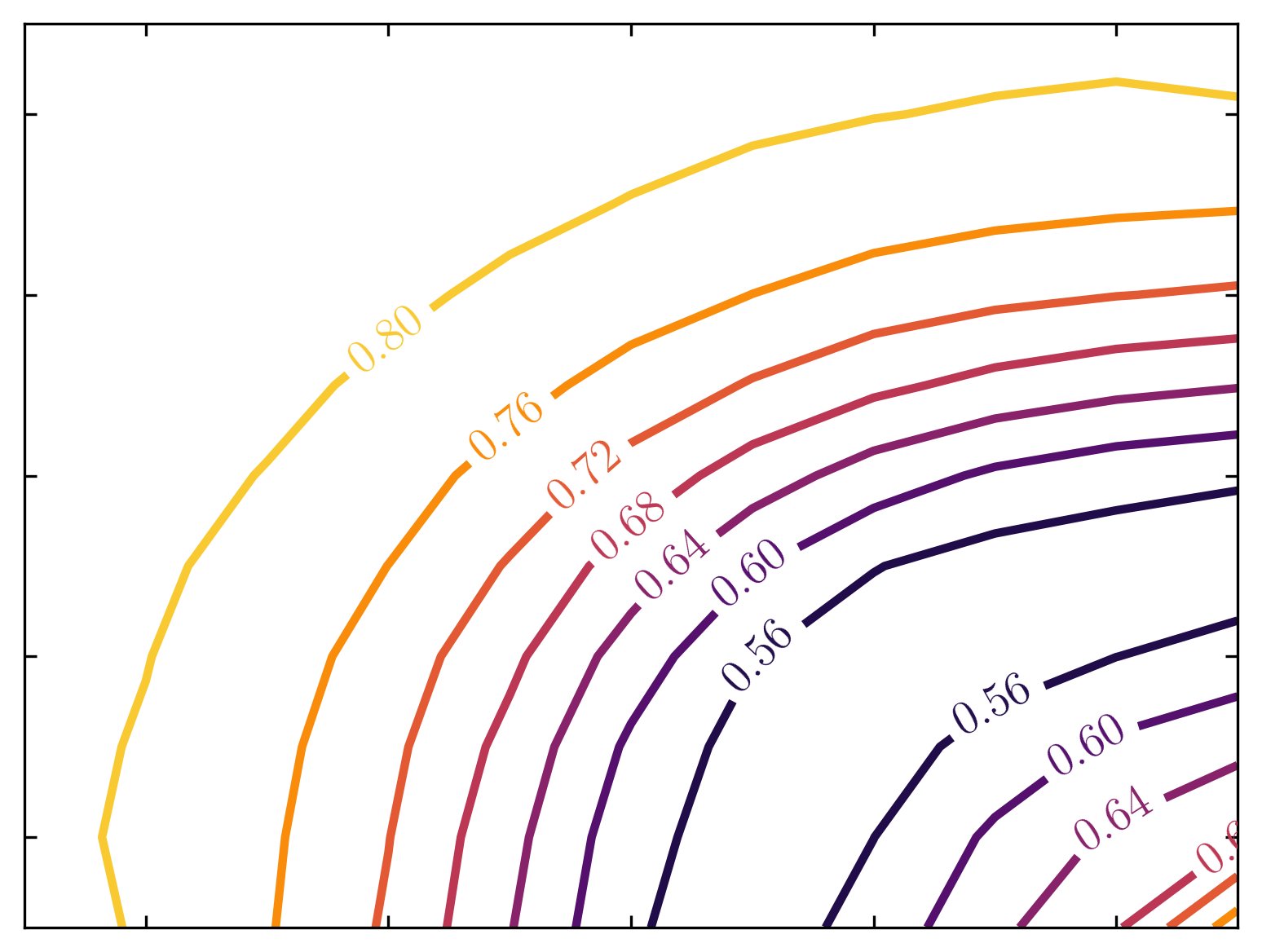}
    \caption{Log-Euclidean: FA} \label{fig:Sym_2D_LogEuclidean_FA}
  \end{subfigure}
  \begin{subfigure}[b]{.34\linewidth}
    \includegraphics[trim=0cm 0cm 0cm 0cm,clip,width=1\textwidth]{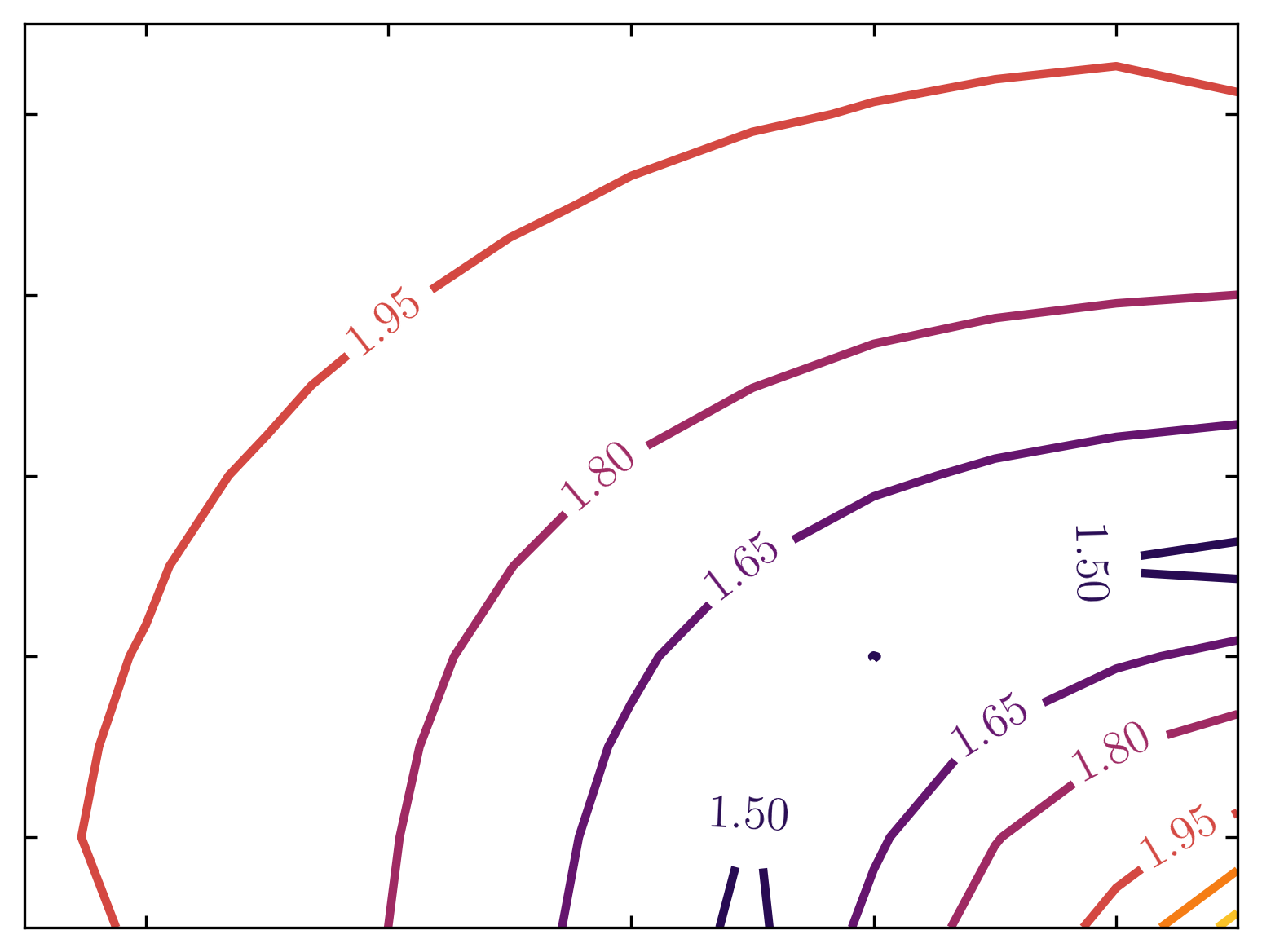}
    \caption{Log-Euclidean: HA} \label{fig:Sym_2D_LogEuclidean_HA} \end{subfigure} \\ \vspace{4pt}
  \centering
  \begin{subfigure}[b]{.30\linewidth} \hspace{15pt}
    \includegraphics[trim=0cm 7cm 0cm 0cm,clip,width=0.85\textwidth]{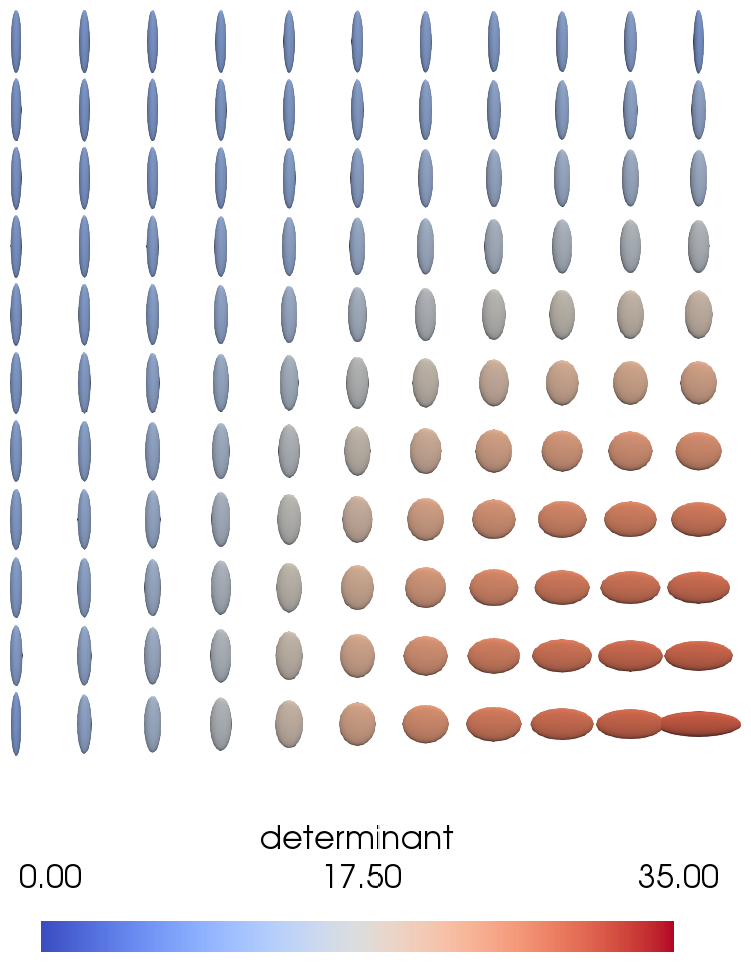}
    \caption{Cholesky} \label{fig:Sym_2D_Cholesky}
  \end{subfigure}
  \begin{subfigure}[b]{.34\linewidth} \hspace{-5pt}
    \includegraphics[trim=0cm 0cm 0cm 0cm,clip,width=1\textwidth]{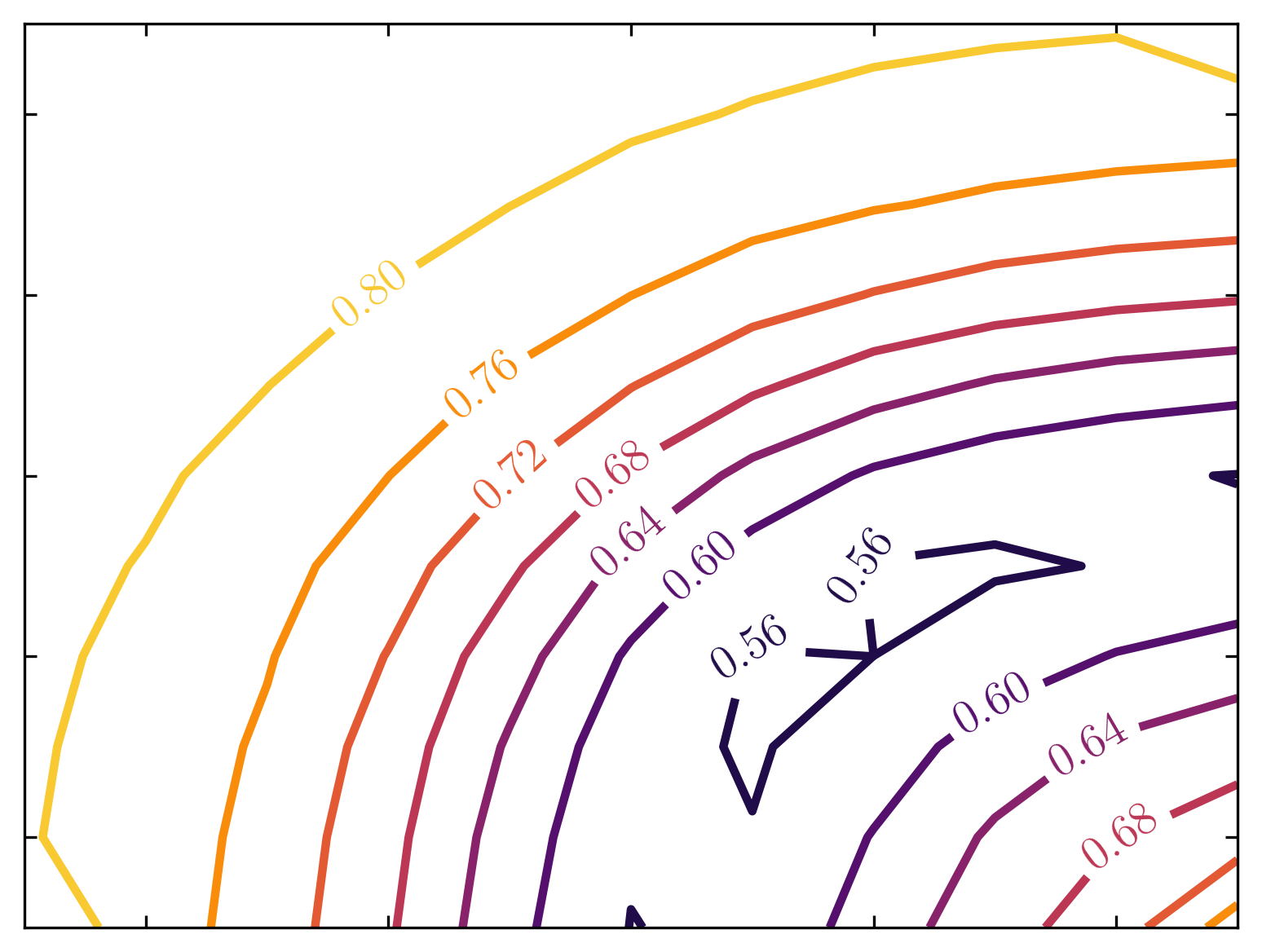}
    \caption{Cholesky: FA} \label{fig:Sym_2D_Cholesky_FA}
  \end{subfigure}
  \begin{subfigure}[b]{.34\linewidth}
    \includegraphics[trim=0cm 0cm 0cm 0cm,clip,width=1\textwidth]{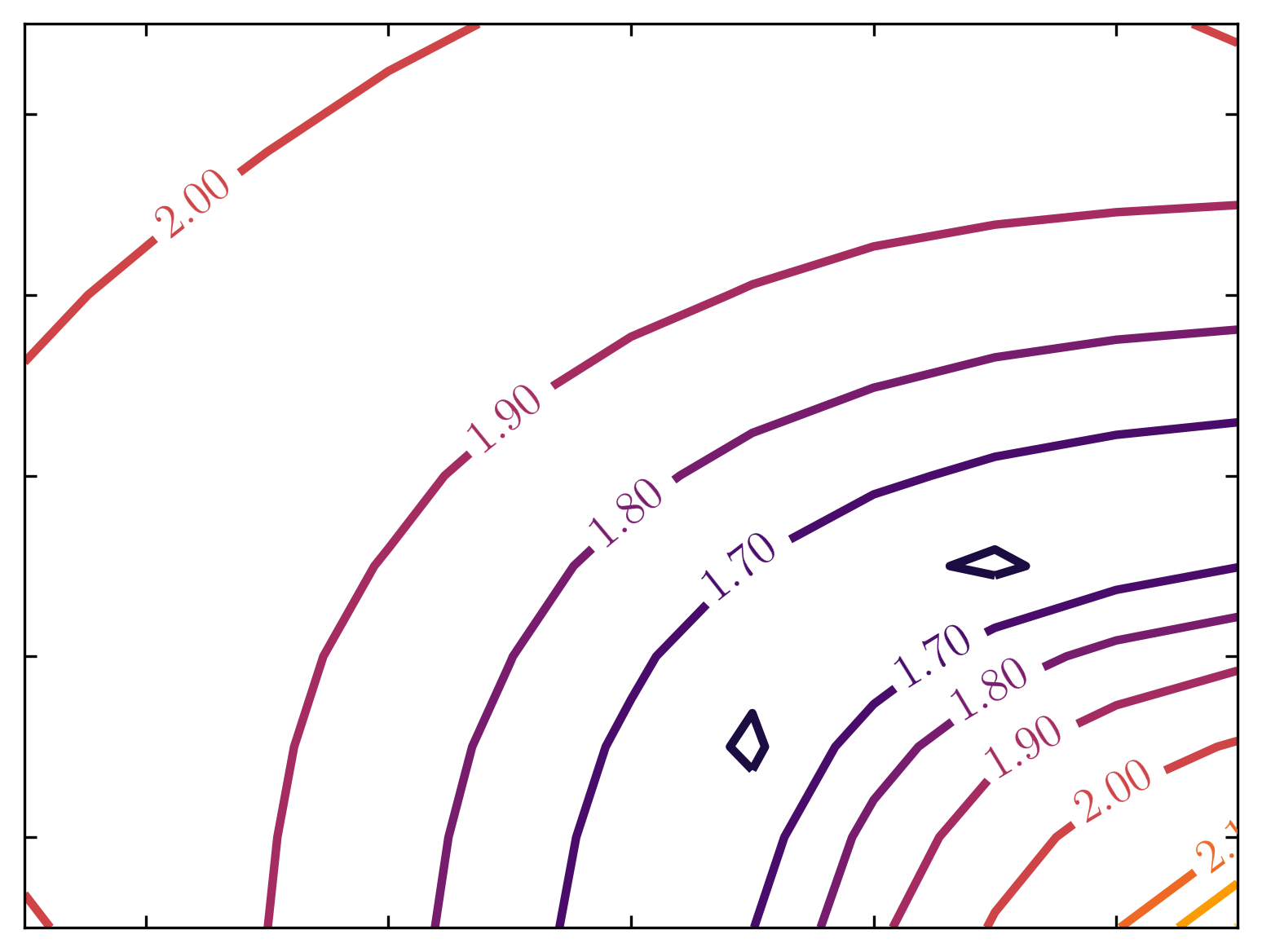}
    \caption{Cholesky: HA} \label{fig:Sym_2D_Cholesky_HA} \end{subfigure} 
    \\  \vspace{-40pt} \hspace{0pt}
      \begin{subfigure}[b]{\linewidth}
        \begin{tikzpicture}
          \coordinate [label = above:{$\bm{e}^1$}](x) at (1,0); \coordinate [label =
          right:{$\bm{e}^2$}](y) at (0,1); \coordinate [](C) at (0, 0); \draw[-latex] (C)--(x);
          \draw[-latex] (C)--(y);
        \end{tikzpicture}
      \end{subfigure}
  \\
  \begin{subfigure}[b]{.33\linewidth} \vspace{5pt}
    \includegraphics[trim=0cm 0cm  0cm 30.5cm,clip,width=0.75\textwidth]{pictures/logeuclidean_2d_symmetric.png}
  \end{subfigure}
  \begin{subfigure}[b]{.33\linewidth}
    \includegraphics[trim=0cm 0cm  0cm 40cm,clip,width=0.75\textwidth]{pictures/logeuclidean_2d_symmetric.png}
  \end{subfigure}
  \begin{subfigure}[b]{.33\linewidth}
    \includegraphics[trim=0cm 0cm  0cm 40cm,clip,width=0.75\textwidth]{pictures/logeuclidean_2d_symmetric.png}
  \end{subfigure}
  \caption{\small Ellipsoidal representation (first column) of tensors featuring the shape and orientation and contour plots of tensor metrics FA and HA (second and third column) for interpolation between four symmetric tensors employing different interpolation methods. The color of the ellipsoid is its determinant. Here  three symmetric tensors ($\bm{T}_{1,2,3}$) with
    eigenvalues $\{\lambda^1_{1,2,3}, \lambda^2_{1,2,3}, \lambda^3_{1,2,3}\}=\{7.5, 1.25, 1.0\}$ and primary eigenvector orientation $\sphericalangle(\bm{e}^1,\hat{\bm{n}}^1_{1,2,3}) \approx \pi/2$ are placed at $\bm{x}_1=(5,5)^T$, $\bm{x}_2=(-5,5)^T$ and $\bm{x}_3=(-5,-5)^T$ (see Figure~\ref{fig:problem_setup_2d}). The fourth tensor $\bm{T}_4$ at  $\bm{x}_4=(5,-5)^T$ is defined by eigenvalues $\{\lambda^1_{4}, \lambda^2_{4}, \lambda^3_{4}\}=\{10, 3, 1.0 \}$ and primary eigenvector orientation $\sphericalangle(\bm{e}^1,\hat{\bm{n}}^1_{4}) =0$.}
  \label{fig:Symmetric_Tensor_interpolation_2D}
\end{figure}

The interpolation results for E, LOG-E, C, LOG-C, R-LOG, and Q-LOG methods in terms of ellipsoidal eigenvalue/eigenvector representation and corresponding FA and HA contour plots
are depicted in \cref{fig:Symmetric_Tensor_interpolation_2D,fig:Symmetric_Tensor_interpolation_2D_2}, which confirm the trends already observed for the interpolation of two tensors. For the Euclidean, Log-Euclidean, Cholesky, and
Log-Cholesky interpolation methods, FA and HA show a non-monotonic evolution (depicted by discontinuous and non-smooth contour lines in second and third columns of \cref{fig:Symmetric_Tensor_interpolation_2D,fig:Symmetric_Tensor_interpolation_2D_2} except for R-LOG and Q-LOG), resulting in a disk-like shape of the ellipsoidal eigenvalue/eigenvector representation, i.e., an isotropic tensor, in some regions of the interpolation domain. In contrast, the proposed schemes (R-LOG and Q-LOG) show a monotonic evolution of FA and HA, preserving anisotropy. Moreover, the quaternion-based (Q-LOG) and the rotation vector-based (R-LOG) rotation interpolation schemes result in very similar tensor orientation fields.

\addtocounter{mycounter}{+1}
\addtocounter{figure}{-1}
\begin{figure}[ht]
  \centering
  \begin{subfigure}[b]{.30\linewidth}\hspace{15pt}\setcounter{subfigure}{9}
    \includegraphics[trim=0cm 7cm 0cm 0cm,clip,width=0.85\textwidth]{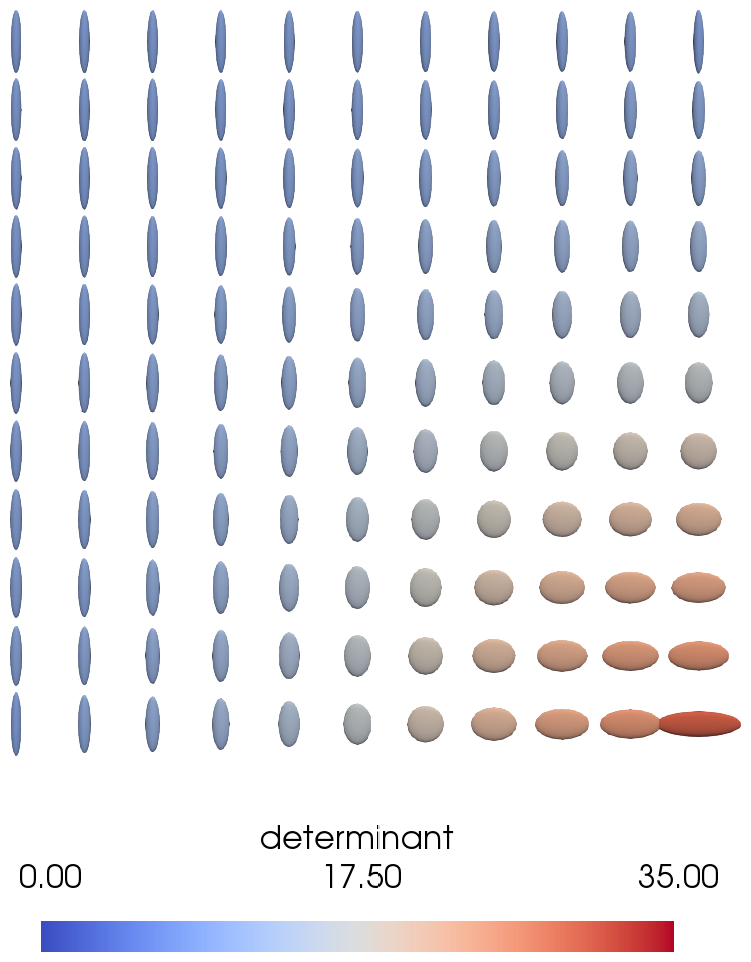}
    \caption{Log-Cholesky} \label{fig:Sym_2D_LogCholesky}
  \end{subfigure}
  \begin{subfigure}[b]{.34\linewidth}\hspace{-5pt}
    \includegraphics[trim=0cm 0cm 0cm 0cm,clip,width=1\textwidth]{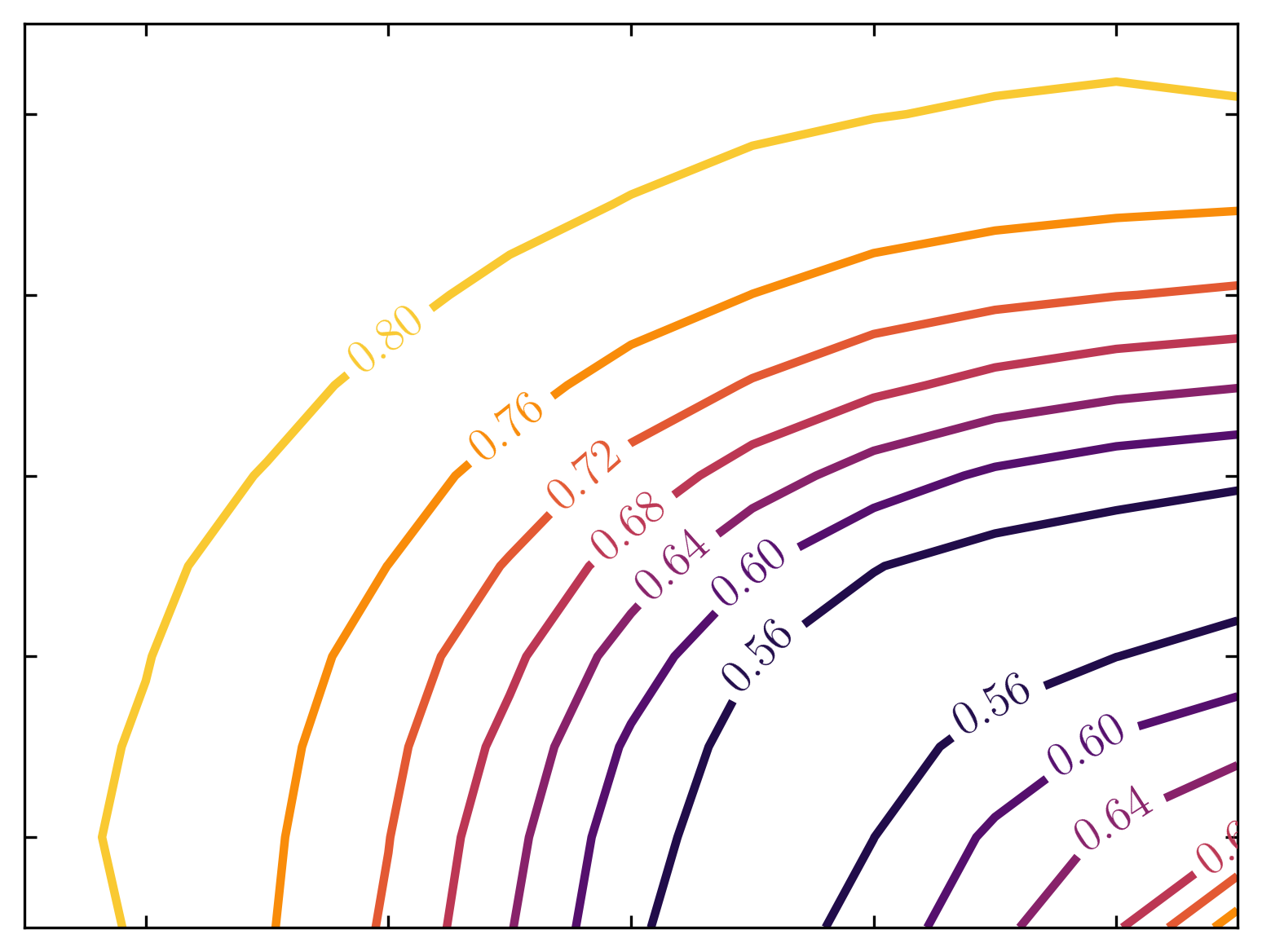}
    \caption{Log-Cholesky: FA} \label{fig:Sym_2D_LogCholesky_FA}
  \end{subfigure}
  \begin{subfigure}[b]{.34\linewidth}
    \includegraphics[trim=0cm 0cm 0cm 0cm,clip,width=1\textwidth]{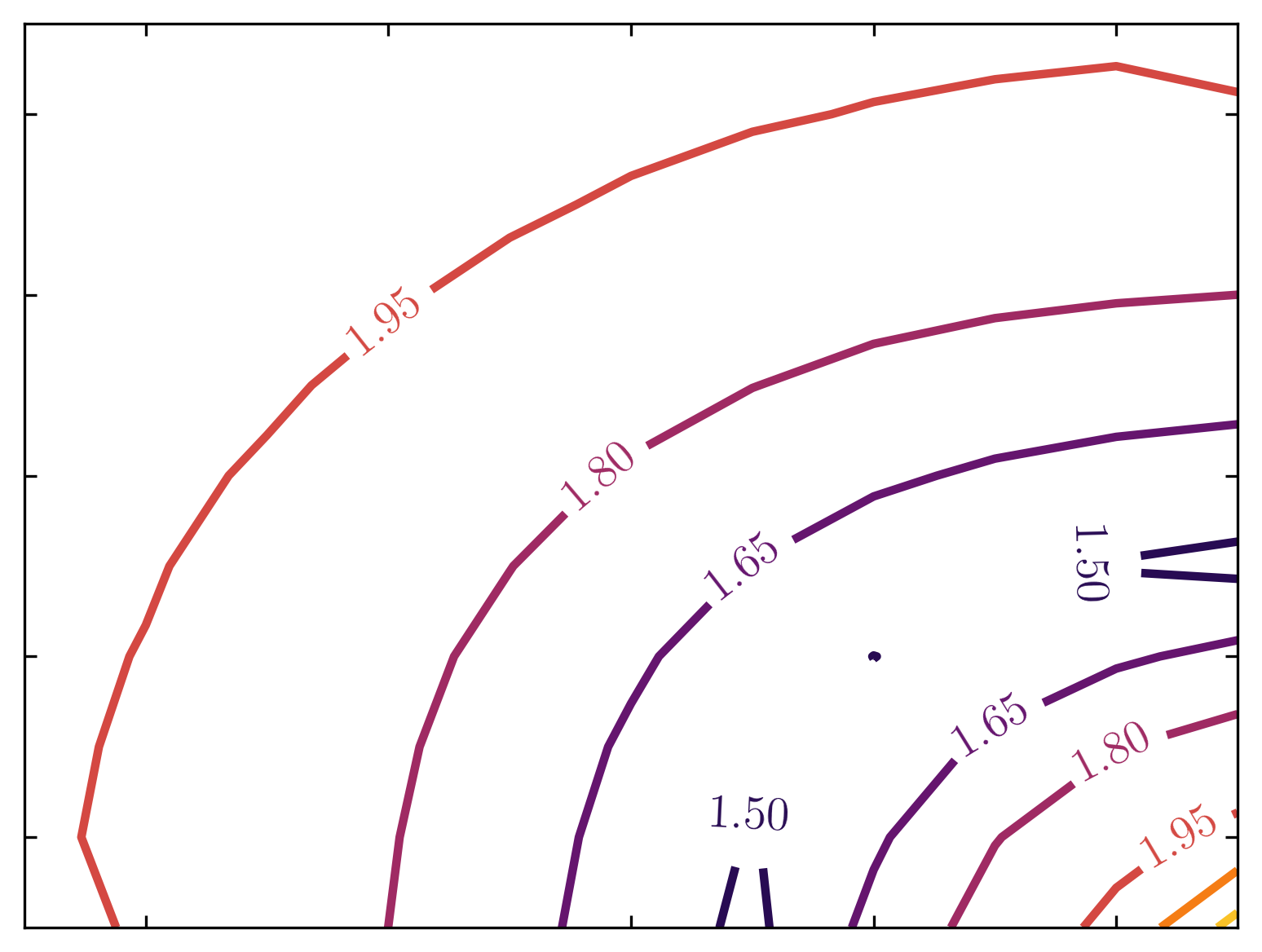}
    \caption{Log-Cholesky: HA} \label{fig:Sym_2D_LogCholesky_HA} \end{subfigure} \\ \vspace{4pt}
  \centering
  \begin{subfigure}[b]{.30\linewidth}\hspace{15pt}
    \includegraphics[trim=0cm 7cm 0cm 0cm,clip,width=0.85\textwidth]{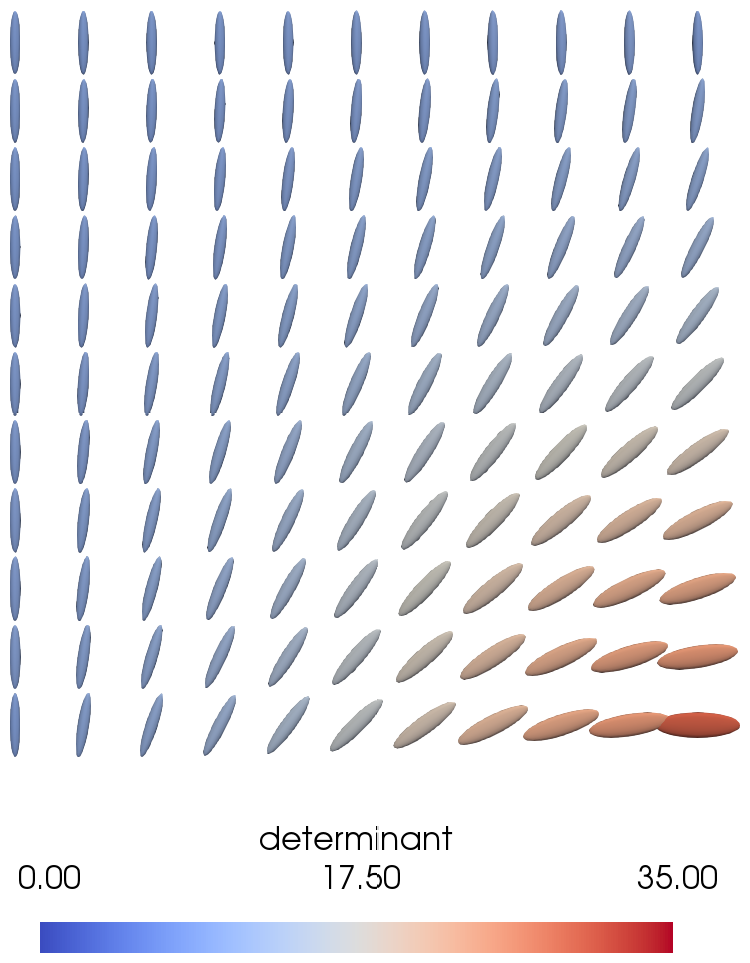}
    \caption{R-LOG} \label{fig:Sym_2D_Rotation_vector}
  \end{subfigure}
  \begin{subfigure}[b]{.34\linewidth}\hspace{-5pt}
    \includegraphics[trim=0cm 0cm 0cm 0cm,clip,width=1\textwidth]{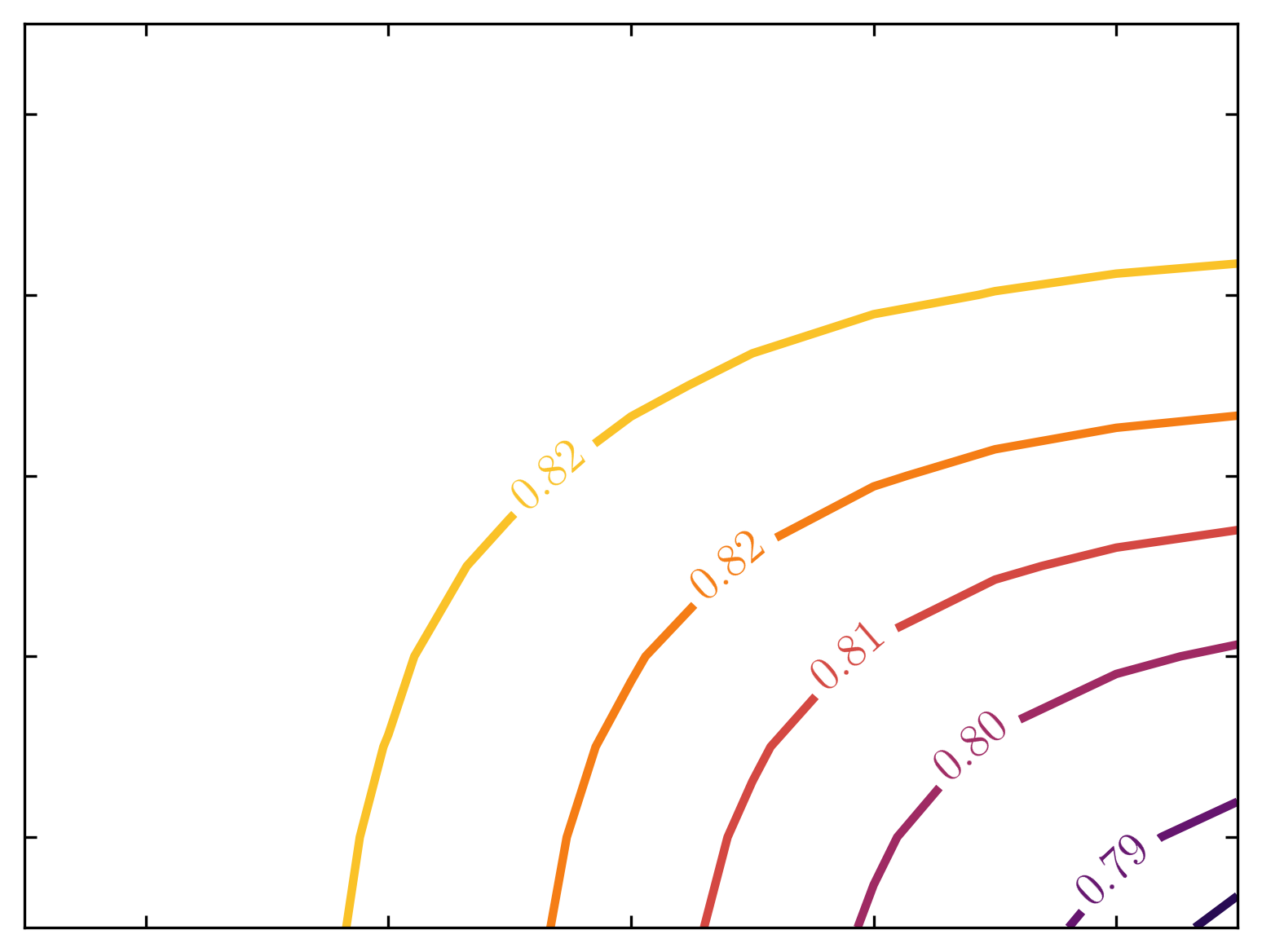}
    \caption{R-LOG: FA} \label{fig:Sym_2D_Rotation_vector_FA}
  \end{subfigure}
  \begin{subfigure}[b]{.34\linewidth}
    \includegraphics[trim=0cm 0cm 0cm 0cm,clip,width=1\textwidth]{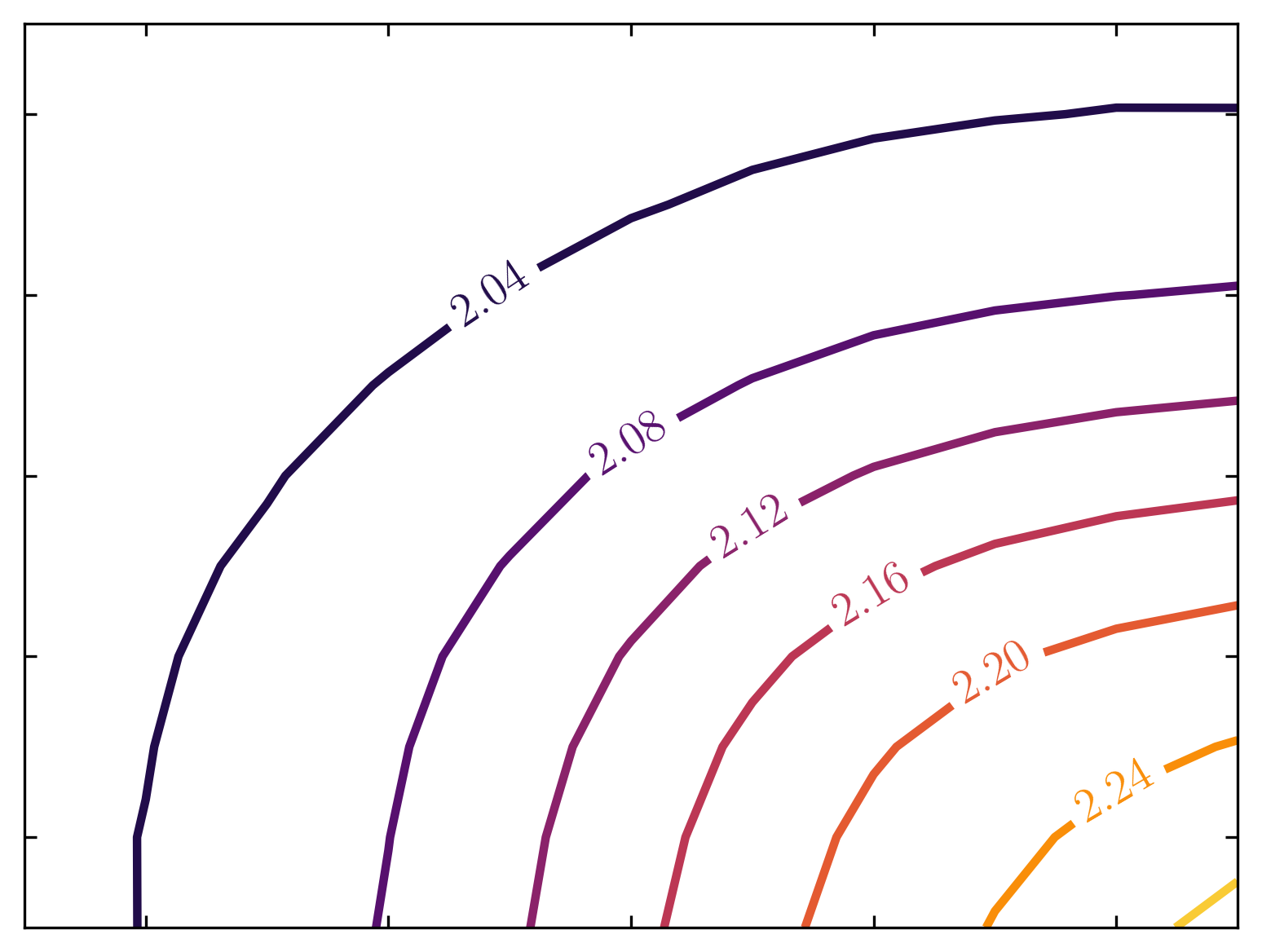}
    \caption{R-LOG: HA} \label{fig:Sym_2D_Rotation_vector_HA} \end{subfigure} \\ \vspace{4pt}
  \centering
  \begin{subfigure}[b]{.30\linewidth}\hspace{15pt}
    \includegraphics[trim=0cm 7cm 0cm 0cm,clip,width=0.85\textwidth]{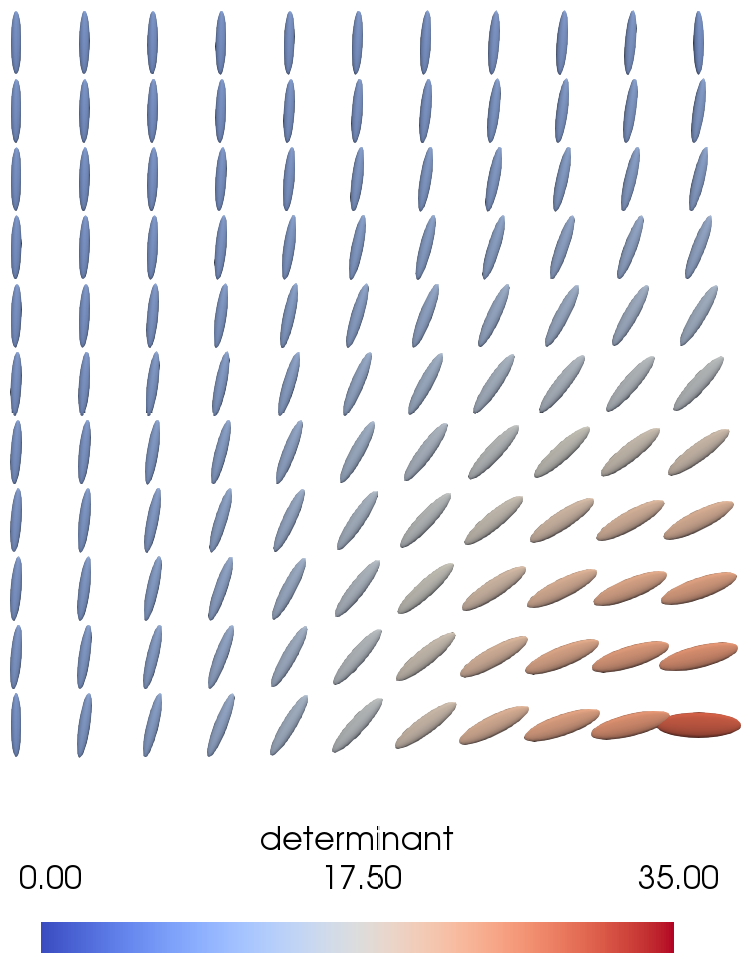}
    \caption{Q-LOG} \label{fig:Sym_2D_quaternion}
  \end{subfigure}
  \begin{subfigure}[b]{.34\linewidth}\hspace{-5pt}
    \includegraphics[trim=0cm 0cm 0cm 0cm,clip,width=1\textwidth]{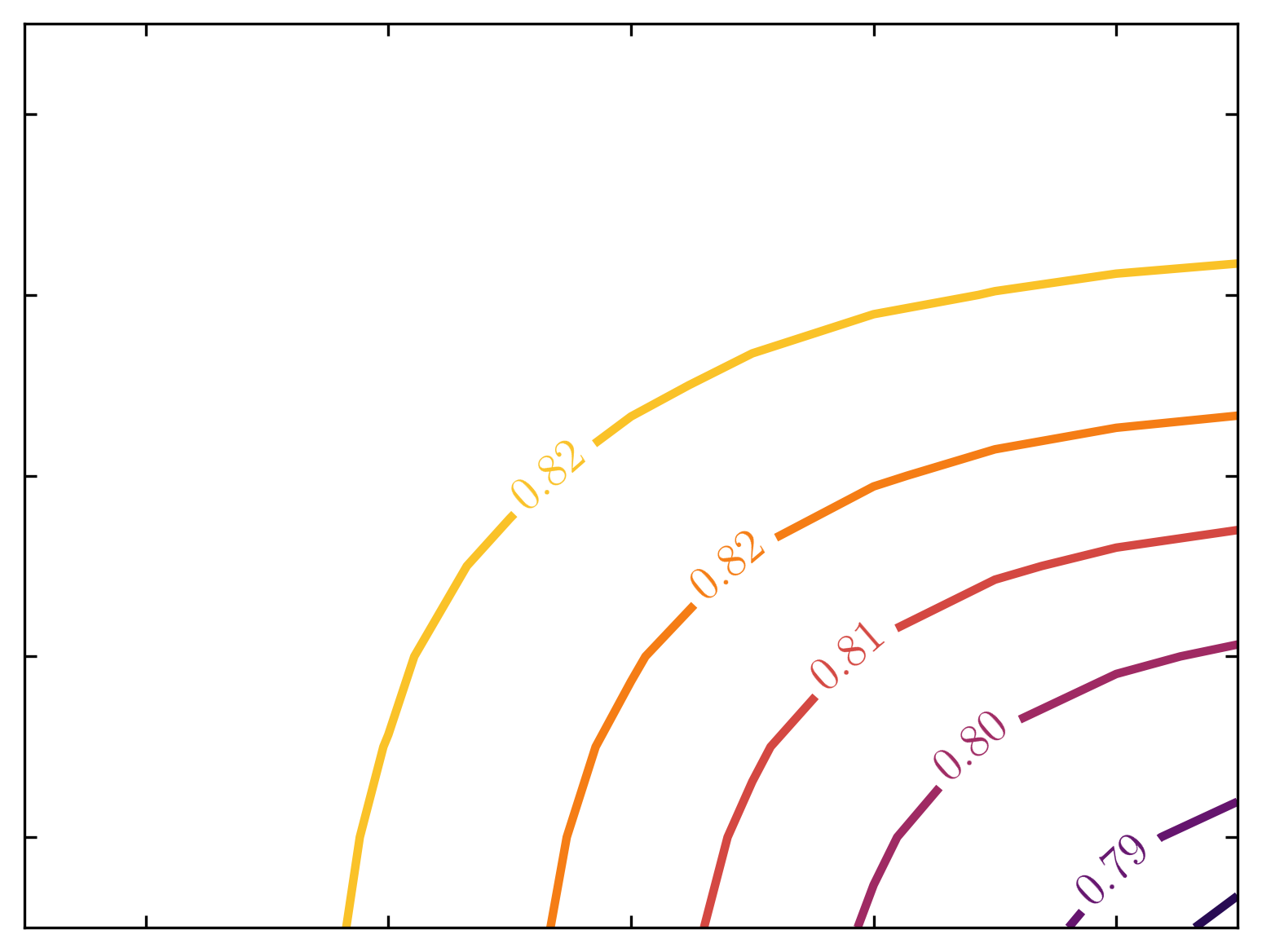}
    \caption{Q-LOG: FA} \label{fig:Sym_2D_quaternion_FA}
  \end{subfigure}
  \begin{subfigure}[b]{.34\linewidth}
    \includegraphics[trim=0cm 0cm 0cm 0cm,clip,width=1\textwidth]{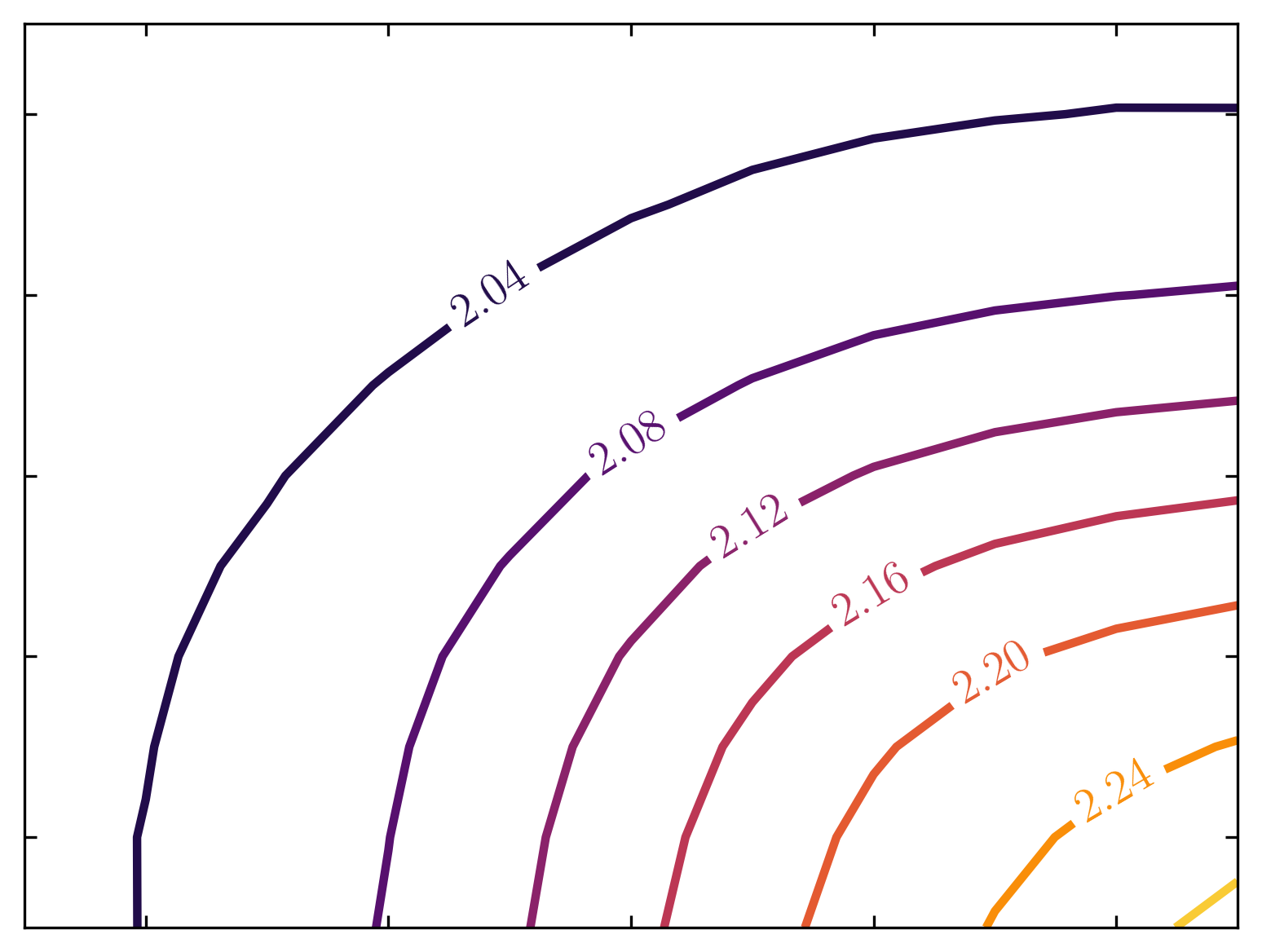}
    \caption{Q-LOG: HA} \label{fig:Sym_2D_quaternion_HA} \end{subfigure}
    \\  \vspace{-40pt} \hspace{0pt}
      \begin{subfigure}[b]{\linewidth}
        \begin{tikzpicture}
          \coordinate [label = above:{$\bm{e}^1$}](x) at (1,0); \coordinate [label =
          right:{$\bm{e}^2$}](y) at (0,1); \coordinate [](C) at (0, 0); \draw[-latex] (C)--(x);
          \draw[-latex] (C)--(y);
        \end{tikzpicture}
      \end{subfigure}
  \\
  \begin{subfigure}[b]{.33\linewidth} \vspace{5pt}
    \includegraphics[trim=0cm 0cm  0cm 30.5cm,clip,width=0.75\textwidth]{pictures/logeuclidean_2d_symmetric.png}
  \end{subfigure}
  \begin{subfigure}[b]{.33\linewidth}
    \includegraphics[trim=0cm 0cm  0cm 40cm,clip,width=0.75\textwidth]{pictures/logeuclidean_2d_symmetric.png}
  \end{subfigure}
  \begin{subfigure}[b]{.33\linewidth}
    \includegraphics[trim=0cm 0cm  0cm 40cm,clip,width=0.75\textwidth]{pictures/logeuclidean_2d_symmetric.png}
  \end{subfigure}
   \caption{(continued from page \pageref{fig:Symmetric_Tensor_interpolation_2D}).}\label{fig:Symmetric_Tensor_interpolation_2D_2}
\end{figure}
\renewcommand{\thefigure}{\arabic{figure}}
\subsection{Interpolation of non-symmetric tensors}
In this section, we explore the interpolation of non-symmetric tensors ($\bm{T}=\bm{R}\bm{U}$). We demonstrate extreme
cases of interpolation similar to the examples presented for symmetric tensor in the previous section. These extreme cases are
achieved by specifying the primary eigenvector orientation ($\sphericalangle(\bm{e}^1,\Hat{\bm{n}}^1_j)$)
of the symmetric part $\bm{U}$ and the angle of rotation for the rotation part $\bm{R}$. While the interpolation of $\bm{U}$ is again visualized by means of ellipsoidal representation, the interpolated rotation tensor field~$\bm{R}$ is represented by its primary eigenvector $\Hat{\tilde{\bm{n}}}^1_j$ (not to be confused with primary eigenvector~$\Hat{{\bm{n}}}^1_j$ of the symmetric  tensor~$\bm{U}$).
\subsubsection{Interpolation of two tensors} \label{example-two-nonsymmetric-tensors}
\begin{figure}[htb]
  \centering
  \begin{subfigure}[b]{.4\linewidth}
    \centering
    \includegraphics[trim=0cm 14.5cm 0cm 0cm,clip,width=\textwidth]{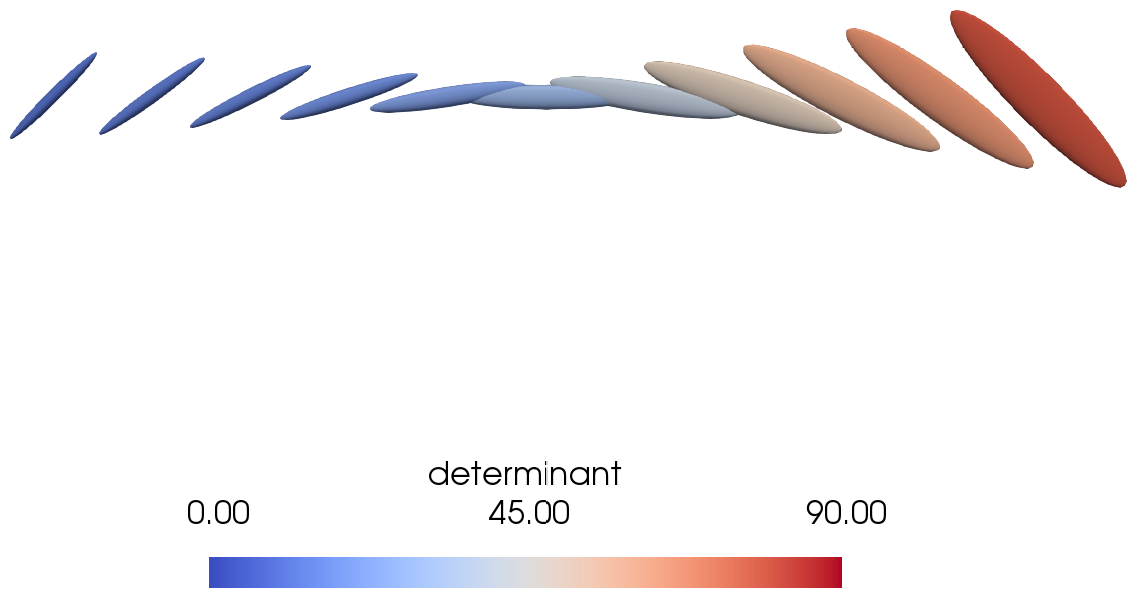}
    \caption{R-LOG: $\bm{U}$ } \label{fig:NonSym_1D_rotationvector_Sym}
  \end{subfigure} \hspace{1cm}
  \begin{subfigure}[b]{.4\linewidth}
    \centering
    \includegraphics[trim=0cm 0cm 0cm 0cm,clip,width=\textwidth]{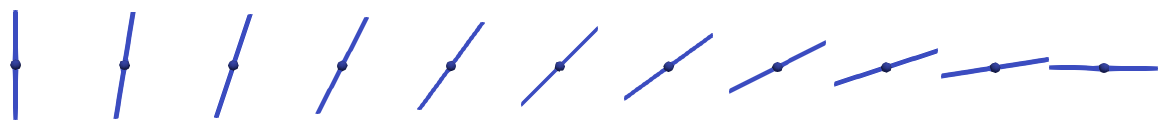}
    \caption{R-LOG: $\bm{R}$ } \label{fig:NonSym_1D_rotationvector_Rot}
  \end{subfigure}\\
  \centering
  \begin{subfigure}[b]{.4\linewidth}\vspace{5pt} 
    \centering
    \includegraphics[trim=0cm 14.5cm 0cm 0cm,clip,width=\textwidth]{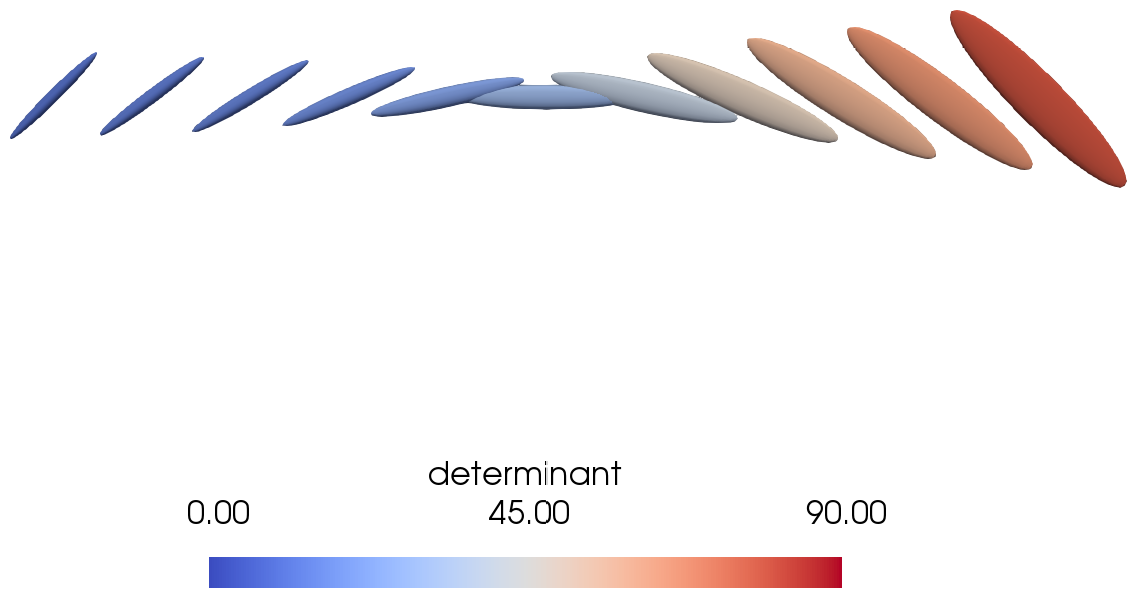}
    \caption{Q-LOG: $\bm{U}$ } \label{fig:NonSym_1D_quaternion_Sym}
  \end{subfigure} \hspace{1cm}
  \begin{subfigure}[b]{.4\linewidth}\vspace{5pt} 
    \centering
    \includegraphics[trim=0cm 0cm 0cm 0cm,clip,width=\textwidth]{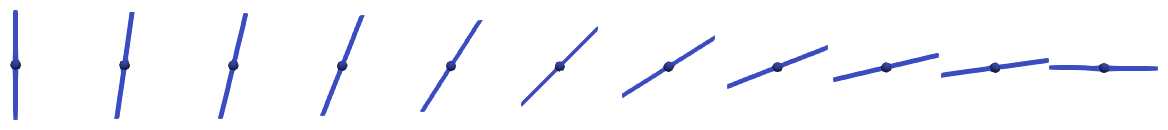}
    \caption{Q-LOG: $\bm{R}$ } \label{fig:NonSym_1D_quaternion_Rot}
  \end{subfigure}\\
  \begin{subfigure}[b]{\linewidth}\vspace{-40pt} \hspace{0pt}
    \begin{tikzpicture}
      \coordinate [label = above:{$\bm{e}^1$}](x) at (1,0); \coordinate [label =
      right:{$\bm{e}^2$}](y) at (0,1); \coordinate [](C) at (0, 0); \draw[-latex] (C)--(x);
      \draw[-latex] (C)--(y);
    \end{tikzpicture}
  \end{subfigure}\\
    \centering
  \begin{subfigure}[b]{0.4\linewidth} \vspace{-40pt}
    \centering
    \includegraphics[trim=0cm 0cm 0cm 17.5cm, ,clip,width=1\textwidth]{pictures/quaternion_1d_nonsym_sym.png}
  \end{subfigure}%
    \centering
  \begin{subfigure}[b]{0.4\linewidth}
    \centering
    \includegraphics[trim=0cm 0cm 0cm 90cm, ,clip,width=1\textwidth]{pictures/quaternion_1d_nonsym_sym.png}
  \end{subfigure}
  \caption{\small Interpolation between non-symmetric and symmetric tensors employing R-LOG and Q-LOG methods. Subfigures (a) and (c) are the ellipsoidal representation (the color denotes the tensor determinant) portraying shape and orientation of the symmetric component  $\bm{U}$ and subfigures (b) and (d) depict the primary eigenvector~$\Hat{\tilde{\bm{n}}}^1_j$ of rotation component~$\bm{R}$ of the tensor. The first tensor
    $\bm{T}_1=\bm{R}_1\bm{U}_1$ is located at $\bm{x}_1=(-5,0)^T$ (see Figure~\ref{fig:problem_setup_1d}). The symmetric component $\bm{U}_1$ is
    constituted by eigenvalues $\{\lambda^1_1, \lambda^2_1, \lambda^3_1\}=\{10, 1.0, 1.0\}$ and
    primary eigenvector orientation $\sphericalangle(\bm{e}^1,\hat{\bm{n}}^1_{1})=\pi/4$ and the rotation component is defined as $\bm{R}_1 = \bm{R}_{\bm{e}^3}(\theta_1 \approx \frac{1}{2}\pi)$ (see~\eqref{R_z}). The second tensor $\bm{T}_2=\bm{U}_2$ at $\bm{x}_2=(5,0)^T$ is defined by eigenvalues
    $\{\lambda^1_2, \lambda^2_2, \lambda^3_2\}=\{20, 4.0, 1.0\}$ and primary eigenvector orientation
    $\sphericalangle(\bm{e}^1,\hat{\bm{n}}^1_{2}) \approx -\pi/4$.}
  \label{fig:Non_Symmetric_Tensor_interpolation_1D}
\end{figure}

In the first step, the interpolation between an invertible, non-symmetric, positive definite tensor ($\bm{T}_1=\bm{R}_1\bm{U}_1$) and a
symmetric, positive-definite tensor ($\bm{T}_2=\bm{U}_2$) is carried out. The tensors $\bm{T}_1$ and  $\bm{T}_2$ are
located at $\bm{x}_1=(-5,0)^T$ and $\bm{x}_2=(5,0)^T$  (see Figure~\ref{fig:problem_setup_1d}). The symmetric part $\bm{U}_1$ is defined by eigenvalues $\{\lambda^1_1, \lambda^2_1, \lambda^3_1\}=\{10, 1.0, 1.0\}$ and a primary eigenvector orientation of $\sphericalangle (\bm{e}^1,\hat{\bm{n}}^1_1)=\pi/4$. The symmetric tensor $\bm{U}_2$ is defined by eigenvalues $\{\lambda^1_2, \lambda^2_2, \lambda^3_2\}=\{20, 4.0,
  1.0\}$ and a primary eigenvector orientation of $\sphericalangle (\bm{e}^1,\hat{\bm{n}}^1_2) \approx -\pi/4$. The rotation part of both tensors is defined as $\bm{R}_j=\bm{R}_{\bm{e}^3}(\theta_j)$ with
  \begin{align}
      \bm{R}_{\bm{e}^3}(\theta_j) =\begin{bmatrix} \cos (\theta_j) & -\sin (\theta_j) & 0 \\
                \sin (\theta_j) & \cos (\theta_j)  & 0 \\
                0           & 0            & 1\end{bmatrix}, \label{R_z}
  \end{align}
which is the rotation about the coordinate axis ${\bm{e}^3}$ by an angle $\theta_j$. The rotation angle of both tensors is given as $\theta_1 \approx \pi/2$ and $\theta_2 = 0$. The relative orientation of the primary eigenvectors of the symmetric parts $\bm{U}_1$ and $\bm{U}_2$ is again nearly $\pi/2$, i.e., $\sphericalangle (\hat{\bm{n}}^1_1,\hat{\bm{n}}^1_2) \approx\pi/2$. The ellipsoidal representation of the interpolated tensor for this extreme case is depicted in Figure~\ref{fig:Non_Symmetric_Tensor_interpolation_1D}.
\begin{figure}[htb]
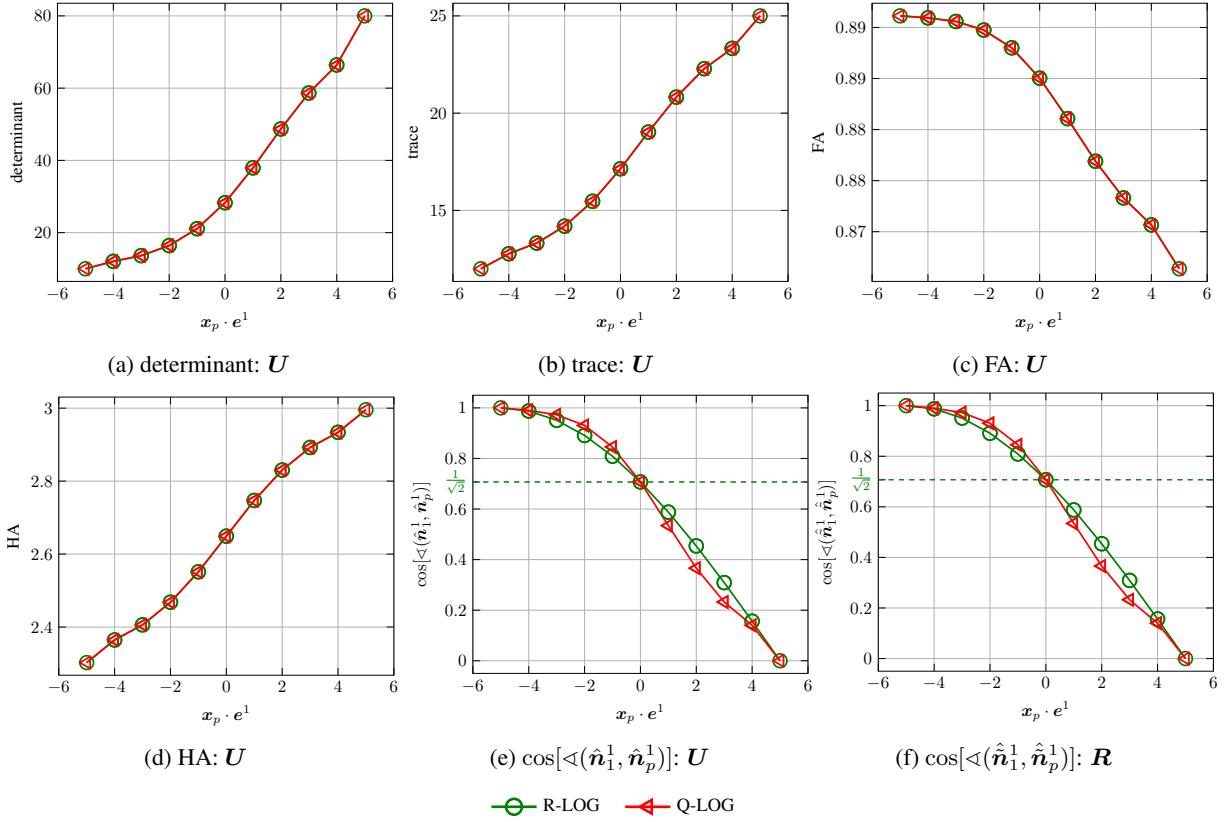

 \centering
  \begin{subfigure}[b]{.32\linewidth}
    \centering
    \scalebox{0.65}{  \input{pictures/invariants__1d_nonsym_determinant.tikz}}
    \caption{determinant: $\bm{U}$} \label{fig:NonSym_1D_determinant} \end{subfigure}%
  \begin{subfigure}[b]{.32\linewidth}
    \centering
    \scalebox{0.65}{   \input{pictures/invariants__1d_nonsym_trace.tikz}}
    \caption{trace: $\bm{U}$} \label{fig:NonSym_1D_trace}
  \end{subfigure}
  \begin{subfigure}[b]{.32\linewidth}
    \centering
    \scalebox{0.65}{  \input{pictures/invariants__1d_nonsym_FA.tikz}}
    \caption{FA: $\bm{U}$ }  \label{fig:NonSym_1D_FA} \end{subfigure} \\ \vspace{4pt}
  \begin{subfigure}[b]{.32\linewidth}
    \centering
    \scalebox{0.65}{  \input{pictures/invariants__1d_nonsym_HA.tikz}}
    \caption{HA: $\bm{U}$} \label{fig:NonSym_1D_HA}
  \end{subfigure}
  \begin{subfigure}[b]{.32\linewidth}
    \centering
    \scalebox{0.65}{  \input{pictures/invariants__1d_nonsym_IA.tikz}}
    \caption{\cosIA: $\bm{U}$} \label{fig:NonSym_1D_cosIA_sym}
  \end{subfigure}
  \begin{subfigure}[b]{.32\linewidth}
    \centering
    \scalebox{0.65}{  \input{pictures/invariants__1d_nonsym_IA_rot.tikz}}
    \caption{$\cos[\sphericalangle(\Hat{\tilde{\bm{n}}}^1_1,\Hat{\tilde{\bm{n}}}^1_p)]$: $\bm{R}$} \label{fig:NonSym_1D_cosIA_rot}
  \end{subfigure}\\
  \vspace{-170pt}
  \begin{subfigure}[b]{1\linewidth}
    \centering
    \input{pictures/invariants__1d_nonsym_legend_final.tikz}
  \end{subfigure}
  \caption{\small Tensor metrics for interpolation between  non-symmetric and symmetric tensor as
    displayed in Figure \ref{fig:Non_Symmetric_Tensor_interpolation_1D}.}
  \label{fig:Invariants_Non_Symmetric_Tensor_interpolation_1D}
\end{figure}
The orientation is gradually interpolated for both rotation interpolation schemes R-LOG and Q-LOG (Figures~\ref{fig:NonSym_1D_rotationvector_Sym} and~\ref{fig:NonSym_1D_quaternion_Sym}). Moreover,
the evolution of the primary eigenvector of $\bm{R}$ as resulting from these two schemes is illustrated in Figures~\ref{fig:NonSym_1D_rotationvector_Rot} and~\ref{fig:NonSym_1D_quaternion_Rot}. The characteristic metrics determinant, trace, FA, and HA of the symmetric part $\bm{U}$ are plotted in Figures~\ref{fig:NonSym_1D_determinant} -~\ref{fig:NonSym_1D_HA}. These plots again demonstrate a monotonic change of anisotropy without swelling,
preserving tensor shape and size. The orientation parameters~\cosIA~and~$\cos[\sphericalangle(\Hat{\tilde{\bm{n}}}^1_1,\Hat{\tilde{\bm{n}}}^1_p)]$~in Figures~\ref{fig:NonSym_1D_cosIA_sym} and~\ref{fig:NonSym_1D_cosIA_rot} describe a gradual and monotonic change of orientation for both tensor components $\bm{U}$ and $\bm{R}$. The interpolated tensors $\bm{U}_p$ and $\bm{R}_p$ at $\bm{x}_p=(0,0)^T$ show a relative orientation of $\approx \pi/4$ with respect to $\bm{U}_1$ and $\bm{R}_1$ at $\bm{x}_1=(-5,0)^T$, i.e.,~$\sphericalangle (\hat{\tilde{\bm{n}}}^1_1,\hat{\tilde{\bm{n}}}^1_p) \approx \pi/4$ and~$\sphericalangle (\hat{{\bm{n}}}^1_1,\hat{{\bm{n}}}^1_p) \approx \pi/4$, respectively (denoted by green dashed lines).
\subsubsection{Interpolation of four tensors} 
\label{example-four-nonsymmetric-tensors}

\begin{figure}[htb]
  \centering
  \begin{subfigure}[b]{.32\linewidth}\hspace{3pt}\vspace{1pt}
    \includegraphics[trim=0cm 5cm 0cm 0cm,clip,width=0.825\textwidth]{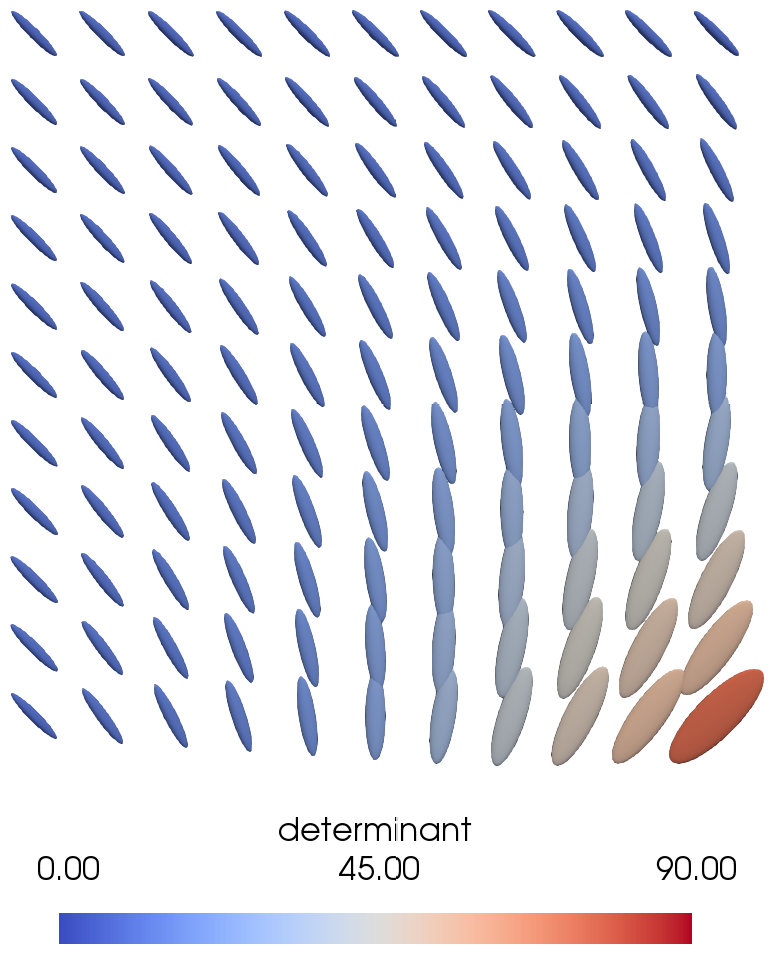}
    \caption{R-LOG:  $\bm{U}$} \label{fig:NonSym_2D_rotationvector_Sym}
  \end{subfigure}
  \centering
  \begin{subfigure}[b]{.32\linewidth}
    \includegraphics[trim=0cm 0cm 0cm 0cm,clip,width=0.825\textwidth]{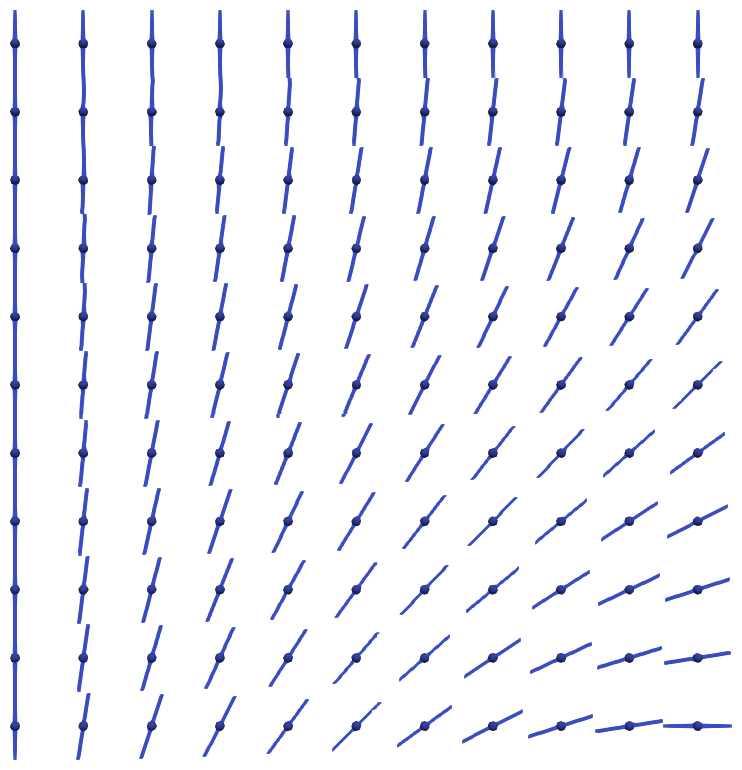}
    \caption{R-LOG:  $\bm{R}$} \label{fig:NonSym_2D_rotationvector_Rot}
  \end{subfigure}
  \centering  \hspace{-1cm}
  \begin{subfigure}[b]{.33\linewidth}
    \includegraphics[trim=0cm 0cm 0cm 0cm,clip,width=1.1\textwidth]{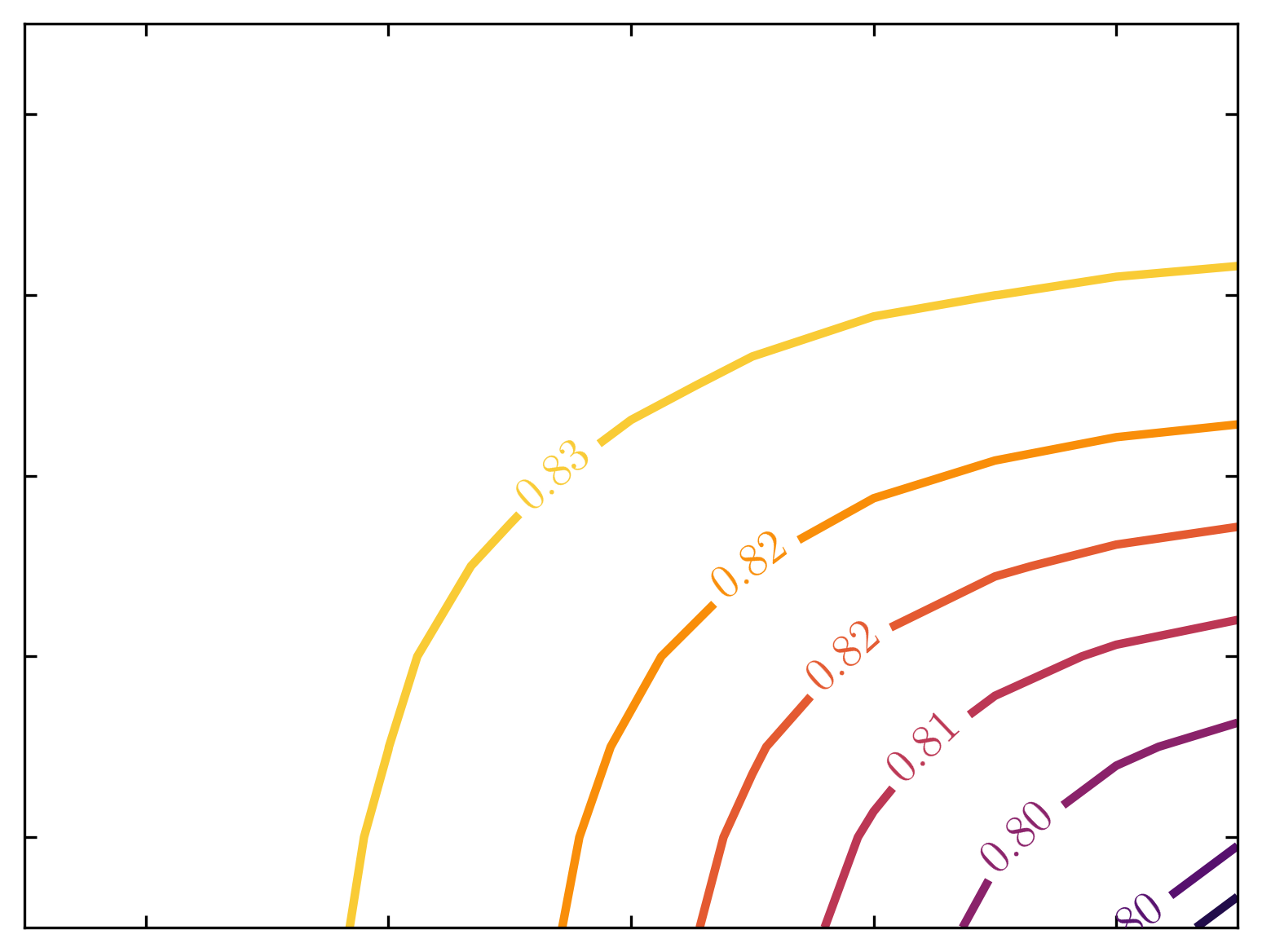}
    \caption{R-LOG: FA - $\bm{U}$} \label{fig:NonSym_2D_rotationvector_FA} \end{subfigure}
  \\ \vspace{4pt}
  \centering
  \begin{subfigure}[b]{.32\linewidth} \hspace{3pt}\vspace{1pt}
    \includegraphics[trim=0cm 5cm 0cm 0cm,clip,width=0.825\textwidth]{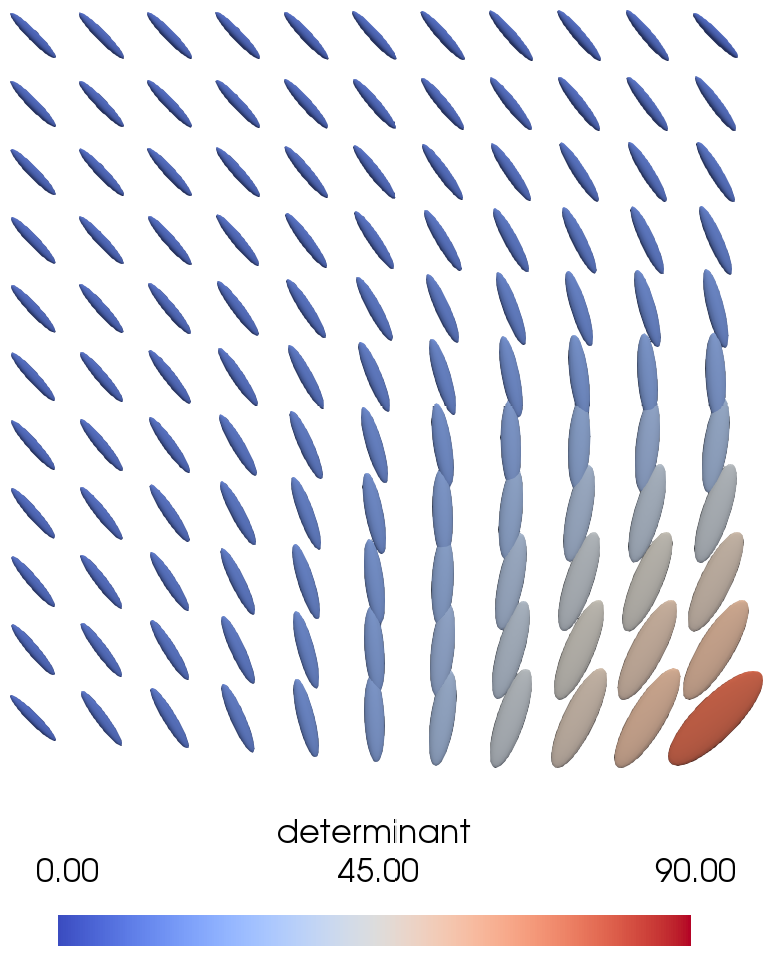}
    \caption{Q-LOG:  $\bm{U}$} \label{fig:NonSym_2D_quaternion_Sym}
  \end{subfigure}
  \centering
  \begin{subfigure}[b]{.32\linewidth}
    \includegraphics[trim=0cm 0cm 0cm 0cm,clip,width=0.825\textwidth]{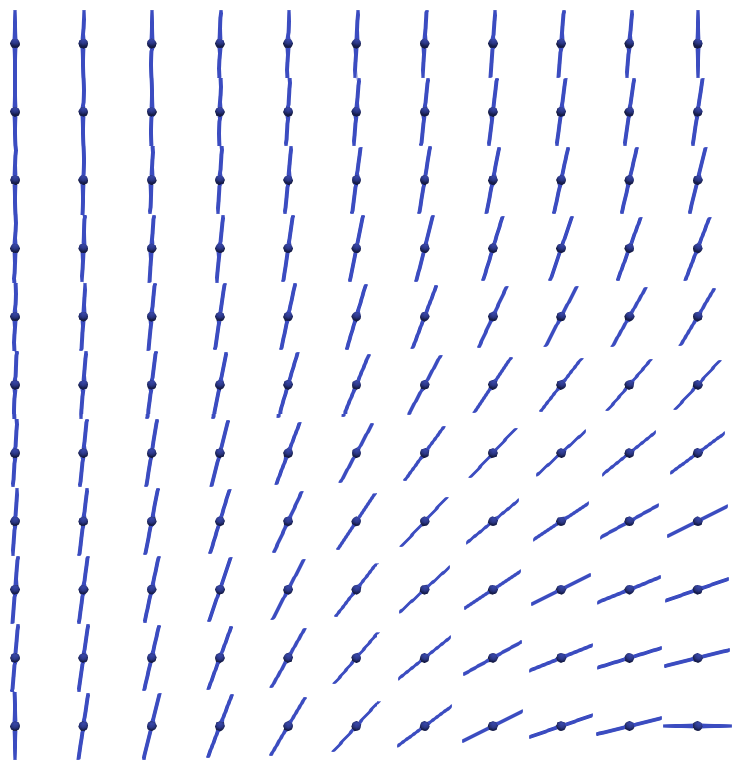}
    \caption{Q-LOG:  $\bm{R}$ } \label{fig:NonSym_2D_quaternion_Rot}
  \end{subfigure}
  \centering
  \hspace{-1cm}
  \begin{subfigure}[b]{.33\linewidth}
    \includegraphics[trim=0cm 0cm 0cm 0cm,clip,width=1.1\textwidth]{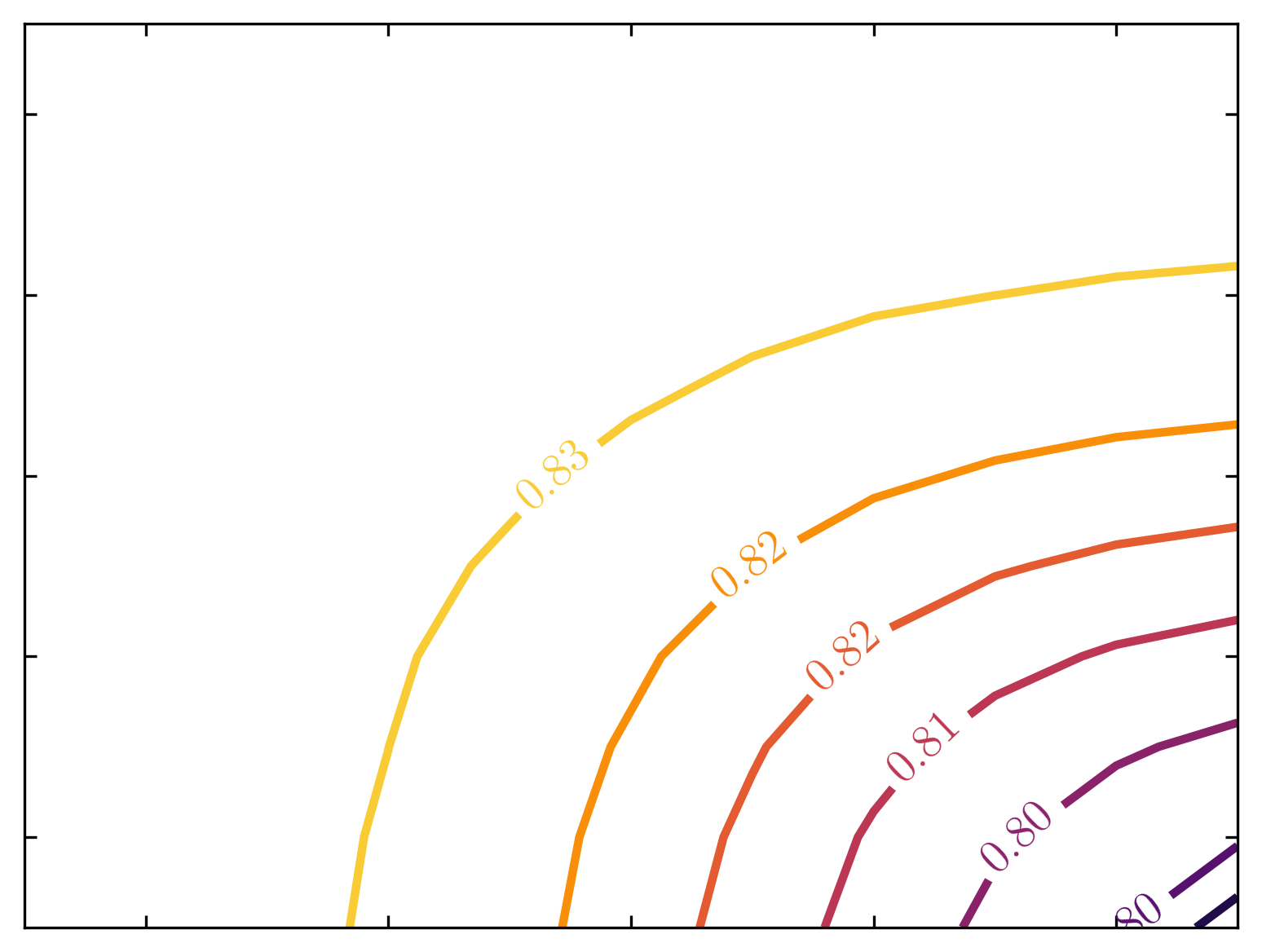}
    \caption{Q-LOG: FA - $\bm{U}$} \label{fig:NonSym_2D_quaternion_FA} \end{subfigure}
    \\
  \begin{subfigure}[b]{\linewidth}\vspace{-90pt} \hspace{0pt}
    \begin{tikzpicture}
      \coordinate [label = above:{$\bm{e}^1$}](x) at (1,0); \coordinate [label =
      right:{$\bm{e}^2$}](y) at (0,1); \coordinate [](C) at (0, 0); \draw[-latex] (C)--(x);
      \draw[-latex] (C)--(y);
    \end{tikzpicture}
  \end{subfigure}\\
  \begin{subfigure}[b]{.3\linewidth}\vspace{5pt}
    \includegraphics[trim=0cm 0cm  0cm 30cm,clip,width=0.9\textwidth]{pictures/quaternion_2d_nonsym_sym.png}
  \end{subfigure}
  \begin{subfigure}[b]{.33\linewidth}
    \includegraphics[trim=0cm 0cm 0cm 100cm,clip,width=1.1\textwidth]{pictures/quaternion_2d_nonsym_sym.png}
  \end{subfigure}
  \begin{subfigure}[b]{.33\linewidth}
    \includegraphics[trim=0cm 0cm 0cm 100cm,clip,width=1.1\textwidth]{pictures/quaternion_2d_nonsym_sym.png}
  \end{subfigure}
  \caption{\small
  Interpolation between three non-symmetric tensors and a symmetric tensor with R-LOG and Q-LOG methods. Subfigures (a) and (d) are the ellipsoidal representation portraying shape and orientation (the color is its determinant) of the symmetric component $\bm{U}$, subfigures (b) and (e) depict the primary eigenvector $\Hat{\tilde{\bm{n}}}^1_j$ of rotation component $\bm{R}$, and subfigures (c) and (f) are the contour plots of the tensor metric FA of $\bm{U}$. Here three identical non-symmetric tensors
  $\bm{T}_{1,2,3} = \bm{R}_{1,2,3} \bm{U}_{1,2,3}$ are placed $\bm{x}_1=(5,5)^T$, $\bm{x}_2=(-5,5)^T$ and
  $\bm{x}_3=(-5,-5)^T$ (see Figure~\ref{fig:problem_setup_2d}). The symmetric part $\bm{U}_{1,2,3}$ is defined by   eigenvalues $\{\lambda^1_{1,2,3}, \lambda^2_{1,2,3}, \lambda^3_{1,2,3}\}=\{7.5, 1.25, 1.0\}$
    with primary eigenvector orientation $\sphericalangle(\bm{e}^1,\hat{\bm{n}}^1_{1,2,3}) \approx
    3\pi/4$ and the rotation part by $\bm{R}_{1,2,3} = \bm{R}_{\bm{e}^3}(\theta_{1,2,3}
    \approx\frac{1}{2}\pi)$ (see~\eqref{R_z}). The fourth tensor $\bm{T}_4$ is symmetric i.e., $\bm{T}_4=\bm{U}_4$ and is located at
    $\bm{x}_4=(5,-5)^T$.
    Tensor $\bm{U}_4$ is defined by 
    eigenvalues $\{\lambda^1_{4}, \lambda^2_{4}, \lambda^3_{4}\}=\{15, 5, 1 \}$ and primary eigenvector orientation
    $\sphericalangle(\bm{e}^1,\hat{\bm{n}}^1_{4}) =\pi/4$.}  \label{fig:NonSym_2D}
\end{figure}

In this section, the interpolation studies for non-symmetric tensors are extended to four data points placed at the four corner points of a square with $x,y \in [-5;5]$  (see Figure~\ref{fig:problem_setup_2d}). Three of these four tensors $\bm{T}_{1,2,3}$ are non-symmetric and identical, which are located at $\bm{x}_1=(5,5)^T$, $\bm{x}_2=(-5,5)^T$ and $\bm{x}_3=(-5,-5)^T$. The symmetric part $\bm{U}_{1,2,3}$ of these tensors is defined with
eigenvalues $\{\lambda^1_{1,2,3}, \lambda^2_{1,2,3}, \lambda^3_{1,2,3}\}=\{7.5, 1.25, 1.0\}$ and primary eigenvector orientation $\sphericalangle (\bm{e}^1,\hat{\bm{n}}^1_{1,2,3})\approx 3\pi/4$. The rotation part of these tensors is given by $\bm{R}_{\bm{e}^3}$ according to~\eqref{R_z} with $\theta_{1,2,3} \approx \pi/2$. The fourth tensor is symmetric $\bm{T}_{4}=\bm{U}_{4}$ and located at $\bm{x}_4=(5,-5)^T$ with eigenvalues $\{\lambda^1_{4}, \lambda^2_{4}, \lambda^3_{4}\}= \{15, 5,1.0\}$ and primary eigenvalue orientation $\sphericalangle (\bm{e}^1,\hat{\bm{n}}^1_{4})=\pi/4$. The ellipsoidal representation of the interpolated tensors is displayed in Figure~\ref{fig:NonSym_2D}. Again, the results from rotation vector-based (R-LOG) and quaternion-based (Q-LOG) rotation interpolation are very similar. The tensor orientation is smoothly interpolated for both, the symmetric part $\bm{U}$ and the rotation part $\bm{R}$. Again, the shape and size of $\bm{U}$ are preserved during interpolation. This observation is confirmed by Figure~\ref{fig:NonSym_2D} (third column), which again shows a monotonic evolution of FA across the interpolation domain. 

\subsection{Convergence study: Application to nonlinear continuum mechanics}\label{sec:application_cont_mechanics}

\subsubsection{General problem setup}

\begin{figure}[htb]
  \centering
\scalebox{0.9} {\input{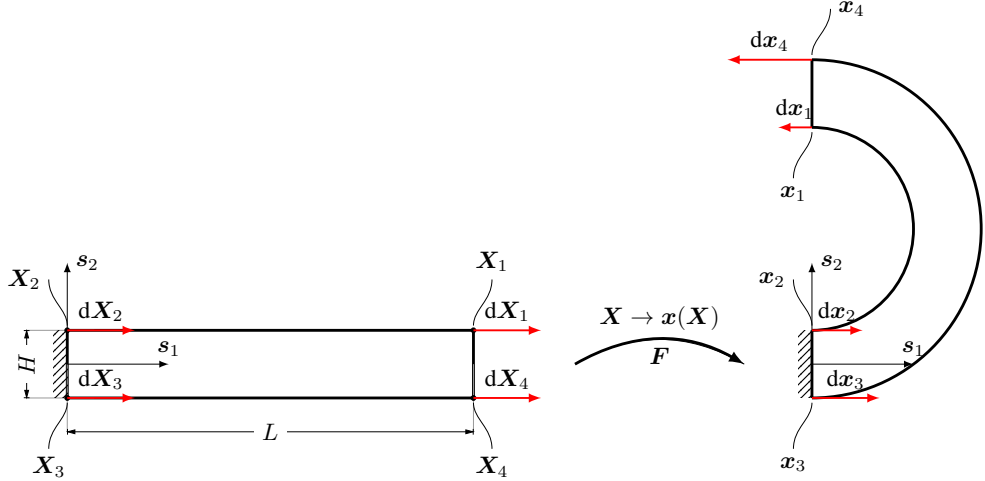}}
  \caption{\small Kinematic quantities of beam deformation. Consider a beam of length $L$ and height $H$ clamped at the edge $s_1=0$. The initial configuration (left) is described by the material coordinates $\bm{X}$. The deformed configuration (right) is represented by the spatial coordinates $\bm{x}$. The red arrows symbolize the fibers where the vectors $\textrm{d} \bm{X}_j$ represent material fibers, whereas the spatial fibers are denoted by $\textrm{d} \bm{x}_j$. Here the deformation gradient $\bm{F}=\frac{\partial \bm{x}}{\partial \bm{X}}$ describes the mapping from initial configuration to deformed configuration governed by  $\textrm{d} \bm{x} = \bm{F} \cdot \textrm{d} \bm{X}$.}
  \label{fig:beam_theory}
\end{figure}

In this section, we explore the application of the proposed tensor interpolation schemes in the context of nonlinear continuum mechanics, and investigate the convergence behavior of the interpolation methods for general non-symmetric tensors. In nonlinear continuum mechanics the deformation gradient $\bm{F}$ is the fundamental kinematic variable describing the mapping (deformation) of an infinitesimal material fiber $\textrm{d}\bm{X}$ (red arrows in Figure \ref{fig:beam_theory}) in the initial state (material configuration) to its new position  $\textrm{d}\bm{x}$ in the current deformed state (spatial configuration): 
\begin{align}
\textrm{d}\bm{x}     = \bm{F} \cdot \textrm{d}\bm{X}  \quad  \textrm{with} \quad   \bm{F} =  \frac{\partial  \bm{x}}{\partial\bm{X}}.
\end{align}
In general, for an arbitrary deformation, $\bm{F}$ is non-symmetric and strictly positive definite. The right polar decomposition of the deformation gradient reads $\bm{F}=\bm{R}\bm{U}$, where $\bm{R}$ represents the rotation and $\bm{U}$ the stretch of a material fiber. Thus, the symmetric positive definite tensor $\bm{U}$ is also denoted as stretch tensor.

In the following, the deformation of a slender body with length $L$ and height $H=L/10$ is considered (see Figure \ref{fig:beam_theory}). In continuum mechanics, the deformation of such bodies can be described by means of beam theories. Based on the so-called geometrically exact beam theory, the deformation gradient associated with such a slender body can be formulated as~\cite{Meier2019}:
\begin{align}
  \bm{F} = [\bm{g}_1(s_1) + \bm{\gamma}(s_1) - k(s_1) \ s_2 \ \bm{g}_1(s_1)] \otimes \bm{g}_{01}(s_1) + \bm{g}_1(s_1) \otimes \bm{g}_{02}(s_1). \label{beam_equation}
\end{align}
Here, $s_1 \in [0;L]$ is an arc-length coordinate describing the beam centerline, and $\bm{g}_{01}$, $\bm{g}_{02}$ as well as $\bm{g}_{1}$, $\bm{g}_{2}$ are local basis vectors in the initial and deformed configuration, respectively. Moreover, $k(s_1)$ is the curvature of the beam centerline and the strain vector $\bm{\gamma}$ is given by
\begin{align}
    \bm{\gamma}(s_1) = \eta(s_1) \bm{g}_1 + \xi(s_1) \bm{g}_2 , \label{beam_equation_2}
\end{align}
where $\eta$ and $\xi$ are the shear and axial strain, respectively.
In this study, we consider four data points $\bm{X}_1,\ \bm{X}_2,\ \bm{X}_3,$ and $\bm{X}_4$ at initial positions $(L,H/2)^T,\ (0,H/2)^T,\ (0,-H/2)^T,$ and $(L,-H/2)^T$ (see Figure~\ref{fig:beam_theory}) defining a rectangle of size $L \times H$. The performance of the proposed interpolation methods is quantified by considering the error between the interpolated deformation gradient and the corresponding analytical value according to~\eqref{beam_equation} evaluated at the center of the beam $(L/2,0)^T$. For comparison, we consider all interpolation methods defined in Section~\ref{sec:tensor_interpolation}, i.e., R-LOG, Q-LOG, R-MLS, Q-MLS, R-LOGMLS, and  Q-LOGMLS.
\subsubsection{Alternative strategy for eigenvector assignment}
In the following numerical studies, deformed configurations will be considered, where the relative rotation angle between the deformation gradients at positions $\bm{X}_2,\ \bm{X}_3$ and the deformation gradients at positions $\bm{X}_1,\ \bm{X}_4$ are identical or close to $\pi$, and therefore the strategy for eigenvector assignment based on eigenvalue magnitude as presented in Section~\ref{sec:impose_uniqueness} is no longer applicable. However, in examples of nonlinear continuum mechanics, in particular when considering slender structures, an alternative strategy for eigenvector assignment seems to be more natural. Thereto, we consider eigenvectors $\hat{\bm{N}}$ associated with the material configuration, which - according to the theory of nonlinear continuum mechanics - are related to the eigenvectors $\hat{\bm{n}}$ in the spatial configuration via the rotation part $\bm{R}$ of the deformation gradient according to $\hat{\bm{N}}=\bm{R}^T\hat{\bm{n}}$. In the initial "straight beam" configuration, there are two directions that can be distinguished from a mechanical point of view: the beam's length direction $\bm{s}_1$, and the beam's transverse direction $\bm{s}_2$. For such slender structures, the normal strains in the beam length direction typically dominate the overall deformation. This means, an eigenvector in the material configuration whose orientation is identical or close to the $\bm{s}_1$-direction exists. Thus, for each data point, the eigenvector that encloses the smaller relative angle with respect to the global $\bm{s}_1$-direction, is denoted as $\bm{n}^1_j$, and the eigenvector with the larger relative angle is denoted as $\bm{n}^2_j$. In summary, when applying the proposed tensor interpolation schemes to slender structures, the basic idea is the following: The relative rotation between material eigenvectors is typically small even when the relative rotation between spatial eigenvectors is very large, which enables a unique eigenvector assignment even for large deformation problems.
\subsubsection{Numerical studies}

In a first step of the numerical studies, a special case of the deformation~\eqref{beam_equation} is considered, where axial and shear strains vanish, i.e., 
 $\eta=0$ and $\xi=0$ (see~\eqref{beam_equation}) and the curvature is constant along the beam centerline, i.e., $k(s_1)=k=\pi/L$. In this special case, the beam centerline in the deformed configuration represents a semi circle as shown in Figure~\ref{fig:beam_theory}. Accordingly, we get the following deformations of the fibers at the four considered data points: $\text{d} \bm{X}_2$ - negative stretch; $\text{d} \bm{X}_3$ - positive stretch; $\text{d} \bm{X}_1$ - negative stretch plus rotation by $\pi$; $\text{d} \bm{X}_4$ - positive stretch plus rotation by $\pi$. Moreover, the analytical solution for the deformation gradient at the beam center point $(L/2,0)^T$ is a pure rotation tensor (i.e., no stretch) with rotation angle $\pi/2$ as shown on the very left of Table~\ref{table:no_shear_axial_strains}. According to this table, the proposed interpolation schemes with MLS interpolation of the eigenvalues, i.e., R-MLS and Q-MLS, can exactly represent this analytical solution. This result is remarkable, considering the large relative distance between the data points and, as a result, the very large relative rotations between the associated eigenvectors at these data points. The proposed interpolation schemes with LOG or LOGMLS interpolation of the eigenvalues, still represent the analytical solution in very good approximation, even though not exactly. For comparison, on the very right of Table~\ref{table:no_shear_axial_strains}, also the interpolated deformation gradient resulting from an Euclidean interpolation is shown. Accordingly, the Euclidean interpolation results in a zero tensor, i.e., a deformation gradient that leads to a vanishing fiber length in the current configuration. This result is non-physical, thus the Euclidean interpolation of data points with such large separations cannot be recommended.
\begin{table}[htb]
    \centering
    \caption{Interpolated deformation gradient $\bm{F}$ at $(L/2,0)^T$ for $k= \pi/L$ and $\gamma = 0$}
    \begin{adjustbox}{width=\columnwidth,center}
    \begin{tabular}{lccccccccc b{2pt}}
    \hline 
    Analytic solution &R-LOG &Q-LOG&R-MLS &Q-MLS &R-LOGMLS &Q-LOGMLS &E\\[0.25ex]
    \hline \\[-2ex]
    $\begin{bmatrix}
        0 & -1.0 \\
        1.0 & 0 
    \end{bmatrix}$&$\begin{bmatrix}
            0 & -1.0 \\
            0.988 & 0 
            \end{bmatrix}$&$\begin{bmatrix}
                    0 & -1.0 \\
                    0.988 & 0 
                    \end{bmatrix}$&$\begin{bmatrix}
                        0 & -1.0 \\
                        1.0 & 0 
                        \end{bmatrix}$&$\begin{bmatrix}
                        0 & -1.0 \\
                        1.0 & 0 
                        \end{bmatrix}$ &$\begin{bmatrix}
                            0 & -1.0 \\
                            0.988 & 0 
                            \end{bmatrix}$ &$\begin{bmatrix}
                                0 & -1.0 \\
                                0.988 & 0 
                                \end{bmatrix}$ &$\begin{bmatrix}
                                    0 & 0 \\
                                    0 & 0
                                    \end{bmatrix}$\\  [2ex]
    \hline 
    \end{tabular}
    \label{table:no_shear_axial_strains}
\end{adjustbox}
\end{table}

In a second step, we study the spatial convergence behavior of the proposed interpolation schemes by subdividing the initial beam domain, denoted as interpolation length $h$, successively by a factor of 2, giving rise to the sample sizes $h \in h*\{\ 2^{-1}\dots,\ 2^{-7} \}$. To achieve a more general deformation state, we consider non-vanishing axial and shear strains according to $\eta,\xi = 0.15 s_1$ and a non-constant curvature according to $k = 0.15 s_1$. Moreover, a 2D bilinear polynomial (see Appendix~\ref{appendix:polynomial_basis_bilin}) is adopted for the least squares approximation of eigenvalues and rotation vector. In the following, the errors of the individual tensors resulting from the decomposition $\bm{F}\!\!=\!\!\bm{R}\bm{Q}^T \!\! \bm{\Lambda} \bm{Q}$ are considered. The error of an interpolated tensor $\bm{T}$ is computed as the $L^2$ norm of the components, i.e., $\sqrt{\sum^3_{i,j=1} (T_{ij} - T^\text{ana}_{ij} )^2}$, where $\bm{T}^\text{ana}$ is the analytical solution. For the eigenvalues, i.e., the entries of the tensor $\bm{\Lambda}$, the absolute value of the difference from the analytical solution is evaluated individually. In Figure~\ref{fig:Convergence_lin}, the double-logarithmic plot of these errors over sampling size $h$ is shown for the proposed interpolation approaches. In particular, the error in the primary interpolation fields $\bm{R}$, $\bm{Q}$ and $\bm{\Lambda}$ is shown in Figure~\ref{fig:convergenace_lin_eigenvalue}-\ref{fig:convergenace_lin_eigenrotation}. The error in the stretch tensor $\bm{U}$ is depicted in Figures~\ref{fig:convergenace_lin_stretch_log}-\ref{fig:convergenace_lin_stretch_logmls}. Finally, the error in the resulting deformation gradient is portrayed in Figures~\ref{fig:convergenace_lin_defgrad_mls}-\ref{fig:convergenace_lin_defgrad_logmls}. Both the quaternion-based and rotation vector-based rotation interpolations combined with either the logarithmic,
moving least squares, or logarithmic moving least squares eigenvalue interpolation result in a quadratic $\mathcal{O}(h^2)$ convergence order, which is the expected result for first-order interpolation schemes.
\begin{figure}[htb]
  \centering
  \begin{subfigure}[b]{.33\linewidth}
    \scalebox{0.6}{   \input{pictures/convergenace-test-eigenvalue-lin.tikz}}
    \caption{$\text{I}(\lambda^i)$: MLS, LOG, and LOGMLS} \label{fig:convergenace_lin_eigenvalue}
  \end{subfigure}
  \centering
  \begin{subfigure}[b]{.33\linewidth}
    \scalebox{0.6}{   \input{pictures/convergenace-test-rotation-lin.tikz}}
    \caption{$\text{I}(\bm{R})$: R and Q} \label{fig:convergenace_lin_rotation}
  \end{subfigure}
  \centering
  \begin{subfigure}[b]{.33\linewidth}
    \scalebox{0.6}{  \input{pictures/convergenace-test-eigenrotation-lin.tikz}}
    \caption{$\text{I}(\bm{Q})$: R and Q} \label{fig:convergenace_lin_eigenrotation} \end{subfigure} \\ \vspace{6pt}
  \centering
  \begin{subfigure}[b]{.33\linewidth}
    \scalebox{0.6}{  \input{pictures/convergenace-test-stretch-MLS-lin.tikz}}
    \caption{$\text{I}(\bm{U})$: R-MLS and Q-MLS} \label{fig:convergenace_lin_stretch_mls}
  \end{subfigure}
  \centering
  \begin{subfigure}[b]{.33\linewidth}
    \scalebox{0.6}{   \input{pictures/convergenace-test-stretch-log-lin.tikz}}
    \caption{$\text{I}(\bm{U})$: R-LOG and Q-LOG} \label{fig:convergenace_lin_stretch_log}
  \end{subfigure}
  \centering
  \begin{subfigure}[b]{.33\linewidth}
    \scalebox{0.6}{   \input{pictures/convergenace-test-stretch-LogMLS-lin.tikz}}
    \caption{$\text{I}(\bm{U})$: R-LOGMLS and Q-LOGMLS} \label{fig:convergenace_lin_stretch_logmls}
  \end{subfigure} \\\vspace{6pt}
  \centering
  \begin{subfigure}[b]{.33\linewidth}
    \scalebox{0.6}{  \input{pictures/convergenace-test-defgrad-MLS-lin.tikz}}
    \caption{$\text{I}(\bm{F})$: R-MLS and Q-MLS} \label{fig:convergenace_lin_defgrad_mls}
  \end{subfigure}
  \centering
  \begin{subfigure}[b]{.33\linewidth}
    \scalebox{0.6}{  \input{pictures/convergenace-test-defgrad-log-lin.tikz}}
    \caption{$\text{I}(\bm{F})$: R-LOG and Q-LOG} \label{fig:convergenace_lin_defgrad_log}
  \end{subfigure}
  \centering
  \begin{subfigure}[b]{.33\linewidth}
    \scalebox{0.6}{   \input{pictures/convergenace-test-defgrad-LogMLS-lin.tikz}}
    \caption{$\text{I}(\bm{F})$: R-LOGMLS and Q-LOGMLS} \label{fig:convergenace_lin_defgrad_logmls}
  \end{subfigure}
  \caption{\small Application to nonlinear continuum mechanics: Convergence of the proposed interpolation schemes based on four data points. Here, the operator~$\text{I}(\cdot)$ denotes the interpolation error of a given quantity in the argument.}
  \label{fig:Convergence_lin}
\end{figure}

Finally, the example is extended by considering 8 data points for interpolation. The additional four data points are located at the edge mid-points of the rectangle formed by the four original data points. Now, we employ 2D quadratic Lagrange polynomials (see Appendix~\ref{appendix:polynomial_basis_quad}) for the moving least squares approaches underlying the eigenvalue and rotation vector-based rotation interpolation. The results for the different interpolation schemes are showcased in Figures~\ref{fig:convergenace_quad_eigenvalue}-~\ref{fig:convergenace_quad_defgrad_logmls}. Importantly, the MLS and LOGMLS approaches used for eigenvalue and rotation interpolation result in a cubic convergence rate $\mathcal{O}(h^3)$ as expected for the employed quadratic polynomials. In contrast, no higher-order schemes are available for the logarithmic weighted average (LOG) used for eigenvalue interpolation and the spherical weighted average used for quaternion-based (Q) rotation interpolation. Thus, also in the case of eight data points these schemes only yield a convergence order of two. Therefore, all combinations of interpolation approaches involving either of these two schemes, i.e., Q-LOG, Q-MLS, Q-LOGMLS, and R-LOG, only result in a quadratic convergence order for $\bm{F}$ and $\bm{U}$. In contrast, rotation vector-based methods in combination with MLS or LOGMLS, i.e. , R-MLS and R-LOGMLS, give cubic convergence.

\begin{figure}[htb]
  \centering
  \begin{subfigure}[b]{.33\linewidth}
    \scalebox{0.6}{   \input{pictures/convergenace-test-eigenvalue-quad.tikz}}
    \caption{$\text{I}(\lambda^i)$: MLS, LOG, and LOGMLS} \label{fig:convergenace_quad_eigenvalue}
  \end{subfigure}
  \centering
  \begin{subfigure}[b]{.33\linewidth}
    \scalebox{0.6}{   \input{pictures/convergenace-test-rotation-quad.tikz}}
    \caption{$\text{I}(\bm{R})$: R and Q} \label{fig:convergenace_quad_rotation}
  \end{subfigure}
  \centering
  \begin{subfigure}[b]{.33\linewidth}
    \scalebox{0.6}{   \input{pictures/convergenace-test-eigenrotation-quad.tikz}}
    \caption{$\text{I}(\bm{Q})$: R and Q} \label{fig:convergenace_quad_eigenrotation}
  \end{subfigure}
  \\\vspace{6pt}
  \centering
  \begin{subfigure}[b]{.33\linewidth}
    \scalebox{0.6}{   \input{pictures/convergenace-test-stretch-MLS-quad.tikz}}
    \caption{$\text{I}(\bm{U})$: R-MLS and Q-MLS} \label{fig:convergenace_quad_stretch_mls}
  \end{subfigure}
  \centering
  \begin{subfigure}[b]{.33\linewidth}
    \scalebox{0.6}{   \input{pictures/convergenace-test-stretch-log-quad.tikz}}
    \caption{$\text{I}(\bm{U})$: R-LOG and Q-LOG} \label{fig:convergenace_quad_stretch_log}
  \end{subfigure}
  \centering
  \begin{subfigure}[b]{.33\linewidth}
    \scalebox{0.6}{   \input{pictures/convergenace-test-stretch-LogMLS-quad.tikz}}
    \caption{$\text{I}(\bm{U})$: R-LOGMLS and Q-LOGMLS} \label{fig:convergenace_quad_stretch_logmls}
  \end{subfigure}
  \\\vspace{6pt}
  \centering
  \begin{subfigure}[b]{.33\linewidth}
    \scalebox{0.6}{   \input{pictures/convergenace-test-defgrad-MLS-quad.tikz}}
    \caption{$\text{I}(\bm{F})$: R-MLS and Q-MLS} \label{fig:convergenace_quad_defgrad_mls}
  \end{subfigure}
  \centering
  \begin{subfigure}[b]{.33\linewidth}
    \scalebox{0.6}{   \input{pictures/convergenace-test-defgrad-log-quad.tikz}}
    \caption{$\text{I}(\bm{F})$: R-LOG and Q-LOG} \label{fig:convergenace_quad_defgrad_log}
  \end{subfigure}
  \centering
  \begin{subfigure}[b]{.33\linewidth}
    \scalebox{0.6}{   \input{pictures/convergenace-test-defgrad-LogMLS-quad.tikz}}
    \caption{$\text{I}(\bm{F})$: R-LOGMLS and Q-LOGMLS} \label{fig:convergenace_quad_defgrad_logmls}
  \end{subfigure}
  \caption{\small Application to nonlinear continuum mechanics: Convergence of the proposed interpolation schemes based on eight data points. Here, the operator~$\text{I}(\cdot)$ denotes the interpolation error of a given quantity in the argument.}
  \label{fig:convergenace_quad}
\end{figure}
\section{Conclusion}\label{sec:Conclusion}
In the present contribution, novel interpolation schemes for general, i.e., symmetric or non-symmetric, invertible square tensors have been proposed, relying on a combined polar and spectral decomposition of the tensor data, followed by an individual interpolation of the resulting rotation and eigenvalue tensors. For rotation interpolation, two different schemes based on either relative rotation vectors (R) or quaternions (Q), have been considered. For eigenvalue interpolation, three different schemes based on either the logarithmic weighted average (LOG), moving least squares (MLS) or a novel approach denoted as logarithmic moving least squares (LOGMLS) have been considered. Altogether, these schemes resulted in six possible interpolation approaches denoted as R-LOG, R-MLS, R-LOGMLS, Q-LOG, Q-MLS, and Q-LOGMLS.

Based on analytical studies and selected numerical examples, the R-LOGMLS approach is recommend for future application, as it provides the following desirable properties:
\begin{enumerate}
    \item The orthonormality of the rotation tensors is preserved.
    \item The positive definiteness of the eigenvalue tensor is preserved.
    \item The anisotropy of tensors is preserved, i.e., swelling is avoided.
    \item Tensor invariants such as trace, determinant, Hilbert's anisotropy (HA) and fractional anisotropy (FA) are smoothly and monotonically interpolated.
    \item Scaling and rotational invariance (objectivity) is guaranteed.
    \item Interpolation of an arbitrary number of data points is possible.
    \item Higher-order interpolation is possible.
    \item Consistent spatial convergence orders are observed.
\end{enumerate}
As an alternative, the R-LOG approach is recommended for examples where a monotonic interpolation of the eigenvalues is important, but higher-order interpolations are not required. This means, the R-LOG approach only fulfills the first six of the aforementioned eight properties. However, it does not only preserve positive definiteness of the eigenvalue interpolation but also monotonicity. Based on selected numerical examples, it is demonstrated that well-established approaches such as Euclidean, Log-Euclidean, Cholesky and Log-Cholesky interpolation do typically not fulfill the important properties 1-5.

The proposed schemes are very general in nature and suitable for the interpolation of general invertible second-order square tensors independent of the specific application. In our future research, we plan to apply the developed approaches in the context of remeshing and adaptive finite element discretizations for complex problems of nonlinear continuum mechanics with inelastic constitutive behavior, which requires the consistent interpolation of tensor-valued history data (e.g., the deformation gradient associated with the inelastic part of the deformation) for the transfer between coarse and fine mesh. 
\section*{Acknowledgments}
The authors acknowledge  the financial support from the European Union's Horizon 2020 research and innovation programme under the Marie Skłodowska-Curie grant agreement No 764636.

\setcounter{equation}{0}
\renewcommand{\theequation}{\Alph{section}.\arabic{equation}}
\setcounter{figure}{0}
\renewcommand{\thefigure}{\Alph{section}.\arabic{figure}}
\appendix
\section{Calculation of geodesic distance}
\label{appendix:d_S3}

In the following, the expression for the geodesic distance $\textrm{d}_{\mathbb{S}^3}(\hat{\bm{q}}_1,\hat{\bm{q}}_2)= || \ln (\hat{\bm{q}}_1^{-1} \hat{\bm{q}}_2) ||$ between two quaterions $\hat{\bm{q}}_1$ and $\hat{\bm{q}}_2$ as defined in~\eqref{SWA} shall be further simplified. Thereto, let us consider the logarithm of a general quaternion $\hat{\bm{q}}$ with scalar part $q$ and vector part $\bm{q}$, which is defined as 
\begin{align}
    \ln{(\hat{\bm{q}})}:=\ln{(||\hat{\bm{q}}||)} + \arccos{(q/||\hat{\bm{q}}||)} \frac{\bm{q}}{||\bm{q}||}. \label{lnq-1}
\end{align}
Next, let us consider a unit quaternion $\hat{\bm{q}} = \cos (\theta/2) + \sin (\theta/2) \bm{e}_{\bm{\theta}}$ (with $||\bm{e}_{\bm{\theta}}||=1$) according to~\eqref{unit-quaternion}, with scalar part $q=\cos (\theta/2)$, vector part $\bm{q}=\sin (\theta/2) \bm{e}_{\bm{\theta}}$, and associated rotation vector $\theta \bm{e}_{\bm{\theta}}$. Since only rotation vectors within $\theta \in [0;\pi]$ can be uniquely extracted from a rotation tensor, we will restrict ourselves to this angular range in the following. When considering a unit quaternion, the norm of~\eqref{lnq-1} simplifies to 
\begin{align}
    ||\ln{(\hat{\bm{q}})}||=\arccos{(\cos (\theta/2))}||\bm{e}_{\bm{\theta}}||= \theta/2 \quad \text{with} \quad \theta \in [0;\pi]. \label{lnq-2}
\end{align}
According to~\eqref{compound-quaternion}, the quaternion product $\hat{\bm{q}}_1^{-1} \hat{\bm{q}}_2$ results in a quaternion $\hat{\bm{q}}_{21}$ associated with the relative rotation tensor $\bm{R}_{21}=\bm{R}_{1}^T\bm{R}_{2}$. When inserting this relation into~\eqref{lnq-2}, we get:
\begin{align}
    ||\ln{(\hat{\bm{q}}_1^{-1} \hat{\bm{q}}_2)}||=||\ln{(\hat{\bm{q}}_{21})}||= \theta_{21}/2 \,\,\, \text{with} \,\,\, \theta_{21}=||\bm{\theta}_{21}|| \in [0;\pi], \,\,\, \bm{R}(\bm{\theta}_{21})=\bm{R}(\hat{\bm{q}}_1)^T\bm{R}(\hat{\bm{q}}_2), \label{lnq-3}
\end{align}
which is the expression stated in~\eqref{SWA-simplification}. To derive an alternative expression for the relative angle $\theta_{21}$ based on the quaternions $\hat{\bm{q}}_1$ and $\hat{\bm{q}}_2$, we first realize that based on~\eqref{quaternion-product} the scalar part of the quaternion product $\hat{\bm{q}}_1^{-1}\hat{\bm{q}}_2$ is given by
  \begin{align}
  \text{Re}(\hat{\bm{q}}_1^{-1}\hat{\bm{q}}_2) = q_1 q_2 + \bm{q}_1 \cdot \bm{q}_2 = \hat{\bm{q}}_1 \cdot \hat{\bm{q}}_2, \label{quaternion-product-Re}
\end{align}
with $\hat{\bm{q}}_1^{-1}=q_1 - \bm{q}_1$ and $\hat{\bm{q}}_2=q_2 + \bm{q}_2$. Here, the operator $\text{Re}()$ is defined to extract the scalar (or real) part of a quaternion. On the other hand, the scalar part of the relative quaternion $\hat{\bm{q}}_{21} = \cos (\theta_{21}/2) + \sin (\theta_{21}/2) \bm{e}_{\bm{\theta}_{21}}$ is given by $\cos (\theta_{21}/2)$. Equalizing these to expressions yields:
  \begin{align}
  \bm{q}_1 \cdot \bm{q}_2 = \cos (\theta_{21}/2). \label{quaternion-product-Re2}
\end{align}
With this result, expression~\eqref{lnq-3} can be alternatively formulated as
\begin{align}
    ||\ln{(\hat{\bm{q}}_1^{-1} \hat{\bm{q}}_2)} ||= \theta_{21}/2 = \arccos{(\bm{q}_1 \cdot \bm{q}_2)} \,\,\, \text{with} \,\,\, \theta_{21}= \in [0;\pi].
\end{align}

\section{2D polynomial basis}\label{appendix:polynomial_basis}
\subsection{Bilinear polynomial basis } \label{appendix:polynomial_basis_bilin}
A 2D bilinear polynomial basis reads
\begin{align}
    \bm{p} = [1\ x\ y\ xy].
\end{align}
A bilinear function is plotted in Figure~\ref{fig:bilinear_basis}. The figure shows a surface created by the function $z(x,y)=1+x+y+xy$ in $x,y \in [-1;1]$. 
\subsection{Quadratic polynomial basis }\label{appendix:polynomial_basis_quad}
\begin{figure}[htb]
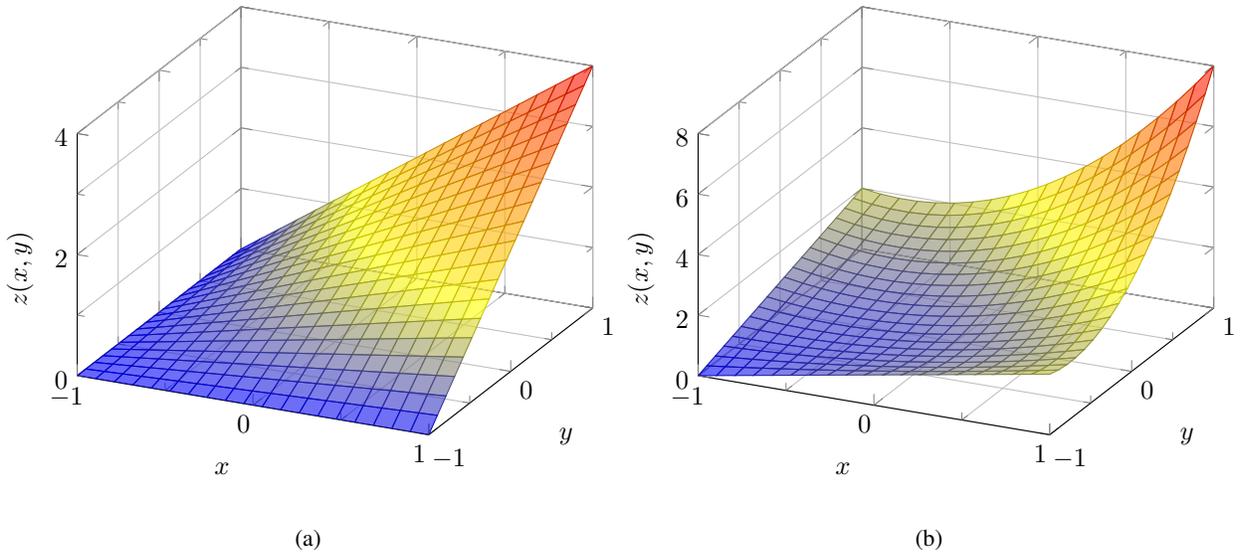

  \centering
  \begin{subfigure}[b]{.5\textwidth}
    \input{pictures/ploynomial_basis_2d_bilinear.tikz}
    \caption{} \label{fig:bilinear_basis} \end{subfigure}%
  \begin{subfigure}[b]{.5\textwidth}
    \input{pictures/ploynomial_basis_2d_qadartic.tikz}
    \caption{} \label{fig:quadratic_basis} \end{subfigure}%
  \caption{\small Polynomial basis functions: (a) bilinear surface defined by the function $z(x,y)=1+x+y+xy$ and (b) quadratic surface generated by the function $z(x,y)=1+x+y+x^2+y^2+xy+x^2y+xy^2$ in $x,y \in [-1;1]$.}
  \label{fig:polynomial_basis}
\end{figure}

A 2D quadratic polynomial basis can be defined as
\begin{align}
    \bm{p} = [1\ x\ y\ x^2\ y^2\  xy\  x^2y\ xy^2].
\end{align}
A quadratic function is shown in Figure~\ref{fig:quadratic_basis}. The figure depicts  surface defined by the function $z(x,y)=1+x+y+x^2+y^2+xy+x^2y+xy^2$ in $x,y \in [-1;1]$.

\bibliographystyle{unsrt}
\bibliography{references}

\end{document}